\documentclass[%
 reprint,
superscriptaddress,
 amsmath,amssymb,
 aps,
]{revtex4-2}

\usepackage{amsmath}
\usepackage{graphicx,epstopdf}
\usepackage{dcolumn}
\usepackage{bm}
\usepackage{xcolor}
\usepackage{float}
\usepackage{enumerate}
\usepackage{natbib}
\usepackage{multirow}
\begin{document}

\title{Significant four-phonon scattering and its heat transfer implications in crystalline Ge$_2$Sb$_2$Te$_5$}

\author{Kanka Ghosh}
\email{kanka.ghosh@lspm.cnrs.fr}
\affiliation{CNRS, LSPM UPR3407, Université Sorbonne Paris Nord, 93430 Villetaneuse, France}
\affiliation{University of Bordeaux, CNRS, Arts et Metiers Institute of Technology, Bordeaux INP, INRAE, I2M Bordeaux, F-33400 Talence, France}
\author{Andrzej Kusiak}%
\affiliation{University of Bordeaux, CNRS, Arts et Metiers Institute of Technology, Bordeaux INP, INRAE, I2M Bordeaux, F-33400 Talence, France}
 \author{Jean-Luc Battaglia}%
 \affiliation{University of Bordeaux, CNRS, Arts et Metiers Institute of Technology, Bordeaux INP, INRAE, I2M Bordeaux, F-33400 Talence, France}


\begin{abstract}

\noindent We systematically demonstrate the temperature-dependent thermal transport properties in crystalline Ge$_2$Sb$_2$Te$_5$ via first-principles density functional theory-informed linearized Boltzmann transport equation. The investigation, covering a wide temperature range (30 K-600 K), reports the emergence of an unusual optical phonon-dominated thermal transport in crystalline Ge$_2$Sb$_2$Te$_5$. Further, a significant contribution of four-phonon scattering is recorded which markedly alters the lattice thermal conductivity. Therefore, the combined effect of cubic and quartic phonon anharmonicity is seen to navigate the underlying physical mechanism and open up intriguing phononic interactions in Ge$_2$Sb$_2$Te$_5$ at high temperature. Irrespective of three and four-phonon processes, Umklapp is seen to prevail over normal scattering events. Consequently, four-phonon scattering is found to notably reduce the lattice thermal conductivity of Ge$_2$Sb$_2$Te$_5$ to 28 $\%$ at room temperature and 42 $\%$ at higher temperature. This quartic anharmonicity further manifests in the breakdown of $T^{-1}$ scaling of thermal conductivity and challenges the idea of a universal lower bound to phononic thermal diffusivity at high temperature. The faster decay of thermal diffusivity compared to $T^{-1}$ is rationalized encompassing the quartic anharmonicity via a modified time scale. These results invoke better understanding and precision to the theoretical prediction of thermal transport properties of Ge$_2$Sb$_2$Te$_5$. Concomitantly, this also triggers the possibility to explore the manifestations of the lower bound of thermal diffusivity in materials possessing pronounced four-phonon scattering. 
\end{abstract}

\maketitle


\section{Introduction}

Phase change memory devices (PCM) represent an efficient nonvolatile memory storage mechanism with thermally induced rapid transitions between crystalline and amorphous phases, featuring a wide range of applicability \cite{wuttig2007phase, wuttig2005towards, raoux2010phase}. Amongst these PCM materials, Ge$_2$Sb$_2$Te$_5$ (GST) is one of the most suitable and promising candidates due to its fast switching speed and notable contrast in electrical and optical properties during phase transition \cite{hegedus2008microscopic, wuttig2007phase, pirovano2004electronic, kim2007three, jones2020phase, gan2022new}. Recently, Ge-rich GST materials were also being investigated to study their potency in PCM applications\cite{kusiak2022temperature, abou2021density, chassain2023thermal}. Being such a convenient PCM material, the thermoelectric properties of Ge$_2$Sb$_2$Te$_5$ \cite{lee2012phase, faraclas2014modeling, siegrist2011disorder, ibarra2018ab, wei2019quasi, miao2022remarkable, zhou2023tuned} have been the topic of intense investigations over the years. Ge$_2$Sb$_2$Te$_5$ exists in one amorphous \cite{konstantinou2018ab} and two crystalline phases (cubic and hexagonal). Owing to its low thermal conductivity ($\kappa$), crystalline Ge$_2$Sb$_2$Te$_5$ was found to be suitable for high temperature thermoelectric applications due to its enhanced thermoelectric figure of merit ($ZT$ = $S^{2}\sigma T / \kappa$, $\sigma$: electrical conductivity, S: Seebeck coefficient)\cite{siegrist2011disorder}. Therefore, scrutinizing high temperature thermal transport is essential to provide better understanding of its thermoelectric behavior. Over the years, experimental \cite{lyeo2006thermal, battaglia2010thermal, yanez1995thermal, kuwahara2007measurement, fallica2009thermal, li2023temperature} as well as state-of-the-art \textit{ab initio} theoretical methods \cite{tsafack2011electronic, kheir2023unraveling, mukhopadhyay2016optic, campi2017first, ibarra2018ab, pan2019lattice} were successfully employed to predict and measure thermal conductivity of crystalline Ge$_2$Sb$_2$Te$_5$.

\noindent While Wiedemann-Franz law was quite successful in reproducing the electronic part \cite{risk2009thermal, lyeo2006thermal}, phonon contribution to thermal conductivity of hexagonal Ge$_2$Sb$_2$Te$_5$ needs special attention as an unusual domination of optical phonons was observed \cite{mukhopadhyay2016optic}. Contrary to the conventional heat transport mechanism via acoustic phonons, optical phonon-dominated thermal transport is less common. Also, this feature seems to challenge the conventional Slack model \cite{slack1979thermal, kaviany2014heat} which takes only acoustic phonons into account. Optical phonons gain importance due to various features like large group velocities, anharmonic effects (2H MoS$_2$ \cite{dong2022effect}), and large acoustic-optical band gap (AlSb \cite{lindsay2013ab}), among a few. The large acoustic-optical band gap in AlSb and GaN imposes restriction to the acoustic-optic phonon scattering, resulting in a large thermal conductivity \cite{yang2019stronger}. At higher temperature, optical phonons populate without significant resistances due to forbidden acoustic-optical channels. This is pivotal in sustaining the optical mode-dominance in $\kappa_L$ in these high-$\kappa_L$ materials. On the other hand, for low-$\kappa_L$ materials, moderate contributions of optical phonons (22 $\%$ of total $\kappa_L$ for T = 300-700 K for PbTe \cite{tian2012phonon} and $\sim$ 30 $\%$ at T = 300 K for Mg$_2$Si \cite{li2012thermal}) were seen to stem from relatively dispersive optical branches and their presence at lower frequencies \cite{lindsay2013ab}. The softening of the optical modes in PbTe was identified to be crucial to trigger acoustic-optical scattering channel to produce low $\kappa_L$ \cite{tian2012phonon}. Although featuring acoustic phonon-dominated thermal conduction, non-negligible contributions of optical phonons were observed in a temperature range of 4K-600K ($\sim$ 80 $\%$ acoustic and $\sim$ 20 $\%$ optical modes)  for GeTe \cite{ghosh2020thermal, ghosh2020phonon}and  at T = 300 K ($\sim$ 65 $\%$ acoustic and $\sim$ 35 $\%$ optical modes) for Sb$_2$Te$_3$ \cite{campi2017first}. First principles study by Mukhopadhyay $\textit{et al.}$ \cite{mukhopadhyay2016optic} revealed that around 75 $\%$ of $\kappa_L$ was contributed by the optical modes in stable hexagonal crystalline Ge$_2$Sb$_2$Te$_5$ at room temperature. Also, a recent investigation focused on the coupling between acoustic and optical branches in the phonon spectrum of crystalline Ge$_2$Sb$_2$Te$_5$ \cite{miao2022remarkable}. This coupling between acoustic and optical phonon branches and the dominance of optical phonons at lower frequencies can effectively trigger higher order phonon scattering channels which is our particular interest in this article.

\noindent The strong presence of four-phonon scattering explains the strong temperature dependent thermal conductivity, although the anharmonic renormalization method was also employed \cite{li2022anharmonicity} to account for strong temperature dependent $\kappa_L$. Including four-phonon processes was found to affect the optical phonon contributions to $\kappa_L$ for AlSb and BAs \cite{feng2017four, yang2019stronger}, diminishing the optical phonon dominance in thermal transport. In low-$\kappa_L$ material, the presence of lower frequency optical phonons, their strong contribution to $\kappa_L$ as well as coupling between acoustic-optical branches can incite four-phonon processes as temperature is raised. Recently, low-$\kappa_L$ ionic conductors AgI \cite{wang2023anharmonic} and AgCrSe$_2$ \cite{xie2020first} were found to display strong quartic anharmonicity characterized by flat phonon dispersion and rattling dynamics of weakly bonding atoms. CsPbBr$_3$, a prominent low-$\kappa_L$ chalcogenide perovskite material, was also shown to feature reasonable four-phonon scattering \cite{wang2023role}. However, the implication of the presence of four-phonon scattering on thermal transport in a broad variety of low-$\kappa_L$ materials is intriguing and not well studied. In this spirit, we investigate the phononic behavior of low-$\kappa_L$ chalcogenide Ge$_2$Sb$_2$Te$_5$ to test its higher order scattering features.

\noindent Controlling thickness \cite{reifenberg2007thickness, lee2011thermal, sklenard2021electronic} as well as vacancies \cite{siegert2014impact} or stoichiometric defects \cite{caravati2009first} was proved to be efficient in modulating thermal transport in crystalline Ge$_2$Sb$_2$Te$_5$. A small amount of Ge, Sb vacancies and disorder were found to lower $\kappa_L$ around 70-80 $\%$ \cite{campi2017first}. However, as a standard practice, computing $\kappa_L$ and comparing it with experimental measurements involve calculating three-phonon scattering along with approximate models to incorporate disorder, vacancies, and grain-boundary scattering perturbatively. Hence, to reproduce experimental results, often higher order phonon scattering is ignored which, if significant, can overestimate the $\kappa_L$ coming from three-phonon scattering alone. Moreover, at smaller length scales of the sample, crystals can be found to exist in a defect-starved pristine state. Consequently, it is important to precisely predict $\kappa_L$ that solely comes from phonon scattering. In this context, we focus our investigation on the phonon transport in crystalline hexagonal Ge$_2$Sb$_2$Te$_5$ to decipher the role of higher order phonon scattering mechanisms in a wide temperature range.

\noindent Therefore we have several goals in this paper:
(a) To check, validate and introspect the dominance of optical phonons in the thermal transport of crystalline Ge$_2$Sb$_2$Te$_5$ as a function of temperature.
(b) To study whether four-phonon scattering is significant to describe the thermal transport in crystalline Ge$_2$Sb$_2$Te$_5$ at high temperature.
(c) If significant, to investigate whether different mode contributions are being influenced due to the inclusion of four-phonon scattering.
(d) To gain precise predictions of the phonon-phonon scattering term in $\kappa_L$ for a better estimation of other scattering terms (effect of disorder, grains) in crystalline Ge$_2$Sb$_2$Te$_5$. To answer these issues, we explore the temperature dependent thermal transport in crystalline Ge$_2$Sb$_2$Te$_5$, ranging from $T$ = 30 K to 600 K using first-principles density functional method coupled with the solution of the linearized Boltzmann transport equation (LBTE). A systematic study is carried out to investigate harmonic and anharmonic phononic properties of GST. The presence of significant four-phonon scattering has been identified and the relative competitions between different scattering channels are discussed. Lattice thermal conductivity and its mode-wise contributions are evaluated with and without four-phonon scattering to mark the contrast. The effect of higher order phonon scattering on the anisotropy in thermal conduction has also been studied. Finally, the deviation from high temperature $T^{-1}$ scaling of lattice thermal conductivity is analyzed. This resolved picture of thermal conduction attempts to bring more precision to the microscopic understanding for Ge$_2$Sb$_2$Te$_5$, which can improve thermal management of related chalcogenide PCM materials.

\section{Method: Density functional theory-informed Boltzmann transport equation}{\label{section:computational}}

\noindent The structural parameters of the crystalline, hexagonal phase of Ge$_2$Sb$_2$Te$_5$ (space group $P\overline{3}m_1$) are optimized via first-principles density functional theoretical (DFT) calculations, and the obtained lattice parameters in comparison to other related works are presented in Table I. Phonon dispersion and phonon density of states (PDOS) of two different crystalline, hexagonal Ge$_2$Sb$_2$Te$_5$ structures (Kooi and Petrov (see supplementary Fig S4 for PDOS and dispersion relation for Petrov structure)) have been calculated using the density functional perturbation theory (DFPT) \cite{baroni2001phonons} as employed in the 
$\textsc{quantum-espresso}$\cite{giannozzi2009quantum} suite of programs. Self-consistent field calculations, within the framework of density functional theory (DFT), are carried out to compute the total ground state energy of the crystalline Ge$_2$Sb$_2$Te$_5$. For this purpose, Perdew-Burke-Ernzerhof (PBE) \cite{perdew1996generalized} generalized gradient approximation (GGA) is used as the exchange-correlation functional. Electron-ion interactions are represented by pseudopotentials using the framework of projector-augmented-wave (PAW) method \cite{blochl1994projector}. The Kohn-Sham (KS) orbitals are expanded in a plane-wave (PW) basis with a kinetic cutoff of 35 Ry (476.2 eV) and a charge density cutoff of 140 Ry whereas for relax calculation the plane-wave (PW) kinetic cutoff is taken as 80 Ry (1088.45 eV). The semi-empirical correction due to Grimme (D2) is considered to incorporate the van der Waals (vdW) interactions. The Brillouin zone (BZ) integration for self consistent electron density calculations are performed using a 12$\times$12$\times$12 Monkhorst-Pack (MP) \cite{monkhorst1976special} $\textbf{q}$-point grid. 

\noindent We compute the lattice thermal conductivity and associated phonon scattering rates employing the Boltzmann transport equation (BTE). A linearized version of the BTE is solved for phonons using an iterative self-consistent method as well as relaxation-time approximation (RTA) approach via $\textsc{fourphonon}$ \cite{han2022fourphonon}, an updated version of the $\textsc{shengbte}$ code \cite{li2014shengbte}. The numerical solution to the phonon-BTE demands accurate estimates of harmonic and anharmonic force constants. Therefore, DFT-based self-consistent field calculations are carried out to output harmonic and anharmonic force constants. Harmonic force constants are obtained using the DFPT method. We use a 4$\times$4$\times$2 supercell of Ge$_2$Sb$_2$Te$_5$ within the finite displacement approach to calculate the third-order interatomic force constants using finite differences method via $\textsc{thirdorder}$ package \cite{li2014shengbte}, allowing interactions up to the fourth-nearest neighbors. We adopt same parameter values as that of the work by Ibarra $\textit{et al.}$ \cite{ibarra2018ab} for getting the third order interatomic force constants (IFC). 2$\times$2$\times$1 supercells with interactions up to second-nearest neighbors are used to obtain fourth-order IFCs from 1728 different configurations. Within the relaxation-time approximation (RTA), phonon lifetimes ($\tau_{\lambda}^{RTA}$) take the following form

\begin{equation}
    \frac{1}{\tau_{\lambda}^{RTA}} = \frac{1}{\tau_{\lambda}^{3ph}} + \frac{1}{\tau_{\lambda}^{4ph}} + \frac{1}{\tau_{\lambda}^{I}} 
\end{equation}
with 
\begin{equation}
    \frac{1}{\tau_{\lambda}^{3ph}} = \frac{1}{N} \left(\sum_{\lambda^{\prime}\lambda^{\prime\prime}}^{+} \Gamma_{\lambda\lambda^{\prime}\lambda^{\prime\prime}}^{+} + \sum_{\lambda^{\prime}\lambda^{\prime\prime}}^{-}\frac{1}{2} \Gamma_{\lambda\lambda^{\prime}\lambda^{\prime\prime}}^{-} \right)
\end{equation}
\begin{multline}
    \frac{1}{\tau_{\lambda}^{4ph}} = \frac{1}{N} (\sum_{\lambda^{\prime}\lambda^{\prime\prime}\lambda^{\prime\prime\prime}}^{++}\frac{1}{2} \Gamma_{\lambda\lambda^{\prime}\lambda^{\prime\prime}\lambda^{\prime\prime\prime}}^{++} + \sum_{\lambda^{\prime}\lambda^{\prime\prime}\lambda^{\prime\prime\prime}}^{+-}\frac{1}{2} \Gamma_{\lambda\lambda^{\prime}\lambda^{\prime\prime}\lambda^{\prime\prime\prime}}^{+-} + \\
    \sum_{\lambda^{\prime}\lambda^{\prime\prime}\lambda^{\prime\prime\prime}}^{--}\frac{1}{6} \Gamma_{\lambda\lambda^{\prime}\lambda^{\prime\prime}\lambda^{\prime\prime\prime}}^{--})
\end{multline}
and 
\begin{equation}
    \frac{1}{\tau_{\lambda}^{I}} = \frac{1}{N} \sum_{\lambda^{\prime}}^{I}\Gamma_{\lambda\lambda^{\prime}}^{I}
\end{equation}
where $N$ is total number of $\textbf{q}$ points, $+$ and $-$ denote absorption and emission respectively corresponding to three-phonon scattering. $++$, $+-$ and $--$ represent four-phonon recombination, redistribution and splitting processes respectively. Three-phonon, four-phonon and isotope scattering rates are represented via $\Gamma_{\lambda\lambda^{\prime}\lambda^{\prime\prime}}$, $\Gamma_{\lambda\lambda^{\prime}\lambda^{\prime\prime}\lambda^{\prime\prime\prime}}$ and $\Gamma_{\lambda\lambda^{\prime}}^{I}$ respectively. While fourth order IFCs are only implemented in RTA, third-order IFCs are computed using both RTA and iterative method. For a specific mode $\lambda$, linearized phonon-BTE can be expressed as

\begin{equation}
    \textbf{F}_{\lambda} = \tau_{\lambda}^{RTA} (\textbf{v}_{\lambda} + \Delta_{\lambda})
\end{equation}
with
\begin{multline}
    \Delta_{\lambda}^{\beta} = \frac{1}{N} [\sum_{\lambda^{\prime}\lambda^{\prime\prime}}^{+} \Gamma_{\lambda\lambda^{\prime}\lambda^{\prime\prime}}^{+}(\xi_{\lambda\lambda^{\prime\prime}}^{\beta} F_{\lambda^{\prime\prime}}^{\beta} - \xi_{\lambda\lambda^{\prime}}^{\beta} F_{\lambda^{\prime}}^{\beta}) + \\
    \sum_{\lambda^{\prime}\lambda^{\prime\prime}}^{-} \frac{1}{2}\Gamma_{\lambda\lambda^{\prime}\lambda^{\prime\prime}}^{-}(\xi_{\lambda\lambda^{\prime\prime}}^{\beta} F_{\lambda^{\prime\prime}}^{\beta} + \xi_{\lambda\lambda^{\prime}}^{\beta} F_{\lambda^{\prime}}^{\beta})]
\end{multline}
Here, $\Delta_{\lambda}$ accounts for the iterative process and retrieves RTA with $\Delta_{\lambda}$ = 0. $\xi_{\lambda\lambda^{\prime}}^{\beta}$ = $\omega_{\lambda^{\prime}}v_{\lambda^{\prime}}^{\beta}$/ $\omega_{\lambda}v_{\lambda}^{\beta}$, where $\textbf{v}_{\lambda}$ denotes modal group velocity and $\textbf{F}_{\lambda}$ represents the linear deviation of the phonon population ($f_{\lambda}$) from the equilibrium Bose-Einstein distribution ($f_{\lambda}^{0}$) as $f_{\lambda} = f_{\lambda}^{0} - \textbf{F}_{\lambda}\cdot \frac{\partial f_{\lambda}}{\partial T}\nabla T - ...$ . Finally, using third-order and fourth-order IFCs and knowing the phonon distribution function, lattice thermal conductivity tensor ($\kappa_L^{\alpha\beta}$) is computed from the following expression   

\begin{equation}
    \kappa_L^{\alpha\beta} = \frac{1}{k_{B}T^{2}NV_{0}} \sum_{\lambda} f_{0}(f_{0}+1)(\hbar\omega_\lambda)^{2}v_{\lambda}^{\alpha}F_{\lambda}^{\beta}
\end{equation}
where $\lambda$ defines every phonon mode, $N$ denotes the total number of modes in BZ, $V_0$ is the unit cell volume, $f_0$ = $(e^{\hbar\omega/k_{B}T}-1)^{-1}$ represents equilibrium Bose-Einstein phonon distribution function. In the RTA approach, $F_{\lambda}^{\beta} = \tau_{\lambda}^{\beta} v_{\lambda}^{\beta}$, where $\tau_{\lambda}^{\beta}$ stands for the relaxation time for phonons of polarization $\lambda$ propagating along the $\beta$ direction. Phonon-BTE for both three and four-phonon processes are solved in a 16 $\times$ 16 $\times$ 4 $\textbf{q}$-grid over the Brillouin zone (see convergence of three and four-phonon processes in Supplementary Fig S15).

\section{Results and  Discussions\label{section:results}}

\noindent The stable hexagonal phase of crystalline Ge$_2$Sb$_2$Te$_5$ possesses a crystal structure of $P\overline{3}m_1$ symmetry with nine atoms per unit cell stacked along the $c$ axis. However, the stacking of these nine atoms along $c$ is still being debated \cite{campi2017first, ibarra2018ab, kooi2002electron, petrov1968electron, matsunaga2004structures, tominaga2014ferroelectric}. The stacking sequence along the $c$ axis by Petrov $\textit{et al.}$ \cite{petrov1968electron} (defined as Petrov structure hereafter) was proposed as Te-Sb-Te-Ge-Te-Te-Ge-Te-Sb-  while Kooi and De Hosson \cite{kooi2002electron}(defined as Kooi structure hereafter) observed the stacking to be Te-Ge-Te-Sb-Te-Te-Sb-Te-Ge- via electron diffraction experiments. Experiments also suggested a disordered phase 

\begin{table} [H]
\caption{\label{tab:table1} Comparison between equilibrium lattice parameters of differently stacked hexagonal Ge$_2$Sb$_2$Te$_5$ obtained in this work and in earlier \textit{Ab initio} studies.}
\begin{ruledtabular}
\begin{tabular}{cccc}
 &\multicolumn{2}{c}{Kooi}\\
 Lattice parameters (\AA)&Ref. \cite{campi2017first}&Ref. \cite{ibarra2018ab}&This work\\ \hline
 $a$&4.191&4.23 &4.193 \\
 $c$&17.062&16.88 &17.031 \\
&\multicolumn{2}{c}{Petrov}\\
Lattice parameters (\AA)&Ref. \cite{campi2017first}&Ref. \cite{ibarra2018ab}&This work\\ \hline
 $a$&4.178&4.2 &4.17 \\
 $c$&17.41&17.14 &17.37 
\end{tabular}
\end{ruledtabular}
\end{table}

\noindent in Ge/Sb sublattice with layers randomly occupied by Ge and Sb \cite{matsunaga2004structures}. However, as we focus only on the phononic part of the heat transport, investigating the mentioned disordered structure is out of the scope of this paper. Lately, another stacking arrangement called inverted Petrov structure \cite{tominaga2014ferroelectric} was also proposed for hexagonal Ge$_2$Sb$_2$Te$_5$, featuring interchanging adjacent Te and Ge atoms. It was noted in earlier studies \cite{ibarra2018ab, campi2017first} that Kooi structure exhibits lowest energy amongst these differently stacked structures. Therefore, although we characterize both Petrov and Kooi structures (Table \ref{tab:table1}), hereafter all relevant studies are carried out using the Kooi structure. Both Kooi and Petrov stacking structures for hexagonal Ge$_2$Sb$_2$Te$_5$ are shown in Fig \ref{fig:1}(A)(a) and Fig \ref{fig:1}(A)(b) respectively, using two formula units stacked along $c$ and the periodic copies of atoms at the edges in the $ab$ planes. A fairly good agreement on the lattice parameters between earlier first-principles calculations and this work is observed from Table \ref{tab:table1}. The small differences can be attributed to the numerical uncertainties due to the variation of exchange-correlation functionals used in these studies. For example, PBE generalized gradient approximation (GGA) was employed in both Campi $\textit{et al.}$ \cite{campi2017first}  and this work, whereas Ibarra $\textit{et al.}$ \cite{ibarra2018ab} used GGA with a revised version of PBE for solids and surfaces \cite{perdew2008restoring} (PBEsol) as exchange-correlation functional. We note that the experimental values of lattice parameters of the Kooi \cite{kooi2002electron} structure were measured as  $a$ = 4.25 \AA{} and $c$ = 17.27 \AA, which are also consistent with our \textit{ab initio} findings depicted above. After bench-marking our theoretically relaxed structure, we proceed in describing the phonon transport mechanisms via harmonic and anharmonic properties of Ge$_2$Sb$_2$Te$_5$ in the following subsections.

\subsection{Harmonic properties and optical-mode dominance}{\label{subsection:Harmonic}}

\begin{widetext}
\begin{figure}[H]
\centering
\includegraphics[width=1.0\textwidth]{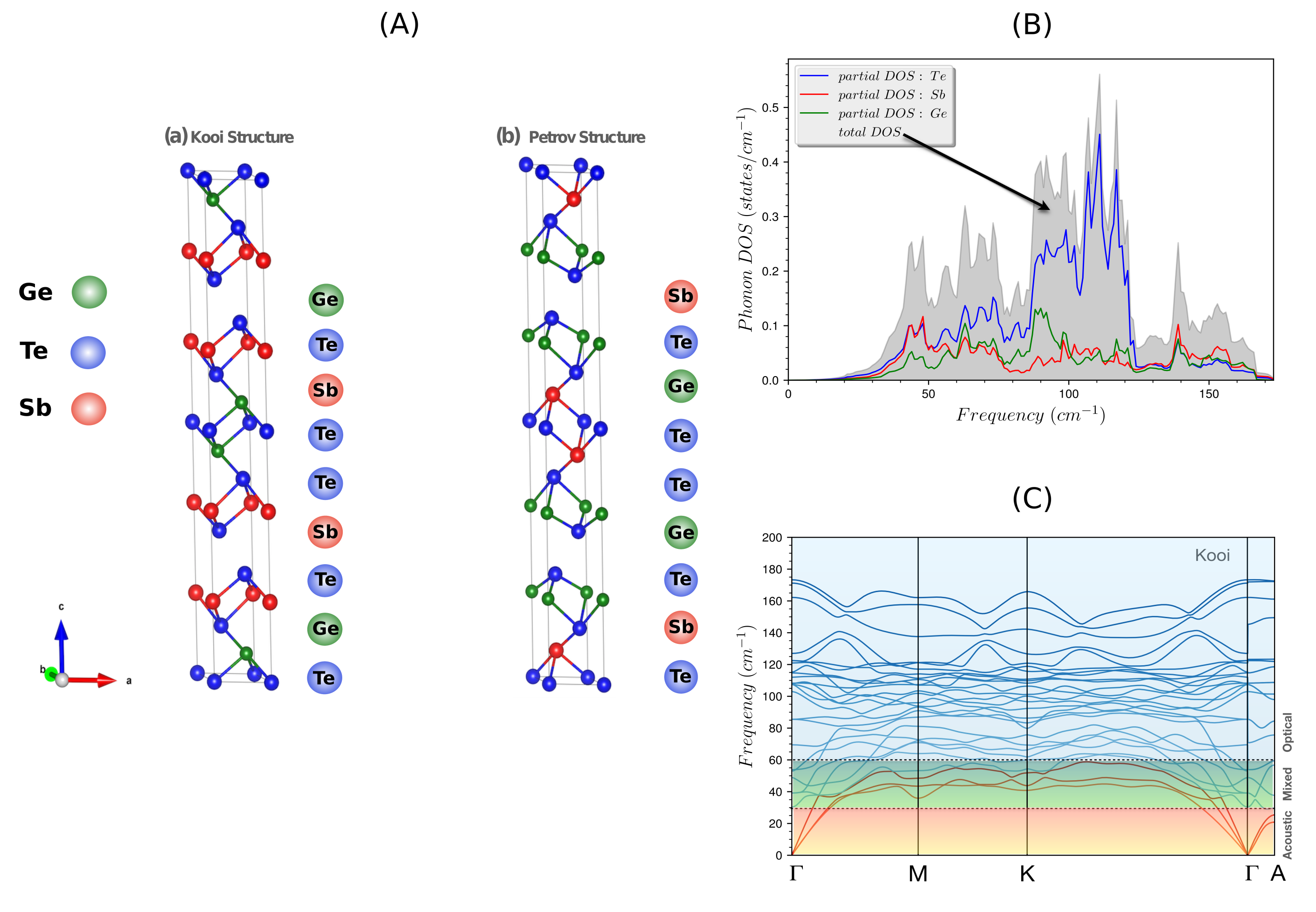}
\caption{(A) The schematic hexagonal structures of Ge$_2$Sb$_2$Te$_5$ in two different stacking arrangements (a) Kooi and (b) Petrov, are presented. $C$ axes are designed to accommodate two formula units of the nine atom hexagonal unit cell. The corresponding nine atom unit cell stacking orders for Ge, Sb, and Te atoms in Kooi and Petrov structures have also been shown in (a) and (b) respectively. The representations are realized through VESTA \cite{momma2011vesta}. (B) Phonon density of states (PDOS) is presented as a function of frequency for crystalline hexagonal Ge$_2$Sb$_2$Te$_5$ stacked in Kooi structure. Corresponding partial DOS for Ge, Sb and Te are also superimposed. (C) Phonon dispersion relations of Ge$_2$Sb$_2$Te$_5$ Kooi structure along high symmetry path $\Gamma$-$M$-$K$-$\Gamma$-$A$. The frequency regimes corresponding to acoustic (A), mixed (M) and optical (O) modes are also distinguished.}
\label{fig:1}
\end{figure}
\end{widetext}

\noindent To understand phonon transport, we start our discussion with the phonon dispersion and phonon density of states (PDOS) realizations for hexagonal Ge$_2$Sb$_2$Te$_5$ Kooi structure, depicted in Fig \ref{fig:1}(C) and Fig \ref{fig:1}(B) respectively. Phonon dispersion relation is shown along the $\Gamma$-$M$-$K$-$\Gamma$-$A$ high symmetry direction in the Brillouin zone (BZ) portraying a stable structure with phonon frequencies going to zero at the $\Gamma$ point in the BZ. The phonon frequencies are seen to span a large frequency regime ranging from 0 to 174 $cm^{-1}$. We mark some specific features in the phonon bands. Firstly, we observe the presence of low-frequency optical phonon bands starting from $\omega$ = 30 $cm^{-1}$. These low frequency optical bands are seen to couple with acoustic bands, indicating strong scattering events to reduce thermal conduction. We will get to this point later. As acoustic and optical phonons feature a strong coupling in an intermediate frequency range (30 $cm^{-1}$ $<$ $\omega$ $<$ 60 $cm^{-1}$), distinguishing individual contribution of the phononic bands to the thermal transport seems unrealistic. However, we can define frequency bands that consist of only acoustic, a mixture of acoustic and optical modes and purely optical modes to comprehend the contributions to thermal transport coming from these frequency bandwidths. Figure \ref{fig:1}(C) distinguishes these frequency regimes that comprise of purely acoustic (0 $<$ $\omega$ $<$ 30 $cm^{-1}$), purely optical (60 $cm^{-1}$ $<$ $\omega$ $<$ 174 $cm^{-1}$) and the mixed modes (30 $cm^{-1}$ $<$ $\omega$ $<$ 60 $cm^{-1}$) and hereafter will be denoted by A, O and M respectively. The dispersion picture is complemented by the phonon density of states representation shown in Fig \ref{fig:1}(B). The total PDOS as well as the separate contributions from three different elements have been projected. The significant contribution of five Te atoms in the Ge$_2$Sb$_2$Te$_5$ unit cell is being manifested as the principal contributor to the PDOS as shown in Fig \ref{fig:1}(B), consistent with earlier studies \cite{mukhopadhyay2016optic}. At small frequency range, heavier atoms (Sb, Te) are seen to dominate the PDOS. Around $\omega$ = 60 $cm^{-1}$, the contribution of Ge atoms starts to be significant. From $\omega$ = 90 $cm^{-1}$ till 120 $cm^{-1}$, the enhancement of PDOS is mostly contributed by the Te atoms. At frequencies higher than 130 $cm^{-1}$, this enhancement dies out and contributions from Ge, Sb, Te become comparable. Relating the element-wise contributions to PDOS with the frequency regimes A, M and O, we observe the following: (a) In regime A (0 $cm^{-1}$ $<$ $\omega$ $<$ 30 $cm^{-1}$), all three elements contribute comparably. (b) Partial DOS of Te atoms starts dominating with a competition between Sb and Ge PDOS in regime M (30 $cm^{-1}$ $<$ $\omega$ $<$ 60 $cm^{-1}$). (c) In regime O (60 $cm^{-1}$ $<$ $\omega$ $<$ 174 $cm^{-1}$), PDOS for Te mostly dominates and controls the total PDOS, except at higher frequencies ($\omega$ $>$ 90 $cm^{-1}$), where all contributions become comparable again. At extreme lower and extreme higher frequencies, all three elements show comparable number of excited vibrational modes. In standard Debye theory, frequency-dependent PDOS can be expressed as PDOS ($\omega$) = $\frac{k^{2}}{2\pi^{2}}\frac{1}{d\omega/dk}$ with the assumption of the linear relation between speed of sound ($c_s$) and the phonon frequencies ($\omega$ = $c_{s}k$). Consequently, lower group velocities ($v_g$ = $d\omega/dk$) of phonons enable higher density of states as a function of phonon frequencies. On the contrary, dispersive phonon branches enhance the phonon group velocities causing a lower PDOS. Near the center of BZ ($\Gamma$ point) in regime A, acoustic branches are seen to

\onecolumngrid

\noindent behave almost in a linear fashion (Fig \ref{fig:1}(C)) which is caused by the vibrational modes of mostly heavy Te and Sb atoms along with little less vibrational contribution from comparably light Ge atoms. In the regime M (30 $cm^{-1}$ $<$ $\omega$ $<$ 60 $cm^{-1}$), dispersion relation shows the coupling between acoustic and optic branches as well as the onset of the comparatively flat bands. These flat bands give rise to an enhancement of PDOS in this regime. Sb and Te atoms seem to contribute equally until $\omega$ = 50 $cm^{-1}$, after which Te modes start dominating and the vibrational modes of Ge atoms become comparable or even greater than that of the Sb atoms (Fig \ref{fig:1}(B)). In regime O (60 $cm^{-1}$ $<$ $\omega$ $<$ 174 $cm^{-1}$), two distinct features are observed. For $\omega$ $<$ 130 $cm^{-1}$, larger contribution of flat phonon optical bands of low $v_g$ (Fig \ref{fig:1}(C)) is seen to develop a larger PDOS with several peaks, mostly contributed by the vibrations of Te atoms (Fig \ref{fig:1}(B)). For $\omega$ $>$ 130 $cm^{-1}$, optical branches becomes quite dispersive warranting a higher $v_g$ and a decrease in PDOS. Small mass difference between Sb, Te ($m_{Sb}$ = 121.75 a.m.u, $m_{Te}$ = 127.6 a.m.u) makes it difficult to feature separated high and low frequency bands as the evolution of the PDOS goes through competitive vibrations between the elements. In contrast, another chalcogenide material GeTe, owing to the large mass difference between its constituents ($m_{Ge}$ = 72.64 a.m.u, $m_{Te}$ = 127.6 a.m.u), was seen \cite{campi2017first, ghosh2020thermal} to possess PDOS with a separated low frequency acoustic bands (due to the vibrations of heavy Te atoms) and high frequency optical bands (due to light Ge atoms). Also, due to their similar masses, distinct contributions from Sb and Te to PDOS mostly rely on their vibrational modes due to the bonding arrangement in Ge$_2$Sb$_2$Te$_5$. In this context, Sosso $\textit{et al.}$ \cite{sosso2009vibrational} studied the Raman spectra of hexagonal Ge$_2$Sb$_2$Te$_5$ (Kooi structure) and A-type Raman-active modes (specifically A$_{1g}$ modes) were identified at high frequencies. These A-type modes indicate vibration along the c-axis due to modulation of the outermost Te-Sb bond \cite{sosso2009vibrational}. The high frequency vibrational modes involving both Sb and Te manifest themselves in comparable and similar phonon DOS at high frequencies ($>$ 125 $cm^{-1}$) for Te and Sb (see Fig \ref{fig:1}(B)).

\begin{figure}[H]
\centering
\includegraphics[width=0.5\textwidth]{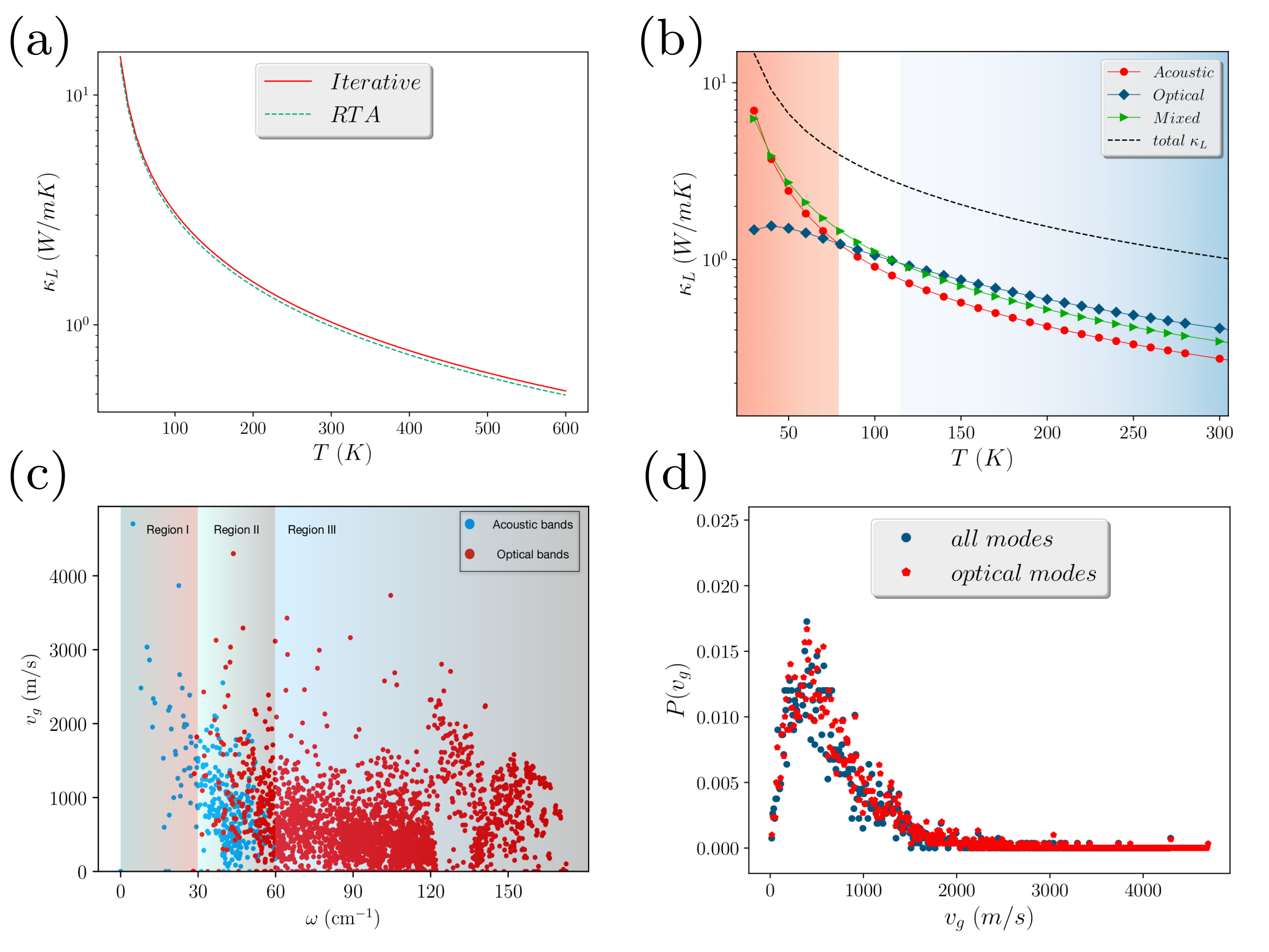}
\caption{(a) The variation of average lattice thermal conductivity ($\kappa_L$) with temperature realized via both iterative and relaxation time approximation (RTA).  (b) The variation of average lattice thermal conductivity, obtained via iterative method, as a function of temperature with distinguished contributions from acoustic (A), mixed (M) and optical frequency regimes (O) as described in the text. The regimes with different colors mark two separate crossings between acoustic-optic contributions and optic-mixed contributions. (c) The variation of phonon group velocity ($v_g$) for both acoustic and optical modes with frequency. Acoustic, mixed and optical regimes are denoted via region I, II, III respectively. (d) Probability distribution ($P(v_g)$) of phonon group velocities ($v_g$) for optical as well as all modes.}
\label{fig:2}
\end{figure}

\noindent To investigate thermal transport, we employ harmonic and third-order anharmonic force constants and therefore the solution of Linearized Boltzmann transport equation (LBTE) takes the form 

\begin{equation}\label{eq:1}
    \kappa_L^{\alpha\beta} = \frac{1}{k_{B}T^{2}NV_{0}} \sum_{\lambda} f_{0}(f_{0}+1)(\hbar\omega_\lambda)^{2}v_{\lambda}^{\alpha}F_{\lambda}^{\beta}
\end{equation}
where $\kappa_L^{\alpha\beta}$ is lattice thermal conductivity tensor, $\lambda$ defines each phonon mode, $N$ is the total number of modes in BZ, $V_0$ is unit cell volume, $f_0$ = $(e^{\hbar\omega/k_{B}T}-1)^{-1}$ denotes Bose-Einstein phonon distribution function, and $F$ is the term in the expansion of the phonon distribution function which is related to the group velocity ($v_g$). We use both relaxation time approximation (RTA) as well as iterative approach for the solution of LBTE to obtain $\kappa_L$. Scattering of phonons by randomly distributed isotopes is also incorporated. The isotope scattering rate ($1/\tau_{\lambda}^{I}(\omega)$), was given by Tamura \cite{tamura1983isotope} via second-order perturbation theory as
\begin{equation}
\resizebox{1.0\hsize}{!}{$
    1/\tau_{\lambda}^{I}(\omega) = \frac{\pi \omega_{\lambda}^{2}}{2N}\sum_{\lambda'} \delta\left(\omega - \omega'_{\lambda} \right) \sum_{k} g_{k}|\sum_{\alpha}\textbf{W}_{\alpha}\left(k,\lambda \right)\textbf{W}_{\alpha}^{*}\left(k,\lambda \right)| ^{2} 
$}
\end{equation}
where $g_k$ is the mass variance parameter, defined as 
\begin{equation}
    g_{k} = \sum_{i} n_{i} \left( 1 - \frac{m_{ik}}{\overline{m}_k}\right)^{2}
\end{equation}
$n_i$ is the mole fraction, $m_{ik}$ is the relative atomic mass of $i$th isotope, $\overline{m}_k$ is the average mass = $\sum_{i} n_{i} m_{ik}$, and $\textbf{W}$ is a polarization vector. The natural abundance data for different elements \cite{de2003atomic} are used to obtain the mass variance parameters. Figure \ref{fig:2}(a) presents $\kappa_L$, averaged over all directions, as a function of temperature for hexagonal Kooi-Ge$_2$Sb$_2$Te$_5$. The difference between iterative and RTA solutions is found to be marginal throughout the whole temperature range studied (30K - 600K). The compatibility between RTA and iterative solution of LBTE indicates that the single uncorrelated phonon gas picture, conceptualised by RTA, is valid throughout the temperature range. This is also a signature of overpowering nature of the resistive phonon Umklapp scattering in dictating the $\kappa_L$ for crystalline Ge$_2$Sb$_2$Te$_5$. Figure \ref{fig:2}(b) describes the contributions to $\kappa_L$ coming from different frequency regimes (acoustic (A), optical (O), mixed (M)) as defined earlier. As temperature increases, we see

\noindent gradually larger contributions from optical and mixed phonon frequency bands compared to those of the acoustic phonons. At low temperature regime till $T$ $\leq$ 80 K (shown via red shaded area in Fig \ref{fig:2}(b)), acoustic as well as coupled acoustic-optical bands seem to dominate the $\kappa_L$. At slightly higher temperature regime (80 K $<$ $T$ $<$ 120 K, shown via white shaded area in Fig \ref{fig:2}(b)), contributions from both acoustic and acoustic-optic mixed modes start decreasing while the significance of purely optical modes starts increasing. At even higher temperature range (120 K $<$ $T$ $<$ 300 K, shown via blue shaded area in Fig \ref{fig:2}(b)), optical modes become the driving force to contribute $\kappa_L$ with higher contributions than both acoustic and mixed phonon modes. To invoke transparency to this feature, we recall the RTA approach, where the solution of LBTE is reduced to 

\begin{equation}{\label{eq:RTA}}
    \kappa_L = \frac{1}{NV_0}\sum_{\lambda} C_{\lambda} v_{g_\lambda}^{2}\tau_\lambda
\end{equation}
which is obtained via using Eq.\ref{eq:1} with zeroth-order approximation to $F$ as $F_{\lambda}$ = $\tau_{\lambda}v_{g{\lambda}}$  and expressing modal heat capacity $C_\lambda$ = $\frac{1}{k_{B}T^{2}} \sum_{\lambda} f_{0} (f_{0}+1) (\hbar\omega_{\lambda})^{2}$. Increasing temperature does not seem to change the group velocity ($v_g$). Similar behavior is also seen for heat capacity as it is known to follow Dulong-Petit law at high temperature (see Supplementary Fig S3). However, the variation of phonon group velocities with frequency, shown in Fig \ref{fig:2}(c), indicates significant contribution from the optical phonons with high group velocities comparable to that of the acoustic phonon modes. Figure \ref{fig:2}(c) also presents $v_g$ coming out of three qualitatively different frequency regimes (Region I (A), Region II (M), Region III (O)). To elucidate this point further, probability distributions of $v_g$, corresponding to optical as well as all the modes are presented in Fig \ref{fig:2}(d). We find the probability distribution ($P(v_g)$) of $v_g$ for all the phonon modes are almost merging with that of the optical modes, asserting the connection between the significant group velocities of optical phonons and the optical phonon-dominated $\kappa_L$ for Ge$_2$Sb$_2$Te$_5$. $P(v_g)$ shows a higher probability of lower group velocities with the maximum is found to be around 500 m/s. However, it also features a long tail distribution of higher group velocities extending up to $\sim$ 5000 m/s. The presence of both high as well as low $v_g$-optical phonons corresponds to the presence of significantly dispersive and comparatively flat optical bands respectively.

\noindent As both group velocity and heat capacities are temperature insensitive, probing the temperature dependent implications in thermal transport in Ge$_2$Sb$_2$Te$_5$ warrants investigation on the variation of the phonon scattering rates with temperature. The presence of lower frequency optic phonons, their strong coupling with acoustic modes as well as the presence of flat bands, observed via sharp PDOS peaks, lead us to investigate the possibility of an existence of four-phonon scattering processes. We discuss this feature in the next subsection.

\subsection{Three and four-phonon processes: competing effects and scattering channels}{\label{subsection:3ph4ph}}

\noindent We start our investigation on phonon scattering rates by comparing the relative importance between three-phonon (third-order anharmonicity, denoted 3ph hereafter) and four-phonon scattering processes (fourth-order anharmonicity, denoted 4ph hereafter) in a wide temperature range (30 K - 600 K). At this point, we mention two important factors that usually imprint the signature of the four-phonon processes: (a) anharmonicity and (b) scattering phase space distribution. The extent of anharmonicity can be captured via the Gruneisen parameter ($\gamma$) which is defined as the rate of change of the vibrational frequency of a particular mode with volume. Thus, $\gamma$ features the departure of crystal from harmonicity. For few modes within ‘M’ regime, corresponding to the strong overlap between acoustic and optical bands, $\gamma$ becomes anomalously large (see supplementary Fig S1(a)). However, mode-averaged Gruneisen parameter smooths out this feature at a particular temperature and therefore, the average $\gamma$ at fixed T behaves in less anomalous way (Figure S1(b)). This average $\gamma$ is found to be 2.49 at 

\begin{figure}[H]
\centering
\includegraphics[width=0.5\textwidth]{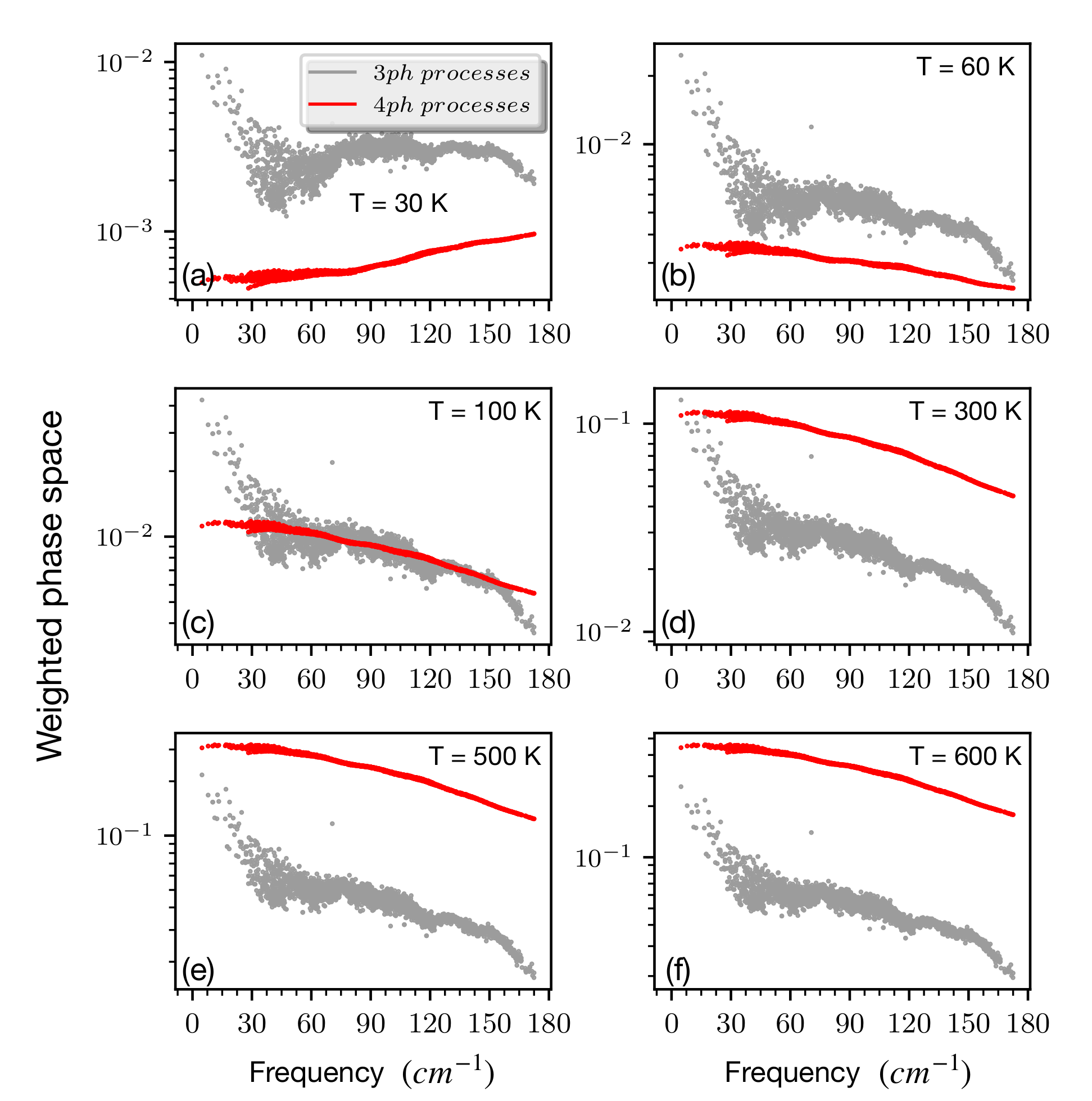}
\caption{The variation of three-phonon (3ph) and four-phonon (4ph) weighted phase space with phonon frequencies at different temperatures: (a) $T$ = 30 K, (b) $T$ = 60 K, (c) $T$ = 100 K, (d) $T$ = 300 K, (e) $T$ = 500 K, and (f) $T$ = 600 K for Ge$_2$Sb$_2$Te$_5$.}
\label{fig:3}
\end{figure} 

\noindent $T$ = 300 K which is similar for other low-$\kappa_L$ materials with comparable total thermal conductivities like PbTe ($\gamma$ = 2.18) and PbS ($\gamma$ = 2.5) at room temperature \cite{zhang2009thermodynamic}.

\noindent Having recorded the strong anharmonicity, we seek to explore the relative dominance between 3ph and 4ph processes via computing their respective weighted phase space as described by Li $\textit{et al.}$ \cite{li2015ultralow}. This weighted phase space is realized via the sum over frequency dependent terms in the 3ph and 4ph transition probabilities without the scattering matrix elements. Higher phase space suggests larger number of available scattering channels, leading to stronger scattering effect and consequently lower lattice thermal conductivity. Figure \ref{fig:3} represents the relative weight-age of 3ph and 4ph scattering phase spaces as a function of temperature ranging from 30 K to 600 K.  As we increase the temperature, a clear indication of overwhelming 4ph scattering phase space compared to its 3ph counterpart is observed. However, we must mention that the weighted phase space (WP) is a qualitative descriptor and their units are different for 3ph and 4ph. Moreover, it also includes the phonon distribution function ($f_0$) while summing over all the frequency-dependent terms. This involves the effect of temperature as in the high temperature limit  $f_0$ $\sim$ $k_{B}T/\hbar\omega$. However, it is evident that as the temperature is raised, 4ph phase space becomes prominent with a significant increment  due to the gradual population of high frequency optical modes at higher temperature (Fig \ref{fig:3}(d),(e), (f)). This feature hints at the connection between optical modes and 4ph 

\begin{figure}[H]
\centering
\includegraphics[width=0.5\textwidth]{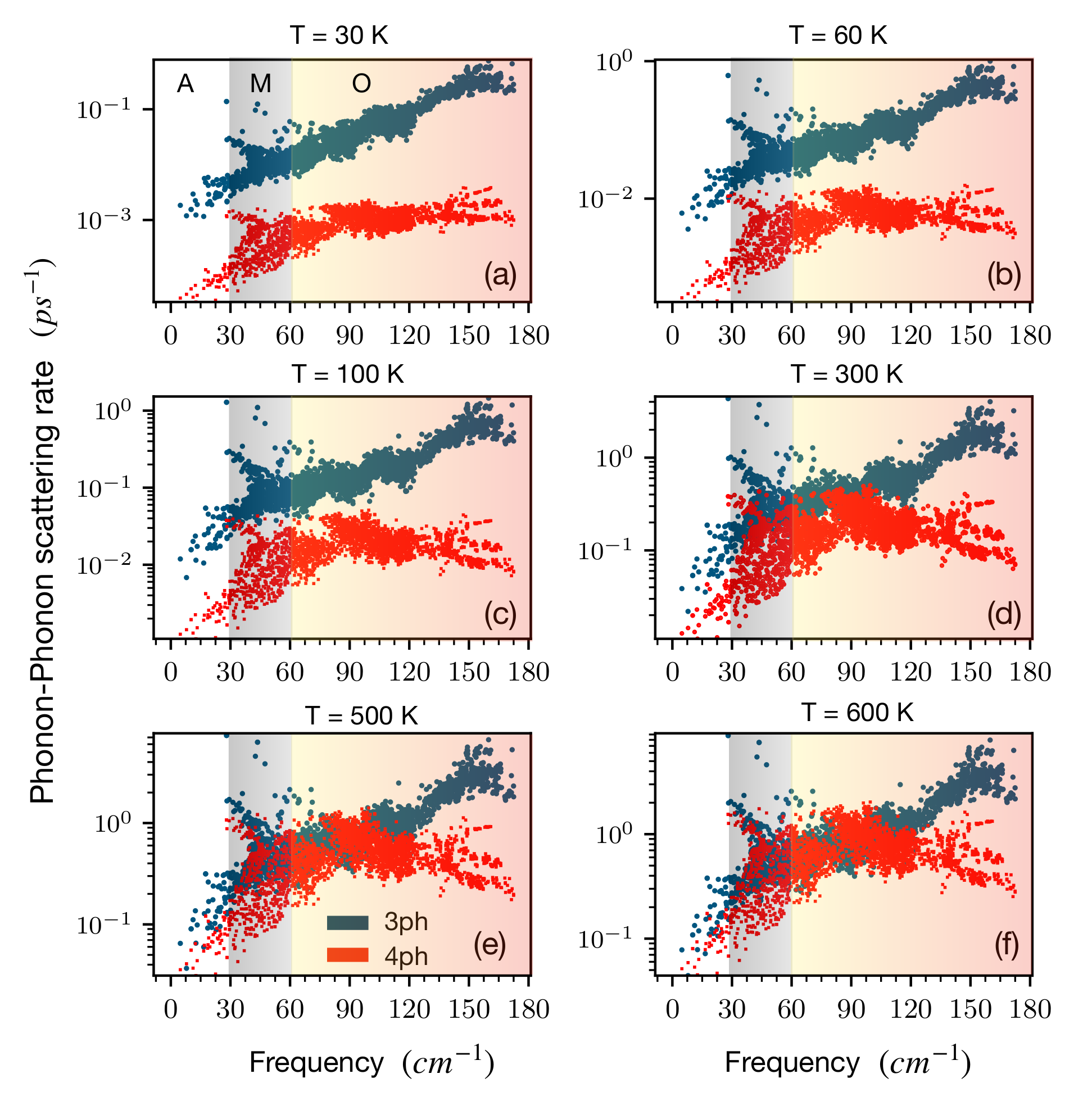}
\caption{The variation of three-phonon (3ph) and four-phonon (4ph) scattering rates with phonon frequencies at different temperatures: (a) $T$ = 30 K, (b) 60 K, (c) 100 K, (d) 300 K, (e) 500 K, and (f) 600 K. Different phonon frequency regimes are marked using white (A), grey (M) and light red (O).}
\label{fig:4}
\end{figure}

\noindent processes in thermal transport of Ge$_2$Sb$_2$Te$_5$.

\noindent Figure \ref{fig:4} describes the frequency variation of both 3ph and 4ph scattering rates for different temperature. Phonon scattering rates have been found to vary roughly 100 and 1000 times for 3ph and 4ph scattering rates respectively within the studied temperature range (30 K - 600 K). The difference between 3ph and 4ph scattering rates are seen to decrease and strong competition between these two scattering rates is realized via the gradual collapse of two scattering rates as we increase the temperature beyond $T$ = 100 K (Fig \ref{fig:4}(d), (e), (f)). We attribute this trend to the significant 4ph scattering phase space beyond $T$ = 100 K, following the weighted phase space analyzes (Fig \ref{fig:3}).  The 4ph scattering rate is seen to gradually eclipse the 3ph scattering rate at high temperature, mostly in the O frequency regime (Fig \ref{fig:4}). In the M regime, both of the scattering rates are comparable as 4ph scattering is triggered due to the coupling between acoustic and optical bands (Fig \ref{fig:4}). In the regime O, till $\omega$ $\sim$ 122 $cm^{-1}$, the dominance of 4ph scattering over 3ph scattering is seen to prevail, especially at higher temperatures. This is the regime in the phonon dispersion which consists of many flat optical bands (see Appendix \ref{appen1} for a detailed description of flat and optical bands and the corresponding allowed 3ph processes). These flat bands are shown to enable significant four phonon scattering with optical bands, consistent with an earlier study \cite{xie2020first}. At frequencies beyond 122 $cm^{-1}$, dispersive optical

\begin{figure}[H]
\centering
\includegraphics[width=0.5\textwidth]{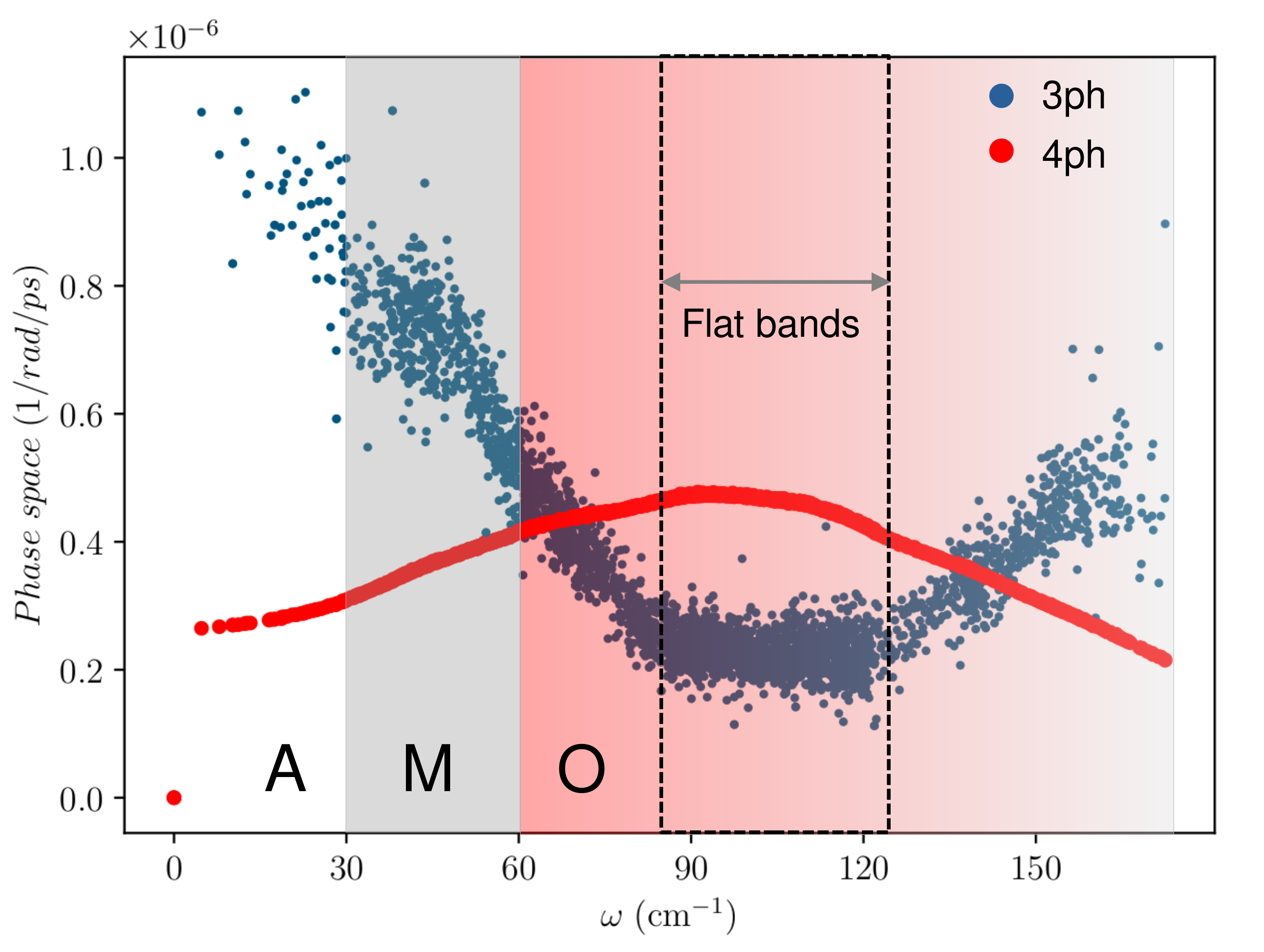}
\caption{The variation of three-phonon (3ph) and four-phonon (4ph) scattering phase spaces with phonon frequencies. Different phonon frequency regimes are marked using white (A) and grey (M) and light red (O). Flat bands within regime `O' are distinguished via the dashed regime with 86 $cm^{-1}$ $<$ $\omega$ $<$ 122 $cm^{-1}$.}
\label{fig:5}
\end{figure}

\noindent  bands are seen to reappear, influencing the 4ph scattering to become weaker compared to the 3ph scattering rates. To elucidate this relative behavior of 3ph and 4ph scattering rates more transparently, we study the phase space evolution of both 3ph and 4ph scattering processes, separating out the temperature effect to make a one to one correspondence between the dispersion and the phase space variation with frequency (Fig \ref{fig:5}). 

\noindent Figure \ref{fig:5} distinguishes two different frequency domains corresponding to the altering hierarchy of 3ph and 4ph phase space. The phonon frequency regime 86 $cm^{-1}$ $<$ $\omega$ $<$ 122 $cm^{-1}$ (denoted as `Flat bands' in Fig \ref{fig:5}) simultaneously hosts the minima and maxima indicating the increment of 4ph scattering phase space at the cost of decreasing 3ph phase space. Recalling our dispersion picture, this regime constitutes closely packed dense flat optical bands which is almost 70 $\%$ of the total optical bands. Sparsely located dispersive optical bands are found to accommodate only 10 $\%$ of the total optical bands, constituting 122 $cm^{-1}$ $<$ $\omega$ $<$ 174 $cm^{-1}$. We demonstrate the allowed 3ph processes due to the flat bands and find that the extremely small bandwidth of flat bands restrict OOO scattering completely and AOO scattering partially in the frequency regime of 86 $cm^{-1}$ $<$ $\omega$ $<$ 122 $cm^{-1}$ (see Appendix \ref{appen1}). While flat optical bands can restrict the OOO 3ph scattering channels due to the conservation of momentum and energy, this restriction does not prevent the 4ph scattering channels concerning AAAO, AAOO, AOOO and OOOO (A, O denote acoustic and optical phonons). This feature hints at the simultaneous occurrence of 3ph-minima and 4ph-maxima in phase space variation with frequency. At higher frequencies 

\begin{figure}[H]
\centering
\includegraphics[width=0.5\textwidth]{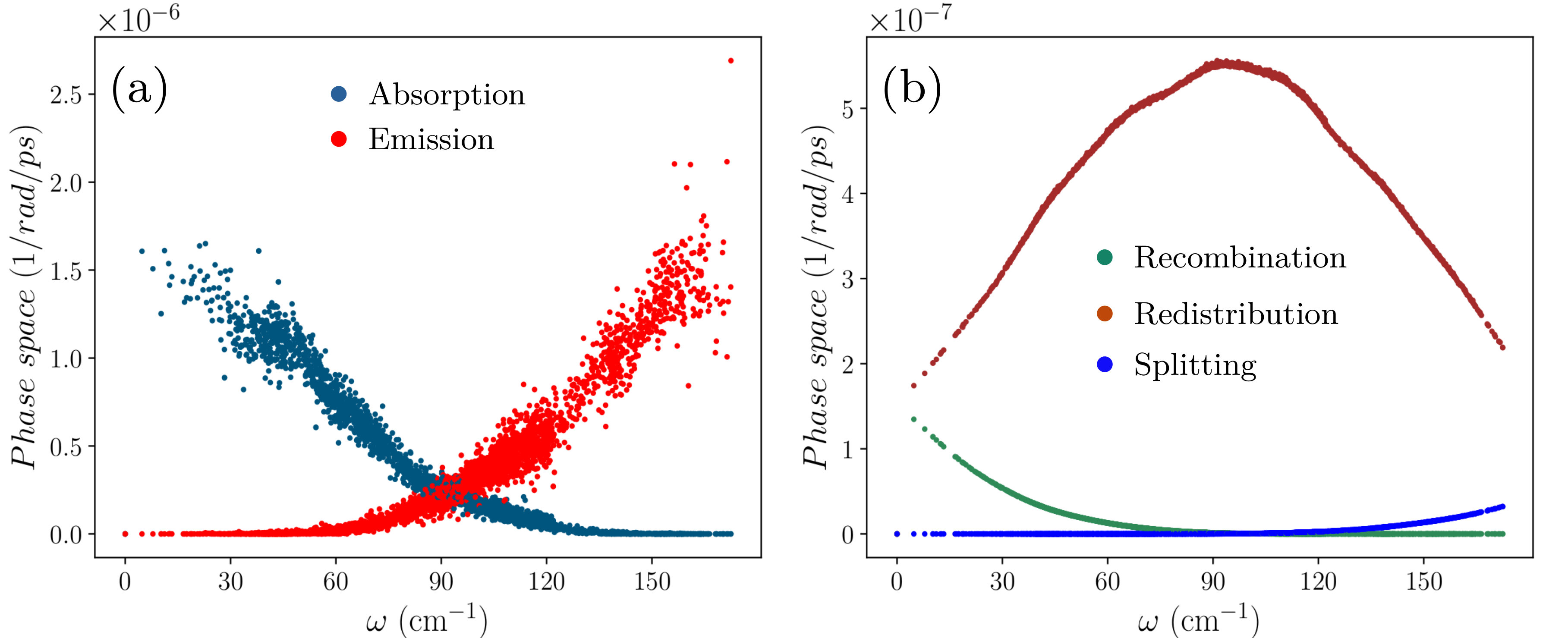}
\caption{The variation of (a) 3ph and (b) 4ph phase spaces with phonon frequencies for different scattering channels for Ge$_2$Sb$_2$Te$_5$.}
\label{fig:6}
\end{figure}  

\noindent beyond 122 $cm^{-1}$, dispersive optical bands reappear which rejuvenate the 3ph processes. As a result 3ph phase space is seen to dominate again over 4ph. The Petrov structure of Ge$_2$Sb$_2$Te$_5$ is also found to possess a dominating 4ph weighted phase space over its 3ph counterpart although in a less pronounced fashion compared to that of the Kooi structure (see Supplementary Fig S5 and Fig S6 for details).

\noindent To further understand the 3ph-minima and 4ph-maxima in their respective phase space, we deconstruct the separate roles of different scattering channels in Fig \ref{fig:6}. 3ph scattering consists of absorption ($\textbf{q}$ + $\textbf{q}_1$ = $\textbf{q}_2$) and emission ($\textbf{q}$ = $\textbf{q}_1$ + $\textbf{q}_2$) channels (Fig \ref{fig:6}(a)) while 4ph scattering (Fig \ref{fig:6}(b)) comprises of three scattering channels namely (a) recombination ($\textbf{q}$ + $\textbf{q}_1$ + $\textbf{q}_2$ = $\textbf{q}_3$), (b) redistribution ($\textbf{q}$ + $\textbf{q}_1$ = $\textbf{q}_2$ + $\textbf{q}_3$) and (c) splitting ($\textbf{q}$ = $\textbf{q}_1$ + $\textbf{q}_2$ + $\textbf{q}_3$) along with their momentum destroying counterparts via the reciprocal lattice vector $\textbf{G}$. Here $\textbf{q}$, $\textbf{q}_1$, $\textbf{q}_2$, $\textbf{q}_3$ define phonon wave vectors. Figure \ref{fig:6}(a) depicts that the acoustic phonons with low frequency modes (regime `A') enable absorption processes which seems to gradually decrease as optical phonons overlap with the acoustic ones in the regime `M'. In regime `O' (60 $cm^{-1}$ $<$ $\omega$ $<$ 174 $cm^{-1}$), optical modes with higher frequencies are more prone to undergo emission processes which dominates at higher frequencies while absorption phase space going to zero (Fig \ref{fig:6}(a)). The aforementioned 3ph-minima are found to originate as a result of the crossing between absorption and emission phase space around 90-100 $cm^{-1}$. Also, Fig \ref{fig:6}(b) shows that 4ph redistribution phase space peaks around the same frequency where 3ph phase space shows minima. This is a result of the occurrence of flat optical bands in this frequency regime (86 $cm^{-1}$ $<$ $\omega$ $<$ 122 $cm^{-1}$) that severely restrict OOO and weaken AOO processes but do not prevent optical phonons to participate in four-phonon processes. Rather flat bands enhance the redistribution processes as energy conservation between four phonons is preferable in this scenario. Similar feature was observed in  AlSb and BAs \cite{yang2019stronger}. This causes a simultaneous dip and peak in 3ph and 4ph phase space respectively. Recombination process is seen to dominate the whole 4ph phase space spectrum of Ge$_2$Sb$_2$Te$_5$ along with a small contribution from recombination and splitting processes at low and high frequencies respectively (Fig \ref{fig:6}(b)). These 3ph and 4ph phase space distributions get mildly modified with temperature as can be seen via their weighted phase space distributions (see Supplementary Fig S7).

\noindent We end this section by studying the importance of phonon-isotope scattering in crystalline Ge$_2$Sb$_2$Te$_5$ compared to both 3ph and 4ph scattering rates (see Fig S2 in the supplementary information). As mentioned in an earlier section, phonon-isotope scattering rates can be computed following the work by Tamura \cite{tamura1983isotope}, where the mode dependent isotope scattering rate varies as $\omega^{2}$ but remains independent of temperature. Therefore, at low temperature (30 K $<$ $T$ $<$ 60 K), phonon-isotope scattering contributes significantly to increase the resistive scattering processes and helps reducing the lattice thermal conductivity. However, as temperature is further raised (see Fig S2(d),(e),(f) in the supplementary), both 3ph and 4ph scattering start dominating the total phonon scattering processes due to their strong temperature dependencies ($\tau_{3ph}^{-1}$ $\sim$ $T$, $\tau_{4ph}^{-1}$ $\sim$ $T^{2}$ \cite{feng2017four}).

\subsection{The role of normal and Umklapp scattering: three and four-phonon processes}{\label{subsection:NU}}

\noindent To further investigate the nature of 3ph and 4ph scattering and their competitive roles in dictating thermal transport, we study each of the 3ph and 4ph scattering channels distinguished as separate normal (N) and Umklapp (U) scattering processes. We compare the relative weights between N and U processes by calculating the thermodynamic average of scattering rates for N and U processes separately for both 3ph and 4ph scattering via 

\begin{equation}{\label{eq:scatrateavg}}
    \Gamma^{i} = \langle \tau_{i}^{-1} \rangle_{ave} = \frac{\sum_{\lambda}C_{\lambda}\tau_{\lambda i}^{-1}}{\sum_{\lambda}C_{\lambda}}
\end{equation}

\begin{widetext}
\begin{figure}[H]
    \centering
    \includegraphics[width=1.0\textwidth]{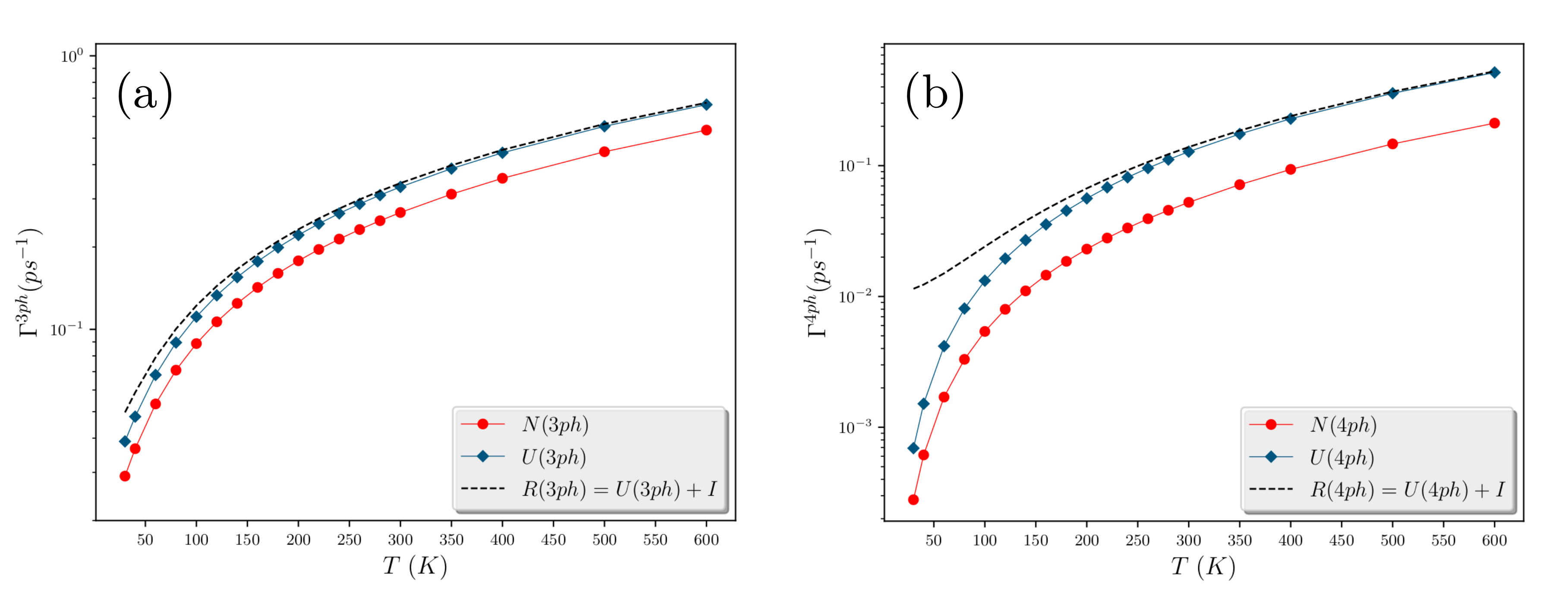}
    \caption{Thermodynamic average of normal (N) and Umklapp (U) phonon scattering rates ($\Gamma^{i} =\langle \tau_{i}^{-1} \rangle_{ave}$) are presented as a function of temperature for both (a) three-phonon ($\Gamma^{3ph}$) and (b) four-phonon ($\Gamma^{4ph}$) scattering events. For each of the cases, $R$ (black dashed line) denotes the total resistive scattering rate involving Umklapp (U) and isotope (I) scattering.}
    \label{fig:7}
\end{figure}
\end{widetext}

\noindent Here, $\lambda$ defines phonon modes ($\textbf{q}$, $j$) comprising wave vector $\textbf{q}$ and branch $j$. Index $i$ denotes either N or U scattering rates used. $C_\lambda$ is the modal heat capacity, given by 
\begin{equation}
    C_\lambda = k_{B} \left(\frac{\hbar\omega_{\lambda}}{k_{B}T}\right)^{2} \frac{exp(\hbar\omega_{\lambda}/k_{B}T)}{[exp(\hbar\omega_{\lambda}/k_{B}T) -1]^2}
\end{equation}
where, $T$ denotes temperature, $\hbar$ is the reduced Planck constant and $k_B$ is the Boltzmann constant. N scattering enables momentum conserving phonon scattering, while U and I scattering bring resistivity in thermal conduction. Therefore, the relative strength of N and U scattering is pivotal to understand the underlying phonon thermal conduction as well as to capture peculiar phenomena like phonon hydrodynamics\cite{guyer1966thermal, ghosh2022phonon}. Figure \ref{fig:7} presents the comparison between the average N and U phonon scattering rates as a function of temperature for both 3ph (Fig \ref{fig:7}(a)) and 4ph scattering (Fig \ref{fig:7}(b)). For both three and four-phonon processes in Ge$_2$Sb$_2$Te$_5$, dominant trend of U scattering over N scattering is found throughout the  

\onecolumngrid

\noindent studied temperature range. Adding isotope scattering is seen to further increase the resistive scattering compared to the N scattering events. The relative strength of U scattering to N scattering is found to be almost constant in the whole temperature range with a factor of 1.3 for 3ph scattering which increases up to around 2.5 for 4ph scattering (see supplementary Fig S8 for details). Therefore, the presence of resistive 4ph scattering is expected to further reduce the thermal conductivity of Ge$_2$Sb$_2$Te$_5$.

\begin{figure}[H]
    \centering
    \includegraphics[width=0.5\textwidth]{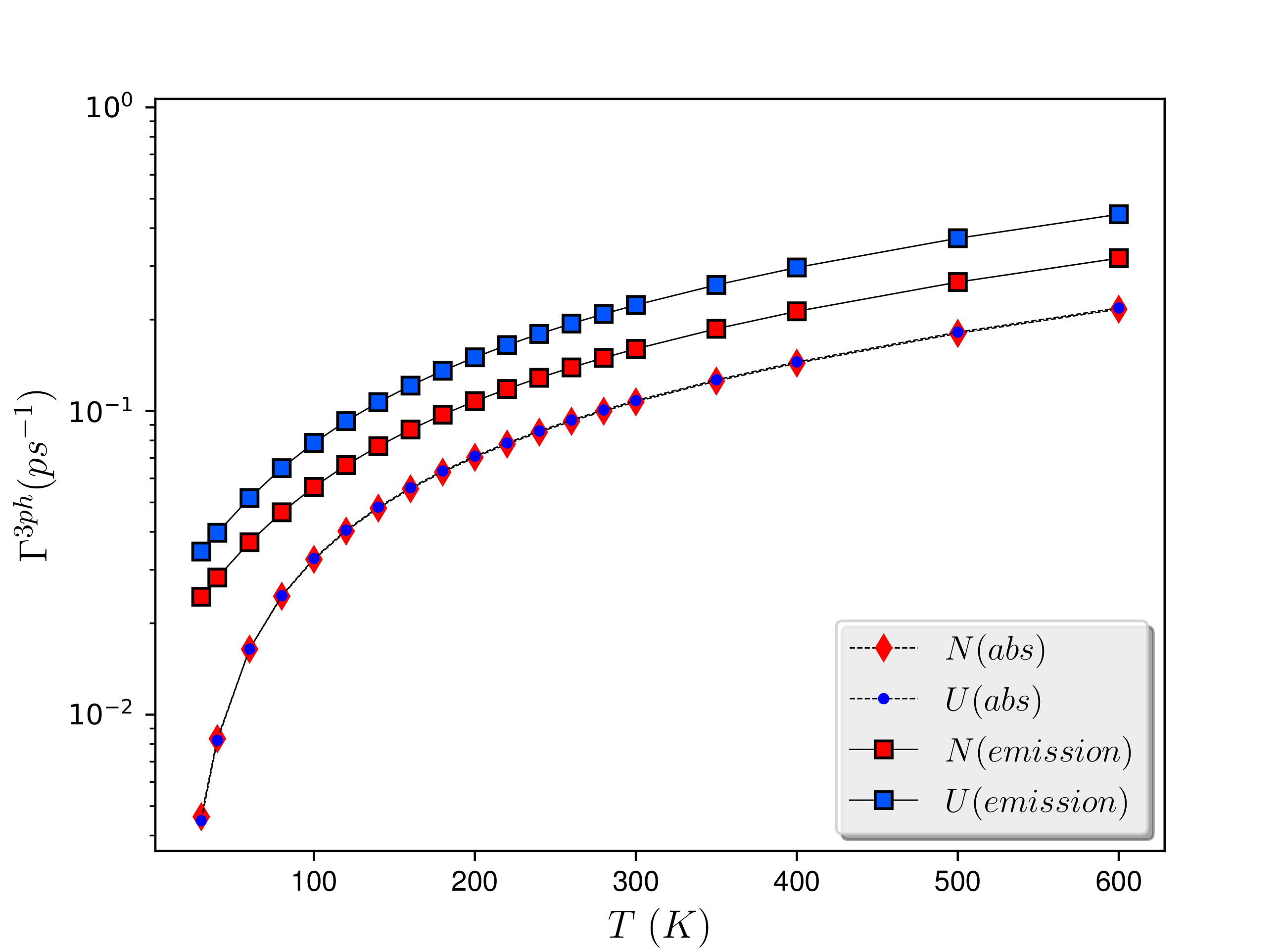}
    \caption{The variation of 3ph average scattering rates ($\Gamma^{3ph}$) with temperature for different scattering channels (absorption and emission) for both normal and Umklapp scattering events.}
    \label{fig:8}
\end{figure}

\noindent Although U scattering is seen dominating over N in both 3ph and 4ph scattering events, it is instructive to check their contributions to different scattering channels to understand the phonon transport mechanism. Figure \ref{fig:8} represents the temperature variation of thermodynamically averaged 3ph absorption and emission scattering rates ($\Gamma^{3ph}$) for both N and U scattering events. N and U absorption scattering rates are seen to overlap and merge together (Fig \ref{fig:8}). Consequently, the emission process determines the relative strength between N and U 3ph scattering. The emission processes corresponding to U scattering events are seen to dominate over the N scattering events throughout the temperature range studied. This stems from the fact that emission processes are mostly activated at high frequency regime, where optical phonons with higher momentum and large wave vectors are more prone to satisfy $\textbf{q}$ = $\textbf{q}_1$ + $\textbf{q}_2$ + $\textbf{G}$, where $\textbf{G}$ is the reciprocal lattice vector.

\begin{widetext}
\begin{figure}[H]
\centering
\includegraphics[width=1.0\textwidth]{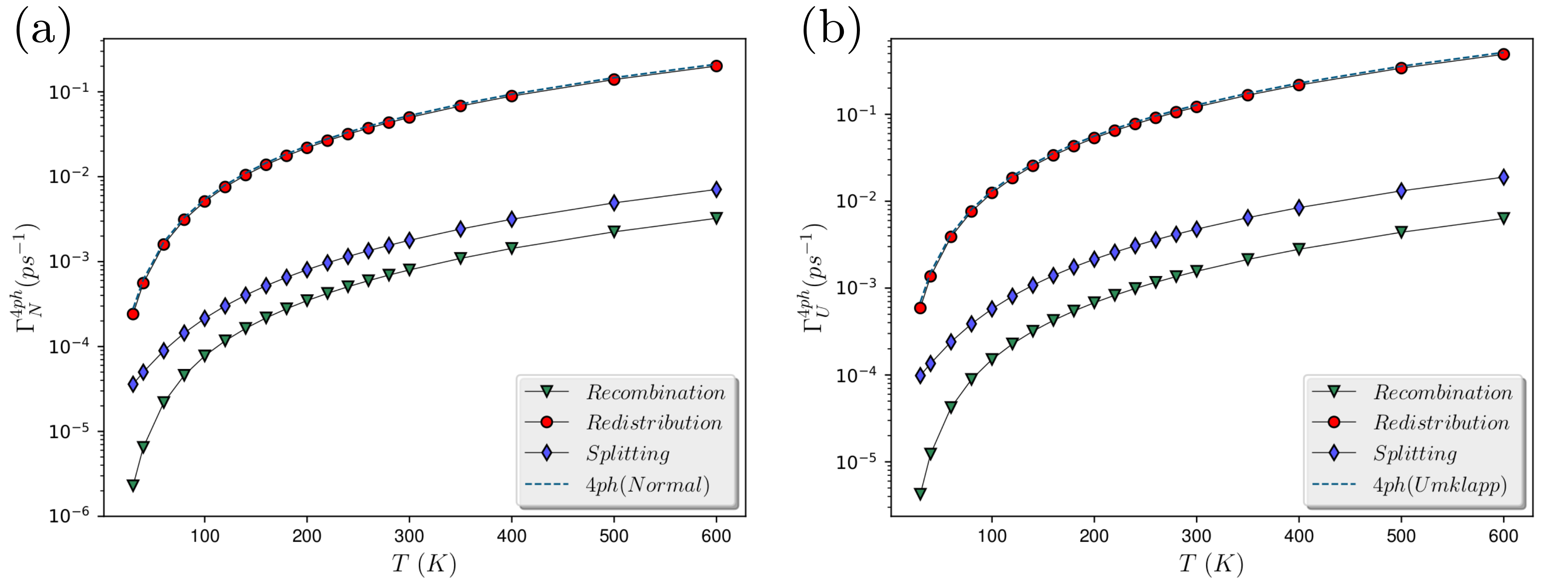}
\caption{The temperature variation of 4ph average scattering rates ($\Gamma^{4ph}$) for different scattering channels (recombination, redistribution, splitting) for (a) normal ($\Gamma_{N}^{4ph}$) and (b) Umklapp scattering ($\Gamma_{U}^{4ph}$). The total 4ph normal and Umklapp scattering rates are denoted via the black dashed lines in (a) and (b) respectively.}
\label{fig:9}
\end{figure}
\end{widetext}

\noindent Figure \ref{fig:9}(a), (b) present the variation of average 4ph scattering rates with $T$ for different scattering channels for N and U scattering respectively. We clearly observe that both N and U scattering rates corresponding to redistribution 4ph processes are strong and they dominate over the two other scattering channels. Moreover, the collapse of the total average scattering rates (black dashed lines in Fig \ref{fig:9}(a) and (b)) to the redistribution scattering (red circles in Fig \ref{fig:9}(a) and (b)) suggests that average N and U scattering rates can be approximated as the contribution coming from redistribution alone with negligible contributions from other two scattering channels. Recombination processes are found to be weakest among these three channels. This dominating 4ph contribution by redistribution processes comes from the large phase space of the redistribution scattering channel as

\onecolumngrid

\noindent observed in Fig \ref{fig:6}(b). Looking at the phonon-mode contributions at different frequencies, we find that the `M' frequency regime is seen to be mostly populated by redistribution as well as recombination processes that exhibit U scattering (supplementary Fig S10). In `O' frequency regime, recombination channels are mostly blocked while splitting processes are found to gain importance along with the dominating redistribution scattering rates. At very high frequencies splitting processes aggravate due to the increasing weighted scattering phase space (see supplementary Fig S7(c), (d)). At higher temperature, marginal change is observed on this frequency distribution of scattering processes (Fig S7(d) and Fig S10(b), (d), (f) in the Supplementary information).

\subsection{Thermal conductivity and its variation with temperature for three and four-phonon processes: Mode analyses}{\label{subsection:kappa}}

\begin{widetext}
\begin{figure}[H]
\centering
\includegraphics[width=1.0\textwidth]{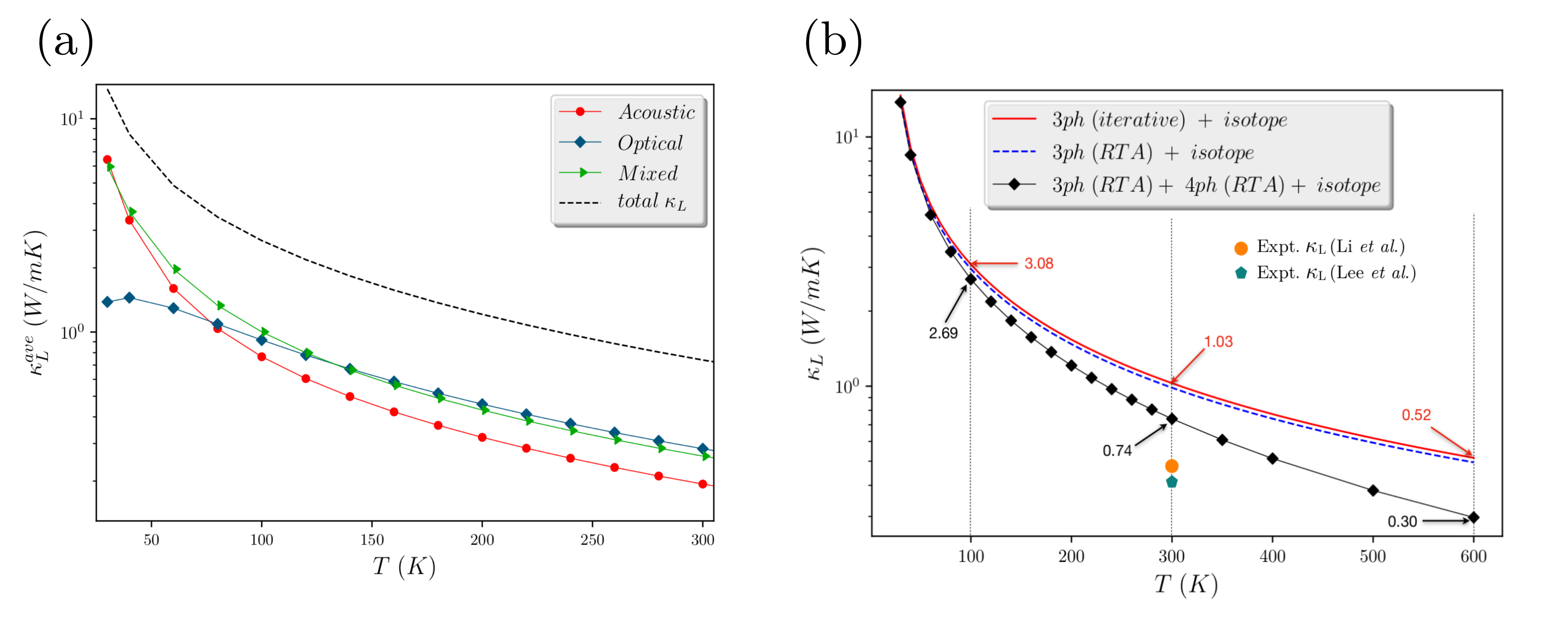}
\caption{(a) The variation of average lattice thermal conductivity with temperature including 3ph, 4ph as well as isotope scattering processes, distinguished via contributions from acoustic, mixed and optical frequency regimes for Ge$_2$Sb$_2$Te$_5$. (b) Average lattice thermal conductivity as a function of temperature for Ge$_2$Sb$_2$Te$_5$. Three different curves correspond to the inclusion of different scattering processes; Red solid line: Inclusion of three-phonon (3ph) iteratively and the phonon-isotope scattering, blue dashed line: three-phonon using RTA and isotope scattering, black solid line with diamonds: three-phonon in RTA,  four-phonon in RTA and isotope scattering. Experimental values of $\kappa_L$, obtained by Li $\textit{et al.}$ \cite{li2023temperature}($\kappa_L$ = 0.47 W/mK) and Lee $\textit{et al.}$ \cite{lee2013phonon}($\kappa_L$ = 0.42 W/mK) are also presented via orange circle and green pentagon respectively.}
\label{fig:10}
\end{figure}
\end{widetext}

\noindent After analyzing the phonon scattering rates and their relative importance, we calculate the variation of lattice thermal conductivity with temperature incorporating both three as well as four-phonon scattering rates for Ge$_2$Sb$_2$Te$_5$. Isotope scattering rates, which are temperature independent, are also included within the scattering terms via Matthiessen’s rule \cite{kaviany2014heat}. 

\begin{equation}{\label{eq:mathiessen}}
    \frac{1}{\tau_{\lambda}} = \frac{1}{\tau_{\lambda}^{3ph}} + \frac{1}{\tau_{\lambda}^{4ph}} + \frac{1}{\tau_{\lambda}^{I}}
\end{equation}

\noindent where $\lambda$ defines each mode and $\tau_{\lambda}^{3ph}$, $\tau_{\lambda}^{4ph}$ and $\tau_{\lambda}^{I}$ denote 3ph, 4ph and phonon-isotope scattering lifetimes respectively. Figure \ref{fig:10}(a) represents the variation of average lattice thermal conductivity ($\kappa_L^{ave}$) with temperature, distinguishing different contributions coming from acoustic (A), optical (O) and mixed (M) frequency regimes defined via dispersion bands. We see a similar optical phonon dominance in $\kappa_L$ as that of the earlier case considering only 3ph and isotope scattering (shown in Fig \ref{fig:2}(b)). At low temperature, as expected, acoustic contribution (red circles in Fig \ref{fig:10}(a)) is large and the contributions from acoustic and mixed modes (regime M, green triangles in Fig \ref{fig:10}(a)) dominate. Beyond 100 K, the dominance of acoustic modes decays while optical contribution (blue diamonds in Fig \ref{fig:10}(a)) takes the lead with almost equal contributions from the mixed regimes. Figure \ref{fig:10}(b) records a value of 1.03 W/mK at 300 K and 0.52 W/mK at 600 K, for the average $\kappa_L$, calculated using only three-phonon and isotope scattering rates (red solid line for iterative and blue dashed line for RTA), consistent with earlier works by Ibarra $\textit{et al.}$ \cite{ibarra2018ab}. Campi $\textit{et al.}$ \cite{campi2017first} and Mukhopadhyay $\textit{et al.}$ \cite{mukhopadhyay2016optic} obtained a value of 1.2 W/mK and 1.4 W/mK respectively, considering an ideal crystal with the inclusion of only 3ph scattering for the Kooi structure at 300 K. Further, the significant implication of incorporating four-phonon scattering in $\kappa_L$ can also be seen in Fig \ref{fig:10}(b). By including four-phonon processes with three-phonon and isotope scattering events, a significant drop in $\kappa_L$ is observed at high temperature. At 100 K, we record a reduction of $\sim$ 13 $\%$ ($\kappa_L$ = 2.69 W/mK), followed by a $\sim$ 28 $\%$ ($\kappa_L$ = 0.74 W/mK) and a striking $\sim$ 42 $\%$ ($\kappa_L$ = 0.3 W/mK) decrement in $\kappa_L$ at 300 K and 600 K respectively.

\begin{figure}[H]
\centering
\includegraphics[width=0.5\textwidth]{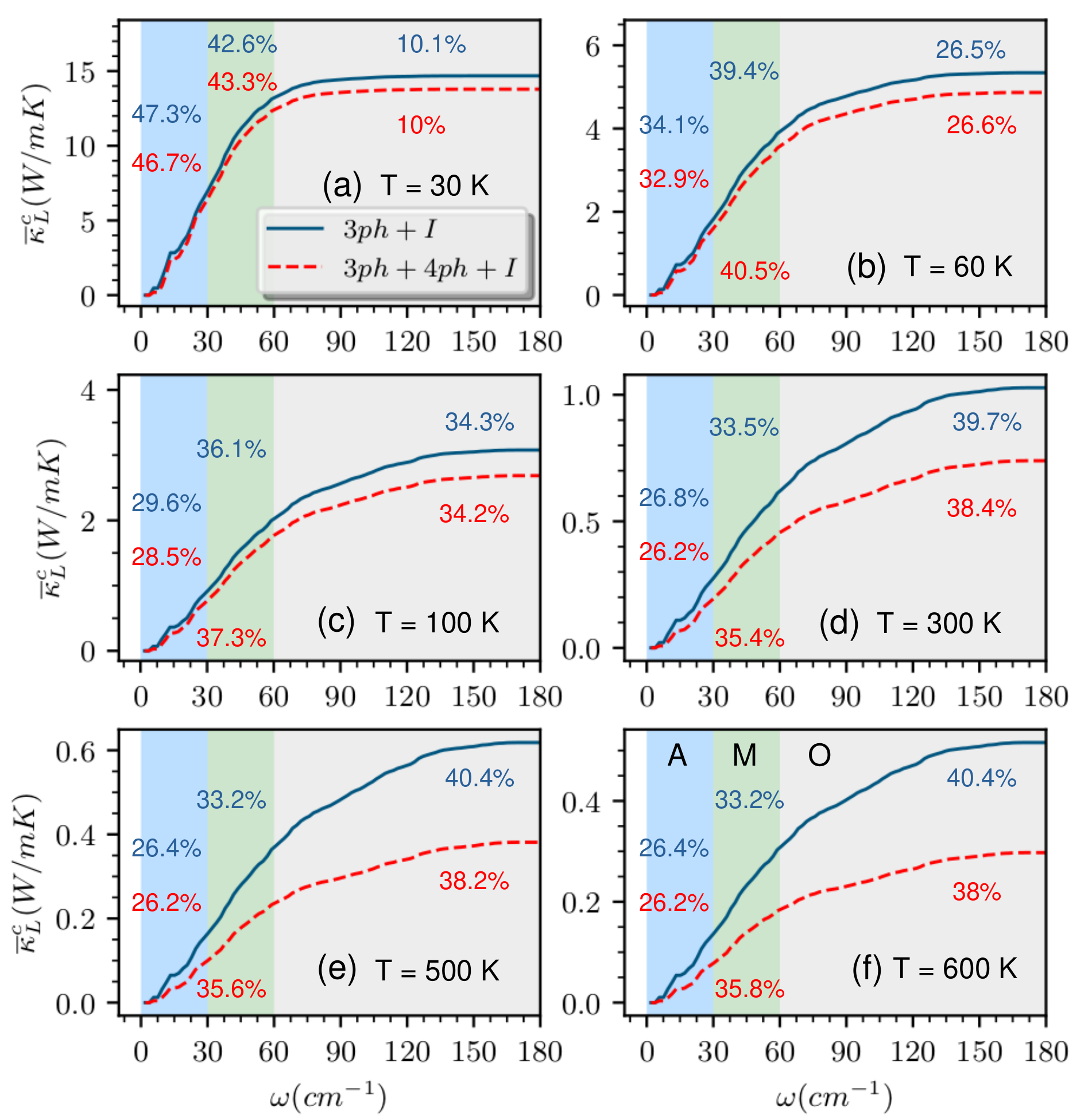}
\caption{Frequency variation of cumulative lattice thermal conductivity ($\overline{\kappa}_{L}^{c}$) using 3ph+I (blue solid lines) as well as 3ph+4ph+I (red dashed lines) for (a) $T$ = 30 K, (b) $T$ = 60 K, (c) $T$ = 100 K, (d) $T$ = 300 K, (e) $T$ = 500 K and (f) $T$ = 600 K. Blue, green and grey shaded regions denote A, M and O frequency regimes. The percentage contributions to $\kappa_L$ coming from these regimes, are presented via blue (3ph+I) and red (3ph+4ph+I) for each $T$.}
\label{fig:11}
\end{figure}

\noindent Further, we study the alteration of relative contributions coming from acoustic (A), mixed (M) and optical (O) frequency regimes with and without 4ph scattering processes. Figure \ref{fig:11} represents the cumulative lattice

\onecolumngrid

\noindent thermal conductivity, averaged over all directions ($\overline{\kappa}_{L}^{c}$), as a function of phonon frequency, including and excluding 4ph scattering along with 3ph and isotope scattering events for different temperature (see supplementary Fig S11, Fig S12 for detailed $\overline{\kappa}_{L}^{c}$ variation with frequency along a-b plane ($\kappa_L^x$) and along c ($\kappa_L^z$), with and without 4ph scattering). Firstly, we describe the evolution of $\overline{\kappa}_{L}^{c}$ using 3ph and I scattering (blue solid lines in Fig \ref{fig:11}(a)-(f)). At $T$ = 30 K, acoustic phonons seem to contribute mostly (47.3 $\%$) to $\kappa_L$ followed by the M regime (42.6 $\%$). As the temperature is increased this trend seems to alter with optical phonons dominating $\kappa_L$ (40 $\%$). We note that the M regime also contributes significantly ($\sim$ 33 $\%$) to $\kappa_L$. At temperatures beyond 300 K, this trend continues with the contributions from different frequency regimes to the total $\kappa_L$ are found to be saturated (regime A: 26.4 $\%$, regime M: 33.2 $\%$, regime O: 40.4 $\%$). To explore the modification of this picture after adding 4ph scattering, modified $\overline{\kappa}_{L}^{c}$ is also presented (red dashed lines in Fig \ref{fig:11}(a)-(f)). Initially, at low temperature ($T$ = 30 K), we see a similar trend of acoustic phonon dominated phonon transport for Ge$_2$Sb$_2$Te$_5$ (regime A: 46.7 $\%$, regime M: 43.3 $\%$, regime O: 10 $\%$). As we increase the temperature, optical modes seem to dominate $\kappa_L$, similar to that of the cubic anharmonicity. However, the contributions from M and O regimes are found to be slightly affected after the inclusion of 4ph scattering events. When temperature is raised beyond 300 K, the contributions from A, M and O frequency regimes are found to be saturated around 26.2 $\%$, 35.8 $\%$ and 38 $\%$ respectively (see red dashed lines in Fig \ref{fig:11}(d), (e), (f)). We observe that at high temperature, including 4ph scattering does not alter the contributions from the regime A. However, the relative contributions of M and O regimes are marginally interchanged. These results indicate that including 4ph scattering processes does not alter the optical mode-dominance in $\kappa_L$ for Ge$_2$Sb$_2$Te$_5$. Rather, including 4ph scattering affects both acoustic and optic modes non-selectively, retaining the significant contribution of optical modes in $\kappa_L$ of Ge$_2$Sb$_2$Te$_5$. This is in sharp contrast to high-$\kappa_L$ material like AlSb \cite{yang2019stronger}, where optical phonon-dominated $\kappa_L$ gets almost washed out while incorporating 4ph scattering processes. 

\begin{figure}[H]
\centering
\includegraphics[width=0.5\textwidth]{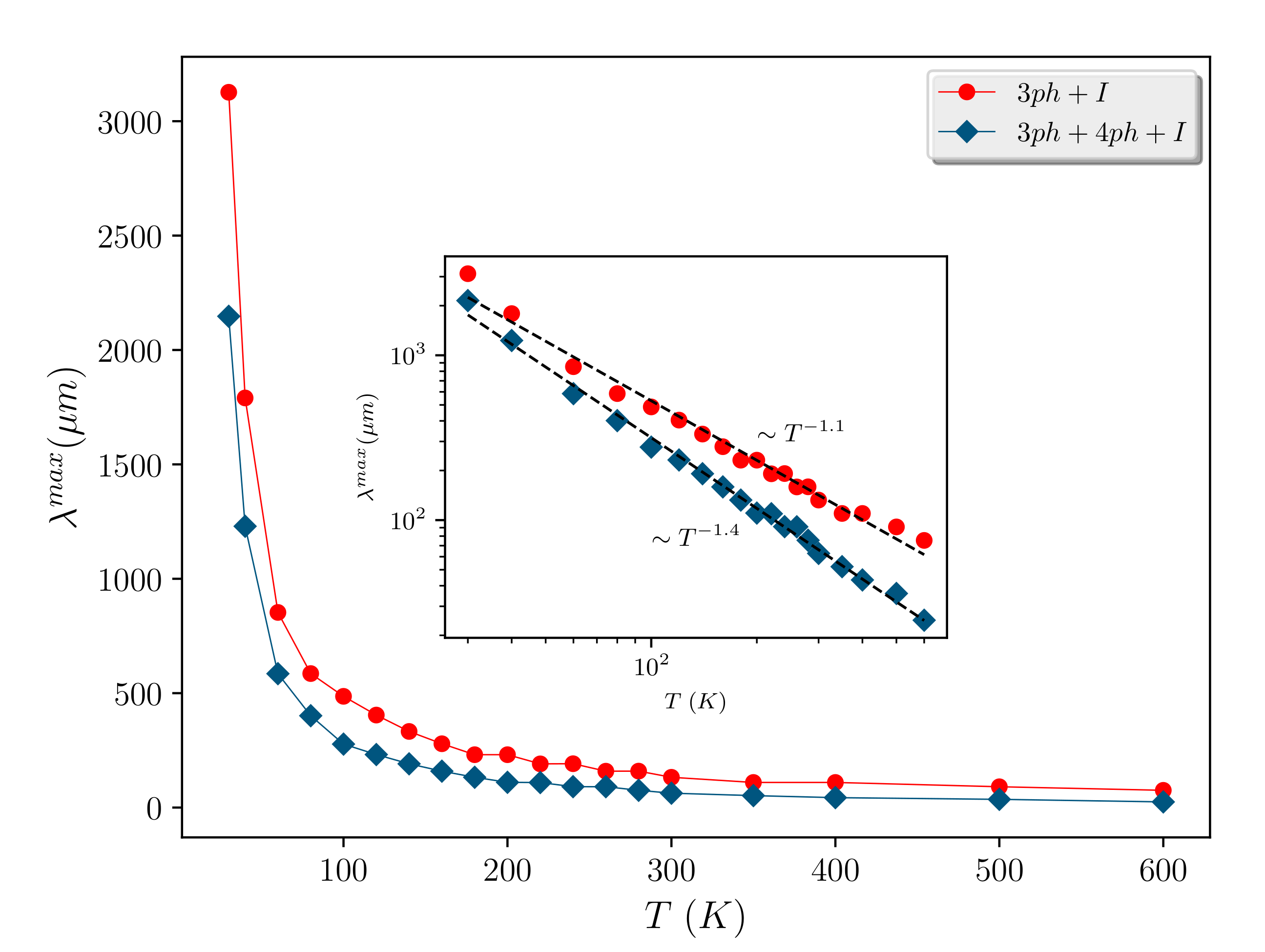}
\caption{The variation of maximum mean free path ($\lambda^{max}$) as a function of temperature for Ge$_2$Sb$_2$Te$_5$ both including and excluding 4ph scattering. Inset shows the scaling relation of $\lambda^{max}$ with $T$ with (3ph+4ph+I) or without (3ph+I) 4ph scattering.}
\label{fig:12}
\end{figure}

\noindent Next, to investigate the size dependence of the thermal conductivity of Ge$_2$Sb$_2$Te$_5$ and its implications due to the presence of 4ph scattering, we present the variation of the maximum mean free path ($\lambda^{max}$) as a function of temperature (Fig \ref{fig:12}). We note that grain boundaries, their scattering with phonons as well as the thickness were found to influence $\kappa_L$ for Ge$_2$Sb$_2$Te$_5$ \cite{reifenberg2007thickness, lee2011thermal}. Therefore, studying size-dependence of thermal conduction serves not only the purpose of understanding the MFP distribution and their contributions to $\kappa_L$, but also of designing nanostructures by assessing the maximum mean free path inside the material (see supplementary Fig S13, Fig S14 for the variation of cumulative $\kappa_L$ with mean free

\noindent path along a-b plane ($\kappa_L^x$) and along c ($\kappa_L^z$), with and without 4ph scattering). Figure \ref{fig:12} shows that $\lambda^{max}$ gradually decreases with temperature. Over the temperature range studied here, $\lambda^{max}$ is seen to vary from 3.1 $\mu m$ to 75.3 nm for 3ph+I and from 2.1 $\mu m$ to 24.7 nm for 3ph+4ph+I. Further, $\lambda^{max}$ is found to vary roughly as $T^{-1}$ ($\lambda^{max}$ $\sim$ $T^{-1.1}$) while including 3ph and I scattering (red circles and their fitted dashed line in the inset of Fig \ref{fig:12}). This trend emerges due to the fact that higher temperature populates more phonons according to phonon distribution (BE distribution), inducing more scattering between phonons to reduce the mean free path between them. However, including 4ph scattering is seen to modify this negative slope with even a higher value of the exponent ($\lambda^{max}$ $\sim$ $T^{-1.4}$, blue diamonds and their fitted dashed line in the inset of Fig \ref{fig:12}). We recall that mean free paths and phonon group velocities are connected as $\lambda$ = $\textbf{v}\tau$. Therefore, this is consistent with a higher order scattering scenario where the scattering rate varies as $\tau^{-1}$ $\sim$ $T^{\alpha}$ where $\alpha$ $>$ 1.

\begin{figure}[H]
\centering
\includegraphics[width=0.5\textwidth]{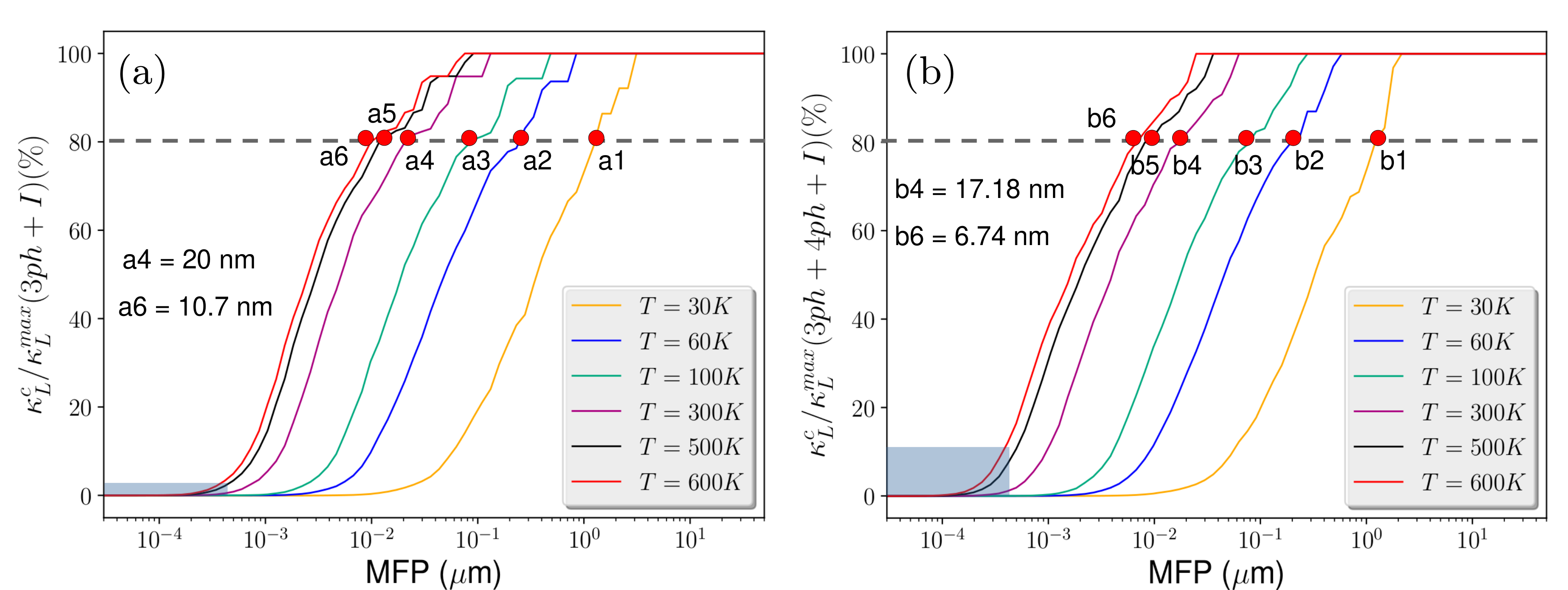}
\caption{The percentage contribution of cumulative lattice thermal conductivity ($\kappa_L^c$) to that of the total $\kappa_L$ is presented as a function of maximum allowed phonon mean free paths (MFP) including (a) 3ph, I scattering and (b) 3ph, 4ph and I scattering for different temperature. The dashed lines in both representations correspond to the 80 $\%$ contribution to $\kappa_L$. The corresponding maximum allowed MFPs for $T$ = 30 K, 60 K, 100 K, 300 K, 500 and 600 K are denoted via a1, a2, a3, a4, a5, a6 in (a) and b1, b2, b3, b4, b5, b6 in (b). For comparison, the values correspond to $T$ = 300 K and 600 K are shown via a4, a6 respectively in (a) and via b4, b6 respectively in (b). The blue shaded regimes in (a) and (b) represent the contributions from MFPs less than or equal to the lattice parameter a (4.193 \AA).}
\label{fig:13}
\end{figure}

\noindent To explain this size effect in a more comprehensive way, we present the percentage of cumulative contribution to $\kappa_L$ as a function of  maximum allowed MFPs for different temperature in Fig \ref{fig:13}. Also, we represent the effect of 

\noindent 3ph+I and 3ph+4ph+I separately in Fig \ref{fig:13}(a) and (b) respectively. Around 80 $\%$ contribution to $\kappa_L$ is seen to come from phonons with MFPs less than 1.26 $\mu$m at 30 K (a1 in Fig \ref{fig:13}(a)), which goes down to 10.7 nm at 600 K (a6 in Fig \ref{fig:13}(a)) for 3ph and I scattering events. Considering 4ph scattering, 80 $\%$ of $\kappa_L$ is found to be contributed by MFPs less than 1.24 $\mu$m at 30 K (b1 in Fig \ref{fig:13}(b)) and 6.74 nm at 600 K (b6 in Fig \ref{fig:13}(b)). As temperature increases, MFP decreases due to vigorous Umklapp collisions between the phonons, causing lower MFP phonons to contribute more significantly to $\kappa_L$. The presence of significant four-phonon scattering at high temperature causes further hindrance to the phonons to propagate and the same contribution to $\kappa_L$ is seen to come from phonons with lower MFPs. At room temperature, MFPs below 20 nm are found responsible for 80 $\%$ of $\kappa_L$ (a4 in Fig \ref{fig:13}(a)) which changes to 17.18 nm while including 4ph scattering (b4 in Fig \ref{fig:13}(b)). This finding relates to a recent discussion on the room temperature experimental thermal conductivity of both cubic and hexagonal Ge$_2$Sb$_2$Te$_5$ \cite{li2023temperature}. Time-domain thermoreflectance measurements with grains of 10-15 nm were found to influence negligibly the room-temperature $\kappa_L$ of hexagonal Ge$_2$Sb$_2$Te$_5$ \cite{li2023temperature}. This behavior can be rationalized using the fact that almost 80 $\%$ of the $\kappa_L$ of hexagonal Ge$_2$Sb$_2$Te$_5$ are seen to come from phonons with mean free paths below $\sim$ 17 nm (20 nm if considering only 3ph+I) as shown in Fig \ref{fig:13}. This suggests that the experimental $\kappa_L$ (0.47 W/mK) at 300 K has negligible effects due to phonon-grain boundary scattering. To realize serious effects of grain boundary scattering in hexagonal Ge$_2$Sb$_2$Te$_5$, grains need to be smaller to enhance grain-boundary scattering (scattering rate $\tau_{B}^{-1}$ = $v_{g}/L$). For example, grains of size 3 nm can effectively 

\noindent reduce 60 $\%$ of the $\kappa_L$ at room temperature as can be seen via Fig \ref{fig:13}(b). Here, we note that the obtained $\kappa_L$ at high T ($\geq$ 600 K) is sensitive to the validity of phonon BTE as we include the 4ph scattering. Figure \ref{fig:13} shows growing contributions of phonons with MFPs comparable to the lattice parameter ($a$) from 3.4 $\%$ to 12.4 $\%$, while including 4ph scattering (blue shaded regimes in Fig \ref{fig:13}). Though $\lambda^{max}$ (24.7 nm) is well above $a$, non-negligible contributions from MFPs ($\leq$ $a$) hint at the closeness of $\kappa_L$ to the amorphous limit ($\kappa_{min}$ $\sim$ 0.27 W/mK \cite{kusiak2022temperature}) where the usage of phonon BTE seems questionable.   

\noindent The negligible effect of grain-boundary scattering as well as the significant presence of four-phonon scattering trigger the discussion concerning the gap between experiments and theory to estimate the thermal conductivity of Ge$_2$Sb$_2$Te$_5$. Experimental lattice thermal conductivity had been found to be around 0.42-0.48 W/mK \cite{lyeo2006thermal, tsafack2011electronic, lee2013phonon, li2023temperature}, which is very close to the minimal thermal conductivity using the Cahill model \cite{cahill1992lower}, considering disorder-dominated structure of Ge$_2$Sb$_2$Te$_5$. This raises concern about the Boltzmann transport equation approach to compute the thermal conductivity as was mentioned earlier \cite{campi2017first}. However, our findings suggest that phonon-phonon scattering has a strong influence with the presence of quartic anharmonicity that can bring down $\kappa_L$ to 28 $\%$ at 300 K and 42 $\%$ at 600 K. More importantly, 4ph and disorder scattering (both phonon-vacancy \cite{ratsifaritana1987scattering} and phonon-isotope scattering \cite{tamura1983isotope}) are two completely different mechanisms following $T^2$ and $T$-independent behavior respectively. Therefore, excluding quartic anharmonicity at room temperature and beyond, can lead to a crucial overestimation of the predicted disorder content at the cost of an underestimation of

\begin{figure}[H]
    \centering
    \includegraphics[width=0.5\textwidth]{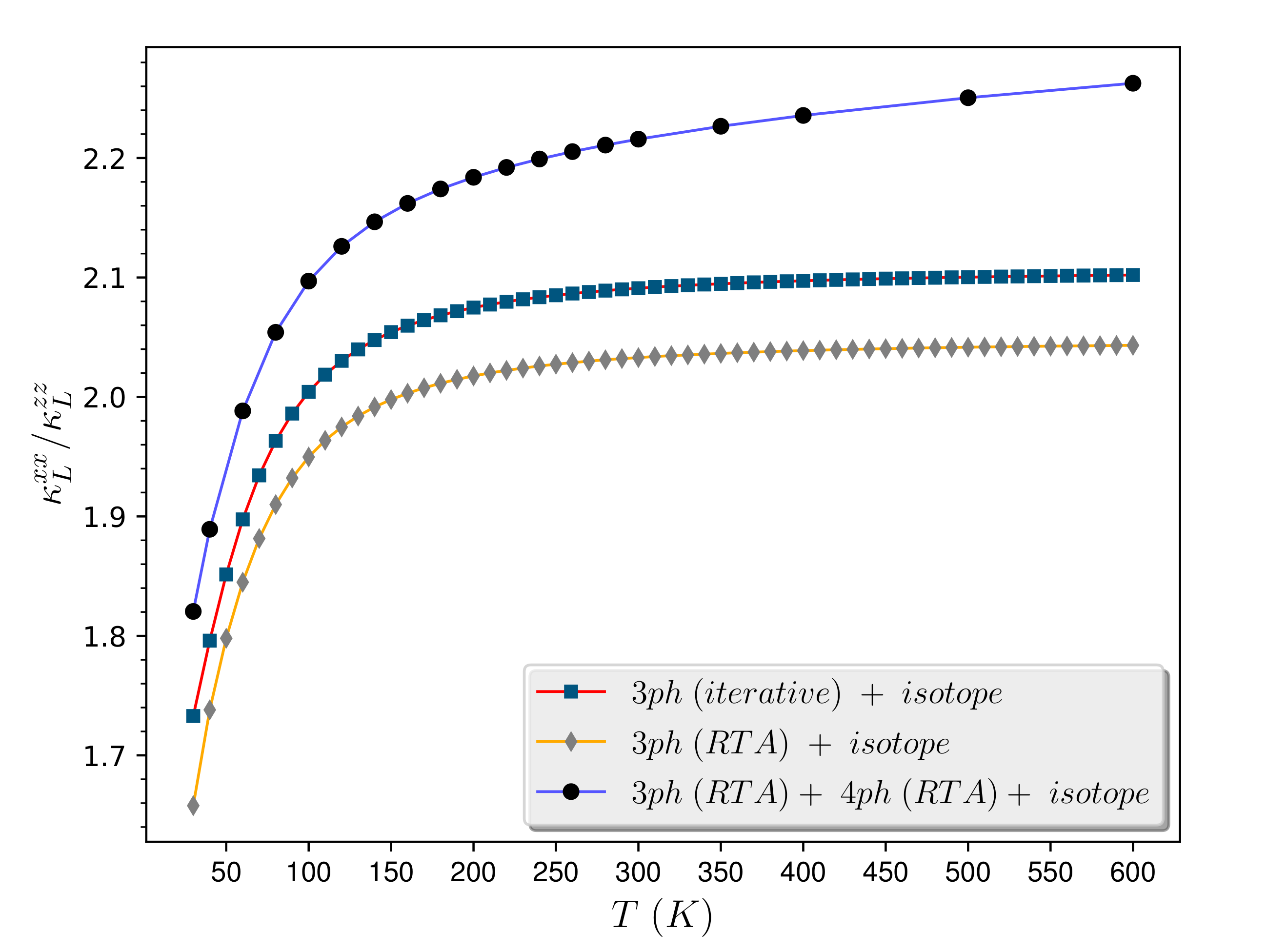}
    \caption{The variation of the anisotropy factor ($\kappa_L^{xx}/\kappa_L^{zz}$) with temperature for crystalline Ge$_2$Sb$_2$Te$_5$ using various scattering processes to calculate $\kappa_L$. Squares: 3ph (iterative) + isotope scattering, diamonds: 3ph (RTA) + isotope scattering, circles: 3ph (RTA) + 4ph (RTA) + isotope scattering.}
    \label{fig:14}
\end{figure}

\noindent phonon-phonon contribution to $\kappa_L$ of Ge$_2$Sb$_2$Te$_5$. Further, experiments \cite{bragaglia2016metal} had also shown to produce fairly ordered structure for Ge$_2$Sb$_2$Te$_5$ by weakening the randomness by ordering the vacancies. Therefore, higher order anharmonicity in Ge$_2$Sb$_2$Te$_5$ is crucial to estimate thermal behavior of these highly ordered materials.

\subsection{Anisotropy under three and four-phonon scattering}{\label{subsection:anisotropy}}

\noindent The anisotropy in the thermal conductivity for GST materials can propose efficient designing of PCM devices that hinders the cross talks between neighboring PCM cells \cite{lee2011thermal}. Indeed, significant differences between $x$ and $z$ component of $\kappa_L$ are recorded while assessing its variation with phonon frequency and mean free path (see supplementary Fig S11 - Fig S14). Figure \ref{fig:14} presents the temperature variation of $\kappa_L$-anisotropy by using the ratio of $\kappa_L$ along $a$-axis to that of the $c$-axis. We see a marginal increment in the anisotropy factor ($\kappa_{L}^{xx}$/$\kappa_{L}^{zz}$) from 2.04 to 2.1 with temperature, employing RTA and iterative method respectively within the three-phonon picture. We notice that the factor remains almost constant ($\sim$ 2) at temperature beyond 300 K, consistent with the work of Mukhopadhyay $\textit{et al.}$ \cite{mukhopadhyay2016optic}. However, various experimental and theoretical studies, having consistent $\kappa_L$, feature a wide variation of the anisotropy factor from 1.9 to 4.6 \cite{mukhopadhyay2016optic, campi2017first, miao2022remarkable, sklenard2021electronic, lee2011thermal} for Ge$_2$Sb$_2$Te$_5$ at 300 K. Further, including 4ph scattering is found to increase the anisotropy factor from 2.04 to 2.21 (Fig \ref{fig:14}) which is a marginal ($\sim$ 8 $\%$) increment. This is an indication of the negligible influence of scattering rates in the 

\begin{figure}[H]
    \centering
    \includegraphics[width=0.5\textwidth]{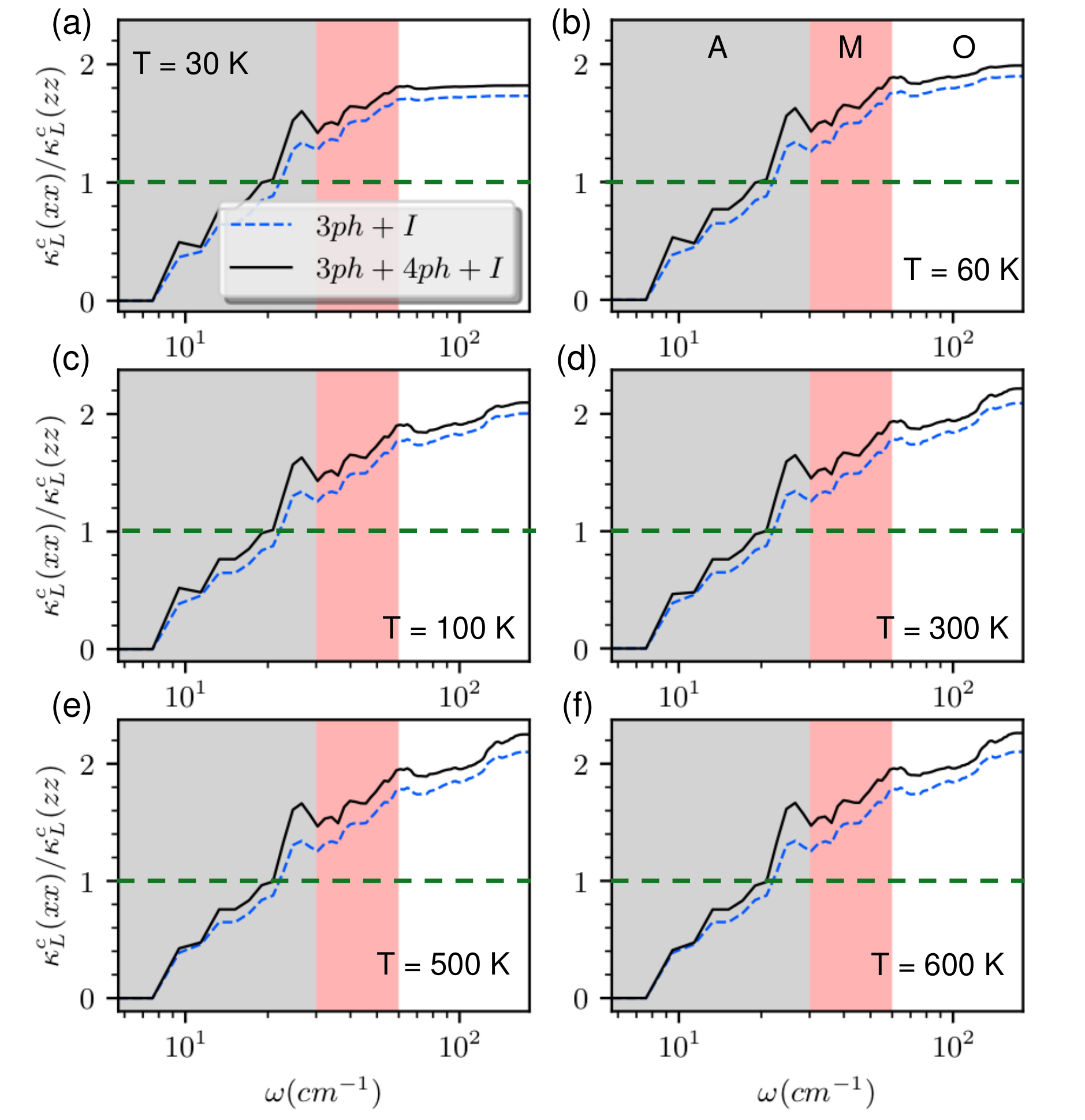}
    \caption{The variation of the anisotropy factor ($\kappa_L^{c}(xx)/\kappa_L^{c}(zz)$) with frequency for crystalline Ge$_2$Sb$_2$Te$_5$ at different temperature. Blue dashed lines and black solid lines denote the scattering processes by 3ph+I and 3ph+4ph+I respectively. Different frequency regimes A, M, O are also shown via shaded regions. The dashed line in each case refers to the isotropic behavior in $\kappa_L^{c}(xx)/\kappa_L^{c}(zz)$ beyond which anisotropy is measured.}
    \label{fig:15}
\end{figure}

\noindent $\kappa_L$-anisotropy. We evaluate the changes in anisotropy as we include 4ph scattering in terms of frequency evolution of the cumulative $\kappa_L$. Figure \ref{fig:15} describes the variation of the anisotropic factor ($\kappa_{L}^{c} (xx)$/$\kappa_{L}^{c} (zz)$) as a function of frequencies both including (black solid lines) and excluding 4ph scattering events (blue dashed lines). We quantify the anisotropic behavior by the increment of the factor beyond 1 (shown via dashed green lines in Fig \ref{fig:15}). A marginally higher value of anisotropy factor is seen to evolve with frequency while including 4ph scattering to the existing 3ph and isotope scattering processes. The increment of this factor towards a final value comprises of contributions from all three frequency regimes. At low temperature of 30 K (Fig \ref{fig:15}(a)), A, M and O regimes contribute to $\sim$ 37 $\%$, $\sim$ 55 $\%$ and $\sim$ 8 $\%$ respectively to the total anisotropy factor, while including 4ph scattering makes these contributions to $\sim$ 49 $\%$, $\sim$ 46 $\%$ and $\sim$ 5 $\%$ respectively. At $T$ = 300 K (Fig \ref{fig:15}(d)), these contributions are found to modify as $\sim$ 24 $\%$, $\sim$ 48 $\%$ and $\sim$ 28 $\%$ respectively for 3ph+I scattering and $\sim$ 39 $\%$, $\sim$ 39 $\%$ and $\sim$ 22 $\%$ respectively for 3ph+4ph+I scattering events. At even higher temperature ($T$ = 600 K, Fig \ref{fig:15}(f)), with and without four-phonon scattering, the anisotropy factor is found to be distributed as follows: (i) excluding four-phonon scattering: $\sim$ 25 $\%$, $\sim$ 48 $\%$ and 

\begin{figure}[H]
\centering
\includegraphics[width=0.5\textwidth]{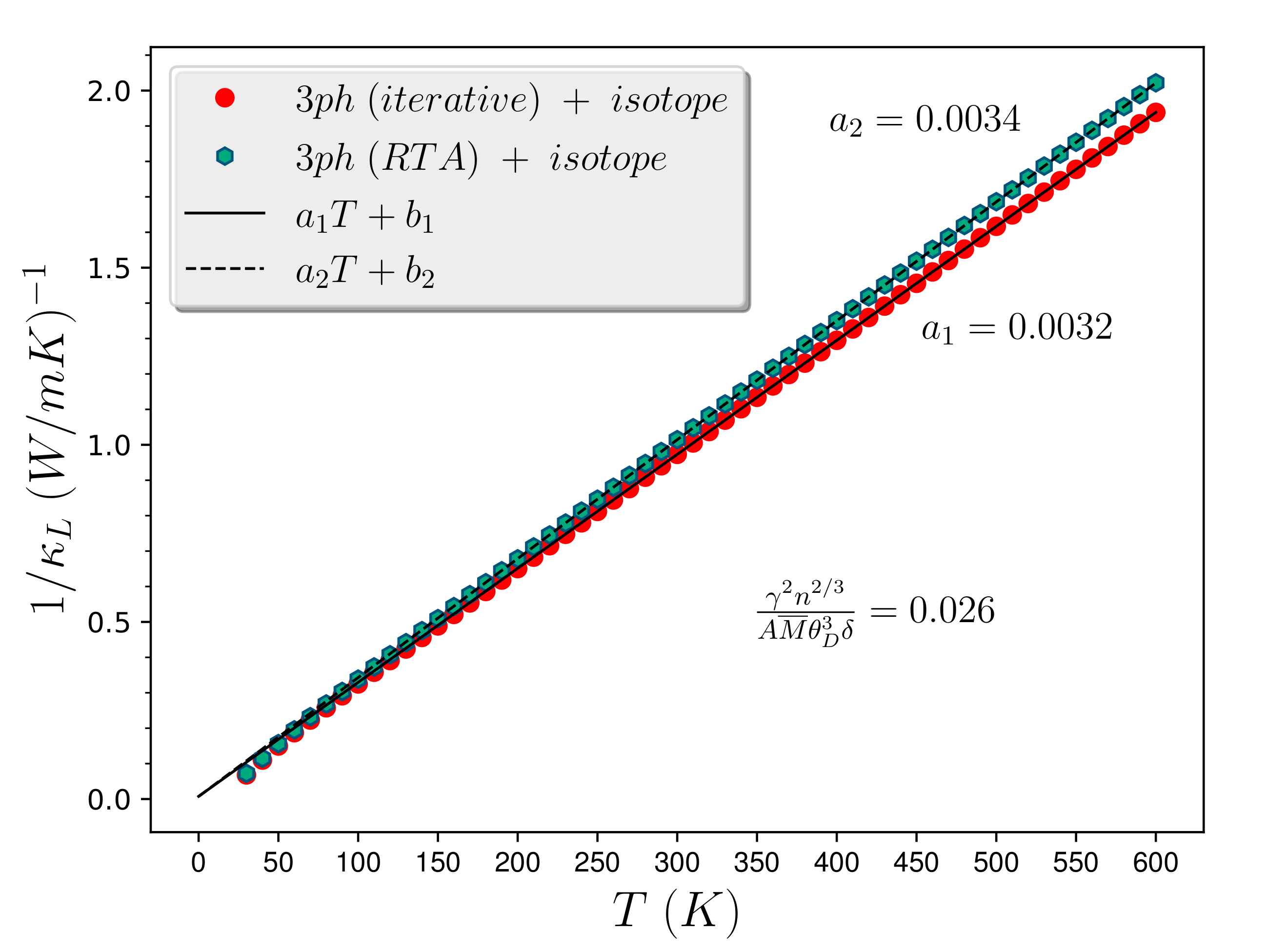}
\caption{The variation of inverse lattice thermal conductivity ($1/\kappa_L$) with temperature for crystalline Ge$_2$Sb$_2$Te$_5$ considering only three-phonon and isotope scattering using iterative (red circles) and RTA method (green hexagons). $T$-linear high temperature regime is obtained with a slope $a_1$ = 0.0032 and $a_2$ = 0.0034 for iterative and RTA methods respectively. The coefficient from the Slack model is also shown (0.026).}
\label{fig:16}
\end{figure}

\noindent $\sim$ 27 $\%$ contributions are coming from A, M and O regimes respectively. (ii) Including four-phonon scattering: $\sim$ 37 $\%$, $\sim$ 40 $\%$ and $\sim$ 23 $\%$ contributions are coming from A, M and O regimes respectively. Irrespective of the included scattering processes, the contribution from optical phonons (regime O) is seen to significantly increase at the cost of a reduced contribution from the A regime. The variation of the relative contributions from different frequency regimes is found to be weakly-dependent on temperature as $\kappa_L$-anisotropy in Ge$_2$Sb$_2$Te$_5$ is predominantly dictated by the anisotropy in phonon group velocity. Mukhopadhyay $\textit{et al.}$\cite{mukhopadhyay2016optic} used a density regression method using PDOS to capture the anisotropy in phonon group velocity along the $a$ and $c$ axes. We observe a consistent behavior with a little steeper acoustic and optical branches along $\Gamma$-$M$ or $\Gamma$-$K$ compared to that of the $\Gamma$-$A$ in the dispersion relation for Ge$_2$Sb$_2$Te$_5$ (Fig \ref{fig:1}(C)), indicating an anisotropy in $v_g$ between its in-plane ($a$-$b$ plane) and out-of-plane (along $c$) components in the hexagonal Brillouin zone.

\subsection{Thermal conductivity scaling at high temperature: Relation between scattering rates and fundamental constants}{\label{subsection:scaling}}

Throughout this study we unfold two significant features of Kooi-Ge$_2$Sb$_2$Te$_5$: (a) significant domination of optical phonon in thermal transport and (b) considerable effect of quartic anharmonicity especially at high temperature. Crucial effect of these factors on the thermal transport leads us to introspect their further implications in the scaling properties of thermal conductivity with temperature for crystalline Ge$_2$Sb$_2$Te$_5$. At high temperature, in the kinetic thermal transport regime, $\kappa_L$ follows a $T^{-1}$ scaling \cite{kaviany2014heat} as a result of dominating Umklapp scattering between phonons. This high temperature behavior in nonmetallic materials had been analytically derived by Slack and Morelli \cite{slack1979thermal, morelli2006high, morelli2002thermal}. This model was quite successful in describing the high temperature $T^{-1}$ behavior of $\kappa_L$ for a wide range of systems for which acoustic phonons in general and Umklapp scattering in particular, play a dominant role to $\kappa_L$ \cite{slack1979thermal}. Therefore, often $T^{-1}$ behavior of nonmetallic systems at high temperature range is attributed to this model. Considering only acoustic phonons as heat transport carriers, Slack and Morelli \cite{morelli2002thermal} developed the intrinsic $\kappa_L$ of a solid in a temperature range where Umklapp scattering is dominant as 

\begin{equation}{\label{eq:slack}}
    \kappa_{L} = A \frac{\overline{M}\theta_{D}^{3}\delta}{\gamma^{2}n^{2/3}T} 
\end{equation}
with \begin{equation}
    A = \frac{2.43\times 10^{-6}}{1-0.514/\gamma+0.228/\gamma^{2}}
\end{equation}
where $n$ is the number of atoms per unit cell, $\delta^{3}$ is volume per atom ($\delta$ is in the unit of \AA), $\overline{M}$ is the average atomic mass of the unit cell of the solid, $\gamma$ is Gruneisen parameter and $\theta_D$ is the Debye temperature of the material. 

\noindent Firstly, we analyze the high temperature $T^{-1}$ behavior of our first-principles result of $\kappa_L$ considering only 3ph and isotope scattering for Ge$_2$Sb$_2$Te$_5$. We recall the optical phonon dominance in $\kappa_L$ at high temperature, revealed by our \textit{ab initio} studies. However, simultaneously we have also found the dominance of Umklapp scattering at high temperature regime. This abundance of Umklapp resistance is seen to manifest itself via a $T^{-1}$ scaling of $\kappa_L$ at high temperature as shown in Figure \ref{fig:16}. The linear fit between 1/$\kappa_L$ and $T$ at high temperature regime yields the coefficients $a_1$ (= 0.0032) or $a_2$ (= 0.0034) considering iterative and RTA method respectively (Figure \ref{fig:16}). To compare these coefficients to that of the Slack-Morelli model, we compute the coefficient from Eq.\ref{eq:slack} using different parameters presented in Table \ref{tab:table2}. However, the Slack-Morelli coefficient yields a value 0.026, which is higher than the fitted \textit{ab initio} results by an order of magnitude. Although the abundance of Umklapp scattering dictates $T^{-1}$ scaling of $\kappa_L$ for both cases, the mismatch between the coefficients can be attributed to the fact that Slack model is analytically derived for acoustic branches while our results show optical phonon dominated thermal transport in Kooi-Ge$_2$Sb$_2$Te$_5$.

\begin{widetext}
    \begin{figure}[H]
    \centering
    \includegraphics[width=1.0\textwidth]{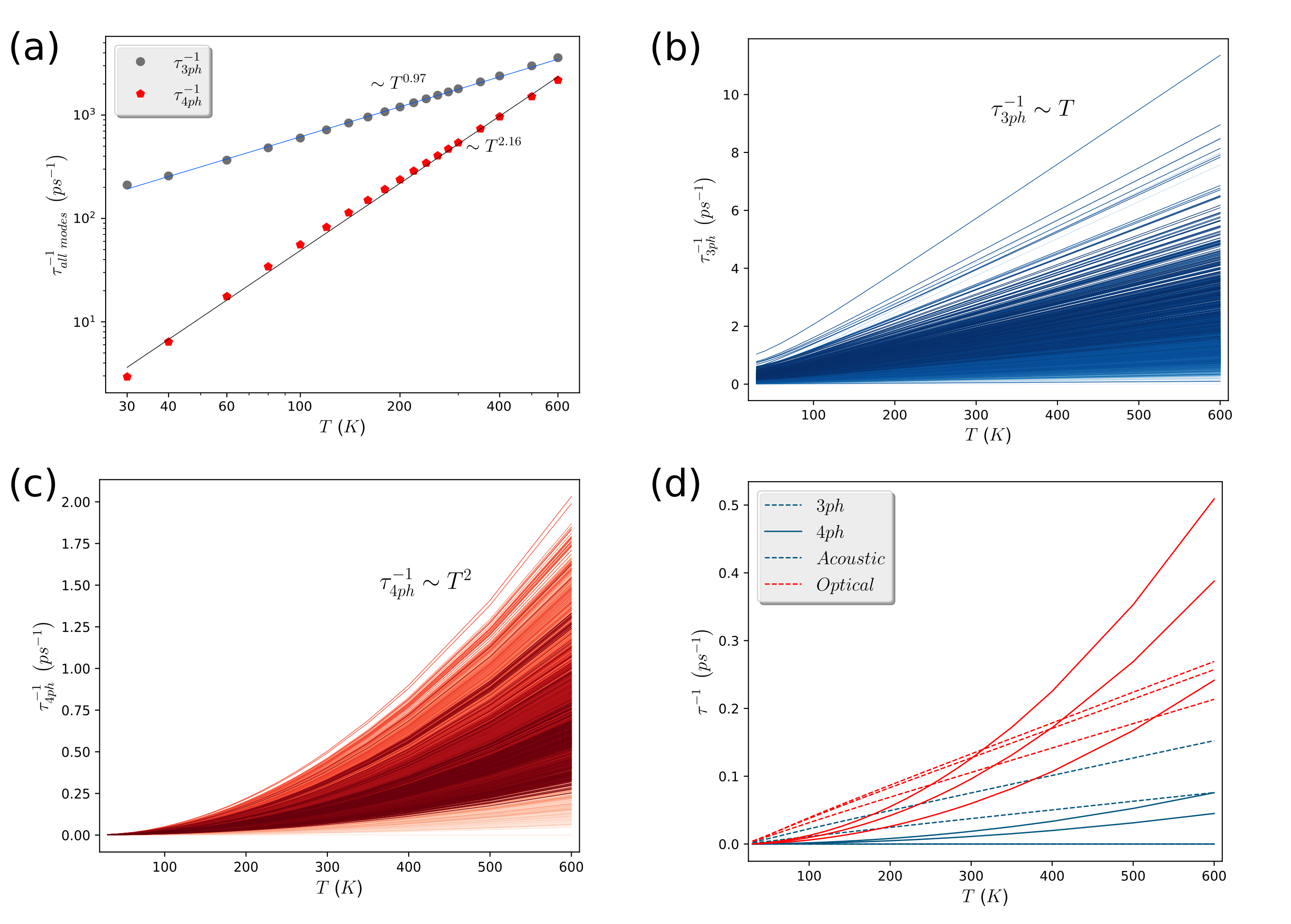}
    \caption{(a) Total three-phonon and four-phonon scattering rates by summing up the contributions from all the modes ($\tau_{all\hspace{0.1cm} modes}^{-1}$), as a function of temperature. $T$-scaling for 3ph and 4ph scattering rates at high temperature are also shown via blue ($T^{0.97}$) and black fitted curves ($T^{2.16}$) respectively. (b) The variation of 3ph scattering rates ($\tau_{3ph}^{-1}$) with temperature for each phonon mode shows $\sim$ $T$ scaling. (c) The variation of 4ph scattering rates ($\tau_{4ph}^{-1}$) with temperature for each phonon mode shows $\sim$ $T^2$ scaling. (d) 3ph (dashed lines) and 4ph (solid lines) scattering rates are presented for first six phonon bands. Scattering rates for both acoustic and optical bands are seen to follow $\sim$ $T$ for 3ph and $\sim$ $T^2$ for 4ph scattering processes. Blue: acoustic phonons, Red: Optical phonons. For the lowest lying acoustic bands for 3ph and 4ph scattering, dashed and solid lines are merged and are indistinguishable.}
    \label{fig:17}
\end{figure}
\end{widetext}

\noindent The dominance of optical modes in thermal transport was observed to feature a deviation of $\kappa_L$ from its $T^{-1}$ scaling at higher temperature range in monolayer GaN \cite{qin2017anomalously}. Although, $\kappa_L$ of Ge$_2$Sb$_2$Te$_5$ is seen to possess $T^{-1}$ scaling, including four-phonon can radically change the scaling behavior. As both specific heat and group velocity are temperature independent, the temperature scaling of $\kappa_L$ at high temperature relies on the temperature dependencies of $\tau_{3ph}^{-1}$ and $\tau_{4ph}^{-1}$. Figure \ref{fig:17}(a) presents the temperature variation of the total three-phonon and four-phonon scattering rates of all modes. At high temperature regime, an almost $T$-linear scaling is observed for $\tau_{3ph}^{-1}$ ($\sim$ $T^{0.97}$) while $\tau_{4ph}^{-1}$ retrieves a $\sim$ $T^{2.16}$ scaling. Further, we present 3ph and 4ph scattering rates for each mode as a function of temperature to more clearly visualize the $T$ and $T^2$ variation of $\tau_{3ph}^{-1}$ (Figure \ref{fig:17}(b)) and $\tau_{4ph}^{-1}$ (Figure \ref{fig:17}(c)) respectively, featuring a generic trend observed in other materials\cite{feng2017four}. Figure \ref{fig:17}(d) shows that regardless of acoustic or optical phonons, specific scaling relations are followed by 3ph and 4ph scattering processes. 3ph and 4ph scattering rates for the first six phonon bands are seen to feature linear (dashed lines in Fig \ref{fig:17}(d)) and quadratic (solid lines in Fig \ref{fig:17}(d)) temperature variation respectively, regardless of acoustic (blue lines in Fig \ref{fig:17}(d)) or optical modes (red lines in Fig \ref{fig:17}(d)).

\begin{figure}[H]
\centering
\includegraphics[width=0.5\textwidth]{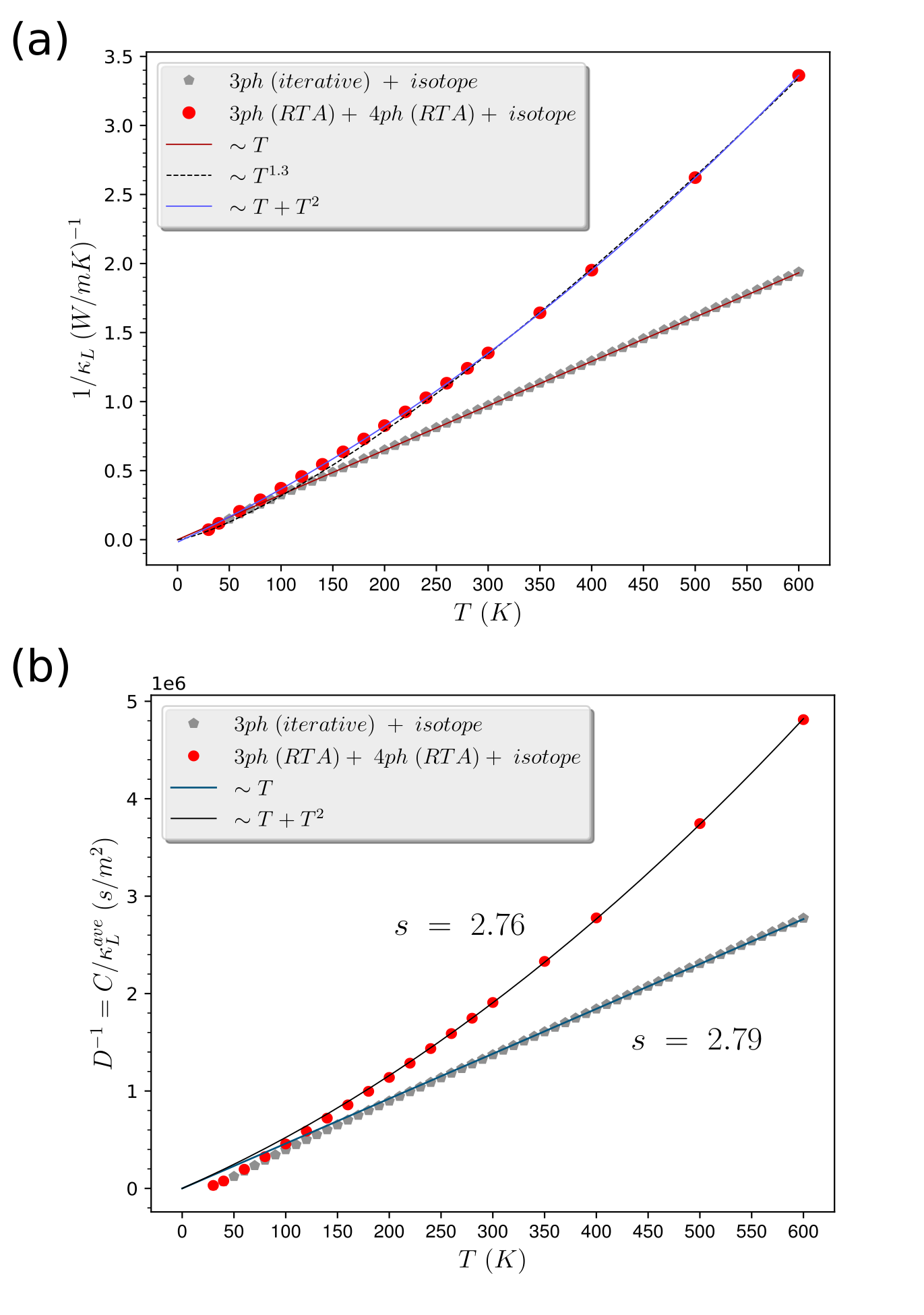}
\caption{(a) The variation of inverse lattice thermal conductivity (1/$\kappa_L$) with temperature both including (red circles) and excluding four-phonon scattering processes (grey pentagons). The linear and non-linear trends of 1/$\kappa_L$ as a function of temperature are represented via brown solid ($\sim$ $T$), blue solid ($\sim$ $T$ + $T^2$) and black dashed lines ($\sim$ $T^{1.3}$). (b) The variation of inverse lattice thermal diffusivity ($D^{-1}$) with temperature both including (red circles) and excluding four-phonon scattering processes (grey pentagons). The linear and non-linear trends of $D^{-1}$ as a function of temperature are represented via blue ($\sim$ $T$) and black solid ($\sim$ $T$ + $T^2$) lines.}
\label{fig:18}
\end{figure}

\noindent The nonlinear temperature scaling of the four-phonon scattering rate indicates an altering temperature dependence of the lattice thermal conductivity in crystalline Ge$_2$Sb$_2$Te$_5$. Figure \ref{fig:18}(a) shows that the inclusion of 4ph scattering processes leads to a deviation of the typical $T^{-1}$ scaling of $\kappa_L$ at high temperature ($T$ $>$ $\theta_D$). The variation of 1/$\kappa_L$ with $T$ including 4ph scattering yields an exponent 1.3 (black dashed line in Fig \ref{fig:18}(a)). As the temperature variation of $\kappa_L$ is directly dependent on the $T$ dependence of the phonon scattering rates, the departure from $\kappa_L$ $\sim$ 1/$T$ can be understood as $\kappa_L$ $\propto$ 1/$\tau^{-1}$ $\propto$ 1/($\tau_{3ph}^{-1}$ + $\tau_{4ph}^{-1}$) $\sim$ 1/($aT$+$bT^2$). Indeed, Figure \ref{fig:18}(a) shows that 1/$\kappa_L$ varies as $aT$+$bT^2$, where $a$ and $b$ are constants.

\noindent The departure of $\kappa_L$ from typical 1/$T$ dependence for Ge$_2$Sb$_2$Te$_5$ at high $T$ evokes complexities to the physical understanding of the universality of the lower bound to the thermal diffusivity of materials, as discussed recently \cite{behnia2019lower}. Behnia and Kapitulnik \cite{behnia2019lower} extended the idea of a universal lower bound to thermal diffusivity in electronic transport in metals \cite{hartnoll2015theory, hartnoll2022colloquium} to the non-metallic phonon transport in materials. At high temperature, phononic thermal diffusivity $D$ was found to exhibit a lower bound, governed by the speed of sound and a Planckian scattering time ($\tau_{p}$) via \cite{behnia2019lower}
\begin{equation}{\label{eq:D}}
    D = sv_{s}^{2}\tau_{p}
\end{equation}
where, thermal diffusion coefficient or diffusivity $D$ = $\kappa_{L}/C$, $C$ denotes heat capacity, $\tau_p$ = $\frac{\hbar}{k_{B}T}$, $v_s$ is average sound speed obtained from the longitudinal acoustic (LA) mode and $s$ ($>$ 1) denotes a dimensionless parameter that is constant for a specific material. This bound had also been realized to have a quantum-mechanical origin which needs further investigations \cite{behnia2019lower}. We find a consistent $D^{-1}$ $\propto$ $T$ behavior at higher temperature  

\begin{table*}[t]
\caption{\label{tab:table2} Theoretical and experimental values of various parameters used to calculate lattice thermal conductivity using the model proposed by Slack and Morelli \cite{slack1979thermal, morelli2006high}.}
\begin{ruledtabular}
\begin{tabular}{ccc}
 Ge$_2$Sb$_2$Te$_5$ (Kooi) &Parameter details& Value(s)\\ 
 \hline
 \\
 $V_{0}$ (\AA$^{3}$)&Volume of the unit cell&259.28 [this study] \\
 $\delta$(\AA)&$(V_{0}/atom)^{1/3}$&3.066 [this study]\\
 $\theta_{D}(K)$&Debye temperature&100 \cite{lee2011thermal}\\
 $\gamma$&Gruneisen parameter&2.49 [this study]\\
 $n$&Number of atoms per unit cell&9\\
 $M_{Ge}$ (g/mol)&Atomic mass of Ge&72.64\\
 $M_{Sb}$ (g/mol)&Atomic mass of Sb&121.76\\
 $M_{Te}$ (g/mol)&Atomic mass of Te&127.6\\
 $\overline{M}$ (g/mol)&Average atomic mass&114.09\\
 &of Ge$_2$Sb$_2$Te$_5$&
\end{tabular}
\end{ruledtabular}
\end{table*}

\onecolumngrid

\noindent in the kinetic transport regime of Ge$_2$Sb$_2$Te$_5$ considering three-phonon and isotope scattering as shown in Fig \ref{fig:18}(b). Fitting the lower bound approximation using $v_s$ (= 3187.65 m/s) and $\tau_p$ as known parameters for Ge$_2$Sb$_2$Te$_5$ yields $s$ = 2.79 (see blue solid line in Fig \ref{fig:18}(b)). At temperature higher than $\theta_D$ (=100 K for Ge$_2$Sb$_2$Te$_5$), heat capacity is almost constant. Therefore, the lower bound of $D$, featuring a relaxation time which is inverse of temperature, stems from the $T^{-1}$ decay of $\kappa_L$ in the kinetic regime of phonon transport. This suggests a physical picture of phonons performing random walk to create a completely diffusive dynamics under dominating Umklapp scattering in this kinetic regime.

\noindent However, this theory faces a serious challenge while incorporating quartic anharmonicity for Ge$_2$Sb$_2$Te$_5$. The deviation from $T^{-1}$ behavior of $\kappa_L$ for Ge$_2$Sb$_2$Te$_5$ at high temperature severely limits to obtain a universal lower bound of thermal diffusivity using Eq.\ref{eq:D}. Within the domain of electronic transport, electron-doped cuprates were found to violate this Planckian bound owing to their high temperature $T^2$ dependence of resistivity in lieu of a typical $T$-linear resistivity \cite{poniatowski2021counterexample}. This $T^2$ variation of resistivity helps surpassing the Planckian bound featuring a faster decay with $\tau^{-1}$ $\gg$ $\tau_{p}^{-1}$, a representative of super-Planckian behavior \cite{hartnoll2022colloquium} in electronic transport. Borrowing the idea of `super-Planckian' behavior to our phonon-based description, we connect the departure of thermal diffusivity from $T^{-1}$ scaling and the violation of Planckian bound to the combined effect of $T$-linear ($\tau_{3ph}^{-1}$ $\sim$ $T$) and $T^2$ ($\tau_{4ph}^{-1}$ $\sim$ $T^2$) variation of phonon resistance at high temperature thermal transport in crystalline Ge$_2$Sb$_2$Te$_5$. Figure \ref{fig:18}(b) (red circles) shows the deviation of $D^{-1}$ from the $T$-linear behavior while incorporating 4ph scattering processes. This deviation was rationalized for insulators via a velocity bound $v_M$ $\gtrsim$ $v_s$ \cite{mousatov2020planckian}, that leads to super-Planckian scattering. Here $v_M$ is melting velocity defined as $v_M$ = $\frac{k_{B}T_{M}}{\hbar}a$, $T_M$ is crystal melting temperature, and $a$ denotes lattice spacing. Using $T_M$ = 900 K, $v_s$ = 3187.65 m/s, and lattice parameter from Table \ref{tab:table1}, we get $v_M$/$v_s$ $\approx$ 15.5 for Ge$_2$Sb$_2$Te$_5$, confirming a super-Planckian type phononic resistive behavior. Further, to capture this deviation as well as to extend Eq.\ref{eq:D} compatible for phonon scattering with quartic anharmonicity, we use the form 
\begin{equation}{\label{eq:D2}}
    \tau_{p}^{\prime} = \frac{\hbar}{k_{B}(T+bT^{2})}
\end{equation}
where $b$ is a constant fitted parameter. Employing Eq.\ref{eq:D2} in $D$ = $sv_{s}^{2}\tau_{p}^{\prime}$ and using this expression to fit the thermal diffusivity at high temperature of Ge$_2$Sb$_2$Te$_5$ yields a value of $s$ = 2.76. We recall that $s$ was originally defined \cite{behnia2019lower} as a material specific constant. Consequently, thermal diffusivity with and without the inclusion of four-phonon scattering gives consistent $s$ (2.76 and 2.79 respectively) for Ge$_2$Sb$_2$Te$_5$ (Fig \ref{fig:18}(b)). Hence, thermal diffusivity of crystalline Ge$_2$Sb$_2$Te$_5$ with its significant higher order phonon anharmonicity seems to exhibit dissipation of super-Planckian type and the corresponding time scale needs to be modified to capture its high temperature behavior. This feature definitely demands more introspection in the future.

\section{Summary and conclusions}{\label{section:summary}}

\noindent First-principles theoretical investigation has been systematically carried out to unveil the phonon thermal transport in crystalline Ge$_2$Sb$_2$Te$_5$, a supremely useful PCM and thermoelectric material possessing low lattice thermal conductivity. The variation of harmonic and anharmonic thermal transport properties of Ge$_2$Sb$_2$Te$_5$ with phonon frequencies is realized in a wide range of temperature from 30 K to 600 K. Using the Kooi structure of Ge$_2$Sb$_2$Te$_5$, the study reveals two notable thermal transport features: (a) significant optical mode-dominance of phonon transport and (b) crucial contributions of quartic anharmonicity to the lattice thermal conductivity, especially at high temperature. Phonon dispersion relations are found to exhibit three broadly distinguished frequency regimes comprised of acoustic bands (A), mixed acoustic and optical bands (M) and optical bands (O). Low-frequency optical bands are found to lie close to the acoustic branches in the `M' regime, enhancing the scattering of acoustic phonons by the optical ones to decrease $\kappa_L$ of Ge$_2$Sb$_2$Te$_5$. 

\noindent Phase space analysis reveals that four-phonon (4ph) scattering becomes gradually important at higher temperature compared to that of the three-phonon (3ph) scattering phase space. Moreover, 4ph phase space is found to display a maxima simultaneously with a minima in its 3ph counterpart in a frequency domain 86 $cm^{-1}$ $<$ $\omega$ $<$ 122 $cm^{-1}$ within the `O' regime. This regime hosts flat bands that prevent some 3ph scattering (especially OOO type) but enable 4ph redistribution processes. Moreover, this 3ph minima results from a crossover between absorption and emission phase space. Absorption dominates at low frequencies whereas emission becomes significant at higher frequencies. Amongst the 4ph scattering channels, redistribution dominates throughout the whole frequency regime. Both 3ph and 4ph scattering processes are found to be predominantly of Umklapp type in the temperature regime studied here. Eventually, this assists in imparting a strong resistance to phonon transport in Ge$_2$Sb$_2$Te$_5$. The closeness between iterative and RTA solutions of BTE for $\kappa_L$ reinstates this feature. Inevitably, including 4ph processes with 3ph and isotope scattering events are found to significantly reduce $\kappa_L$ of Ge$_2$Sb$_2$Te$_5$, starting from a drop of $\sim$ 13 $\%$ at 100 K ($\kappa_L$ = 2.69 W/mK), followed by $\sim$ 28 $\%$ ($\kappa_L$ = 0.74 W/mK) and a significant $\sim$ 42 $\%$ decrement ($\kappa_L$ = 0.3 W/mK) at 300K and 600 K respectively. The domination of optical phonons in $\kappa_L$ is retained in the presence of 4ph scattering with `M' and `O' regimes contributing $\sim$ 36 $\%$ and $\sim$ 38 $\%$ respectively to $\kappa_L$, at high temperature beyond 300 K. Including 4ph scattering has also been found to systematically decrease the maximum allowed mean free paths with temperature with a different scaling ($\sim$ $T^{-1.4}$) compared to a typical $\sim$ $T^{-1}$ relation. Almost 80 $\%$ of the $\kappa_L$ of hexagonal Ge$_2$Sb$_2$Te$_5$ are borne out of phonons with mean free paths below $\sim$ 17 nm. Recent experiments \cite{li2023temperature} used similar grain size to experimentally obtain $\kappa_L$ for Ge$_2$Sb$_2$Te$_5$ which suggests that at room temperature, grain boundary scattering has negligible contribution to $\kappa_L$.
 
\noindent Our investigation suggests that the damping of $\kappa_L$ in Ge$_2$Sb$_2$Te$_5$ is strongly influenced by four-phonon processes, where a significant drop in $\kappa_L$ is found to occur at 300 K and beyond. Therefore, the earlier theoretical estimates on the $\kappa_L$ of Ge$_2$Sb$_2$Te$_5$, which did not include 4ph scattering but used different disorder concentration to match the experimental $\kappa_L$, need to be scrutinized. This modification is crucial since explaining the experimental results with theoretical predictions without 4ph scattering in $\kappa_L$, can lead to misleading physical understanding. 4ph scattering is also seen to marginally enhance the anisotropy in thermal transport of Ge$_2$Sb$_2$Te$_5$ compared to the 3ph only framework. The anisotropy has a very weak temperature dependence as it is mostly dictated by the anisotropy in phonon group velocities along $a-b$ and $c$ axes.

\noindent However, we must mention some limitations of this study. Firstly, we note here the importance and differences of the theoretical approaches to analyze the Kooi and the Matsunaga \cite{matsunaga2004structures} structure of Ge$_2$Sb$_2$Te$_5$. While theoretical studies on Kooi or Petrov structure assume a pristine initial configuration with disorder included perturbatively, Matsunaga structure \cite{campi2017first} is stabilized with the presence of Sb/Ge disorder within the initial configuration. Therefore, this disorder-rich Ge$_2$Sb$_2$Te$_5$ was seen to possess lower $v_g$ leading to lower $\kappa_L$ compared to that of the Kooi structure \cite{campi2017first}. This was also seen to affect the optical phonon contributions to $\kappa_L$ which goes negligibly small. Four-phonon processes may be of negligible importance in that case. This meager effect can also possibly mask the faster decay of $\kappa_L$ compared to $T^{-1}$ scaling at high temperature. However, ordering vacancies \cite{bragaglia2016metal} or producing disorder-starved crystals via recent advanced experimental methods still demands a better understanding of phonon dynamics realized via Kooi and Petrov structures of Ge$_2$Sb$_2$Te$_5$. Secondly, with current state-of-the-art, four-phonon processes could only be realized in RTA to reduce a serious overburden of computational cost. This feature, at least for Ge$_2$Sb$_2$Te$_5$, seems to be less critical as at high temperature regime four-phonon Umklapp scattering is found to dominate over normal scattering, which suffices the usage of RTA at the four-phonon level. 

\noindent Inclusion of four-phonon processes has further serious implications on the temperature-scaling properties of $\kappa_L$ in the high temperature regime. Optical mode-dominated $\kappa_L$, observed in our study, is found to show its signature through the disagreement with the Slack-Morelli description, captured via the coefficient of the $T^{-1}$ scaling of $\kappa_L$. Including 4ph scattering, a departure of $\kappa_L$ from the conventional $T^{-1}$ scaling is observed and $\kappa_L$ is found to vary as 1/(a$T$+b$T^2$). This is rationalized via the $T$ dependencies of the combined effect of 3ph and 4ph scattering. This deviation from $T^{-1}$ scaling further questions a recently conceived universal lower bound of thermal diffusivity in phononic materials. While 3ph-dominated $\kappa_L$ maintains this bound, we recover a super-Planckian type phononic resistance. Consequently, a modified time scale that sets the lower bound in thermal diffusivity in crystalline Ge$_2$Sb$_2$Te$_5$ is obtained. These results will help predicting lattice thermal conductivity of Ge$_2$Sb$_2$Te$_5$ in a more precise fashion as well as can offer better predictions to disorder-driven phonon scattering which can be controlled and tuned for thermoelectric and PCM applications. Also, further studies in the future should dedicate in elucidating the high temperature lower bound of thermal diffusivity in phononic materials having significant four-phonon scattering. Even more introspection is needed to fully decipher its quantum mechanical underpinnings.

\begin{acknowledgments}
We acknowledge the help of Cluster Curta at the University of Bordeaux for high-performance computing. This project had received funding from the European Union’s Horizon 2020 research and innovation program under Grant Agreement No. 824957 (“BeforeHand:” Boosting Performance of Phase Change Devices by Hetero- and Nanostructure Material Design). 
\end{acknowledgments}

\appendix

\section{Three-phonon scattering for optical phonons: Role of flat and dispersive bands}{\label{appen1}}

\noindent We explicitly mark all 24 optical bands in Ge$_2$Sb$_2$Te$_5$ (see Fig \ref{fig:19} below) and compute their bandwidths ($\Delta \omega_O$) as shown below in Table \ref{tab:table3}. The first seven bands and last four bands are denoted as dispersive optical bands while the bands in-between have been found to exhibit lower bandwidth ($\Delta \omega_O$ $\ll$ $\omega_{O}^{min}$) and denoted as flat bands (marked via blue in Table \ref{tab:table3}). Consequently, following \cite{ravichandran2020phonon}, we present the selection rules for 3ph processes concerning optical phonons (AAO, AOO, OOO) below:

\begin{enumerate}
    \item AAO: These 3ph processes can occur if there is an involvement of at least one acoustic phonon whose frequency is higher than the acoustic-optical band gap. As dispersion relation (Fig \ref{fig:1}(C)) suggests that there is no apparent A-O band gap, rather an overlap between A and O phonons, this criterion is satisfied. However, there is also a cut off frequency\cite{ravichandran2020phonon} for O phonons (2$\Delta \omega_A$), beyond which O phonons can not engage in AAO processes. We note that three acoustic branches have bandwidths ($\Delta \omega_A$) of 44.4 $cm^{-1}$ (TA1), 49.3 $cm^{-1}$ (TA2), and 59.1 $cm^{-1}$ (LA). Therefore bands 21-24 (122 $cm^{-1}$ $<$ $\omega$ $<$ 174 $cm^{-1}$) only have weak AAO scattering due to the fact that only LA branch can participate.
    \item AOO: It is possible if $\omega_A$ $\leq$ $\Delta \omega_O$. Therefore, few optical bands will always allow this process as shown in Column 3 of Table \ref{tab:table3}. Obviously, due to extremely small $\Delta \omega_O$, flat bands will allow lesser AOO processes than the highly dispersive bands.
    \item OOO: These 3ph processes will activate if $\Delta \omega_O$ $\geq$  $\omega_{O}^{min}$. Therefore, these processes will be absent in flat bands (bands from 8-20) and will be present marginally for dispersive bands at high frequencies (bands from 21-24).
\end{enumerate}

\begin{table*}[htb!]
\caption{\label{tab:table3}Distinguishing optical phonon bands in Ge$_2$Sb$_2$Te$_5$ as dispersive and flat bands and their corresponding selection rules due to energy conservation \cite{ravichandran2020phonon} to activate different three-phonon (3ph) processes. $\omega_A$ and $\omega_O$ denote acoustic and optical mode frequencies respectively. Here $\omega_{O}^{min}$ $\sim$ 30 $cm^{-1}$, denotes lowest available optical phonon frequency. The three acoustic branches have bandwidths ($\Delta \omega_A$) of 44.4 $cm^{-1}$ (TA1), 49.3 $cm^{-1}$ (TA2), and 59.1 $cm^{-1}$ (LA). Optical band ids and band widths corresponding to flat bands are designated via blue in Table \ref{tab:table3}.}
\begin{ruledtabular}
\begin{tabular}{ccc}
Optical bands id &Band width($\Delta \omega_{O}$ ($cm^{-1}$))& Allowed 3ph processes\\ 
\hline\hline
 \\
$1$&34.09&AAO\\
$2$&34.35&\\
$3$&33.84&AOO (only for $\omega_{A}$ $\leq$ $\Delta \omega_O$)\\
$4$&35.64&\\
$5$&33.84&\\
$6$&28.15&OOO\\
$7$&35.64&\\
\hline
\textcolor{blue}{$8$}&\textcolor{blue}{20.14}&\\
\textcolor{blue}{$9$}&\textcolor{blue}{13.43}&AAO (only for $\omega_{O}$ $\leq$ 2$\Delta \omega_A$) \\
\textcolor{blue}{$10$}&\textcolor{blue}{22.74}&\\
\textcolor{blue}{$11$}&\textcolor{blue}{14.73}&\\
\textcolor{blue}{$12$}&\textcolor{blue}{11.37}&\\
\textcolor{blue}{$13$}&\textcolor{blue}{9.04}&\\
\textcolor{blue}{$14$}&\textcolor{blue}{7.23}&AOO (weak, only for $\omega_{A}$ $\leq$ $\Delta \omega_O$)\\
\textcolor{blue}{$15$}&\textcolor{blue}{7.49}&\\
\textcolor{blue}{$16$}&\textcolor{blue}{10.59}&\\
\textcolor{blue}{$17$}&\textcolor{blue}{9.82}&\\
\textcolor{blue}{$18$}&\textcolor{blue}{10.07}&\\
\textcolor{blue}{$19$}&\textcolor{blue}{7.86}&\\
\textcolor{blue}{$20$}&\textcolor{blue}{19.31}&\\
\hline
\\
&&AAO (weak)\\
$21$&28.73&\\
$22$&25.27&AOO (only for $\omega_{A}$ $\leq$ $\Delta \omega_O$)\\
$23$&29.71&\\
$24$&27.56&OOO(only for $\Delta\omega_{O}$ $\geq$ $\omega_O^{min}$)\\
\end{tabular}
\end{ruledtabular}
\end{table*}


\onecolumngrid
\begin{widetext}
\begin{figure}[H]
\centering
\includegraphics[width=0.7\textwidth]{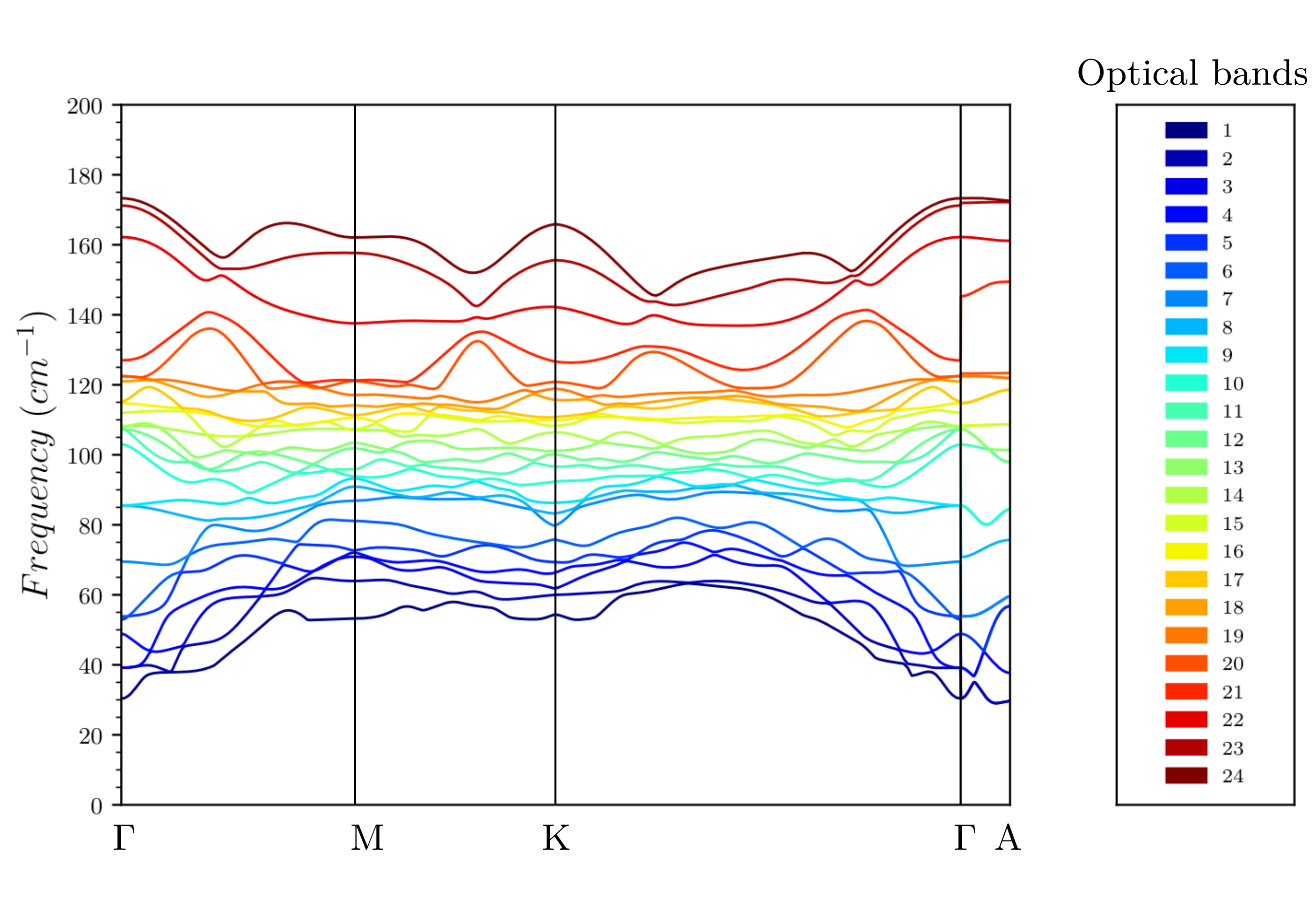}
\caption{Dispersion relation for optical bands in hexagonal-Kooi Ge$_2$Sb$_2$Te$_5$. The optical bands are designated from 1 to 24 gradually from lowest to highest frequency.}
\label{fig:19}
\end{figure}
\end{widetext}

\section{Overdamped phonon modes in acoustic-optic mixed (`M') regime}{\label{appen2}}

\noindent We observe very few overdamped \cite{allen2015anharmonic,fransson2023limits} modes in the `M' frequency regime of Ge$_2$Sb$_2$Te$_5$ via the variation of 3ph scattering rates with frequency (Fig \ref{fig:4}). This feature was also present in an earlier first-principles study of hexagonal Ge$_2$Sb$_2$Te$_5$ \cite{mukhopadhyay2016optic} although not explained. At high temperature beyond 100 K, very few sporadic overdamped phonon modes ($\omega_{0}\tau_{3ph}$ $<$ 1) are seen in the `M' regime, while at low temperature the modes become underdamped ($\omega_{0}\tau_{3ph}$ $>$ 1). We link this feature to the strong anharmonicity in hexagonal Ge$_2$Sb$_2$Te$_5$, especially in the `M' regime, as is evident from our frequency variation of mode Gruneisen parameter ($\gamma$) (shown in supplementary Fig S1(a)). For these few modes within `M' regime, owing to the strong overlap between acoustic and optical bands, $\gamma$ becomes anomalously large. However, mode-averaged $\gamma$ smooths out this feature at a particular temperature and therefore, the average $\gamma$ at fixed $T$ behaves in less anomalous way (supplementary Fig S1(b)). This anharmonicity seems to originate from the weak Te-Te bonds in Ge$_2$Sb$_2$Te$_5$, which allow substantial frequency modulation of Ge$_2$Sb$_2$Te$_5$ with volume. Indeed, rattling dynamics of loosely bounded atoms were discussed by Xie $\textit{et al.}$ \cite{xie2020first} for low-$\kappa$ material AgCrSe$_2$, where four-phonon scattering rates due to TA modes were found to satisfy the overdamped limit. Further, scattering channel distinction reveals that these modes result from the 3ph absorption process via normal scattering (see supplementary Fig S9(b)). 

\noindent Although, the sporadic overdamped modes seem to pose a question on the quasiparticle picture, its validity also depends on the order of the phonon mean free paths. As was mentioned by Allen \cite{allen2015anharmonic}, quasiparticle theory requires larger mean free path compared to the lattice constant. The variation of the maximum mean free path ($\lambda^{max}$) as a function of temperature is shown in Fig \ref{fig:12}. $\lambda^{max}$ is seen to vary from 3 $\mu m$ to 75 nm for 3ph processes which is several order larger compared to the lattice parameters of Ge$_2$Sb$_2$Te$_5$. Moreover, within the three-phonon framework, the variation of $\lambda^{max}$ shows 1/$T$ trend ($\lambda^{max}$ $\sim$ $T^{-1.1}$, inset of Fig \ref{fig:12}). This comes from the fact that higher temperature populates more   

\onecolumngrid
\begin{widetext}
\makeatletter 
\renewcommand{\thefigure}{S\@arabic\c@figure}
\makeatother
\setcounter{figure}{0}
\begin{figure}[H]
\centering
\includegraphics[width=1.0\textwidth]{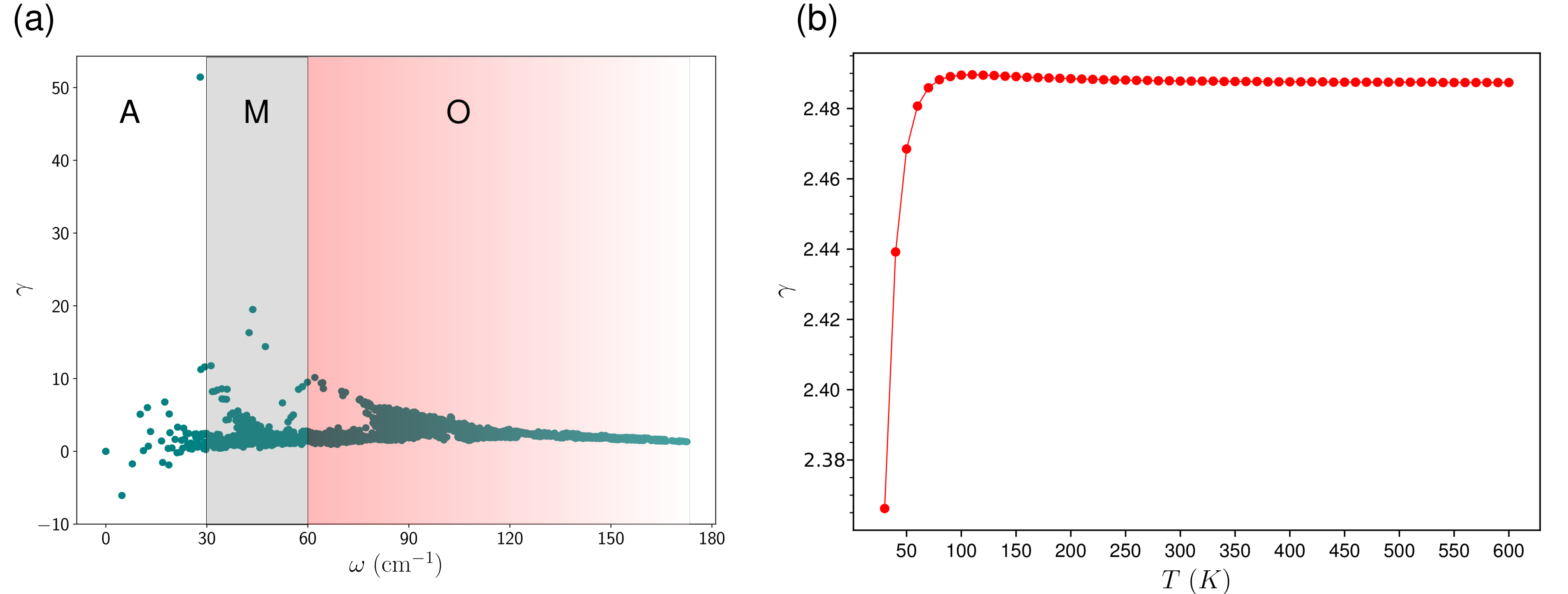}
\caption{(a) Gruneisen parameter ($\gamma$) for the Kooi-Ge$_2$Sb$_2$Te$_5$ as a function of phonon frequencies. Acoustic, mixed and optical regimes, discussed in the main text, are also shown via A, M and O respectively. (b) The temperature variation of $\gamma$, calculated as the weighted sum of the mode contributions.}
\label{fig:gruneisen}
\end{figure} 
\end{widetext}

phonons according to phonon distribution (BE distribution), inducing more scattering between phonons to reduce the mean free path. Thus the variation of $\lambda^{max}$ with $T$ also indicates that phonon quasiparticle picture is valid. We note that Campi $\textit{et al.}$\cite{campi2017first} also mentioned about this validity of quasiparticle picture and the usage phonon BTE in the context of Ge$_2$Sb$_2$Te$_5$. However, the validity of perturbation techniques beyond harmonic approximation needs more attention in systems with overdamped modes \cite{allen2015anharmonic} and it definitely demands more understanding. Ge$_2$Sb$_2$Te$_5$ could be a good candidate to test this issue in future.

\section{Supplementary material}

\subsection{Gruneisen parameter of K\MakeLowercase{o}\MakeLowercase{o}\MakeLowercase{i}-G\MakeLowercase{e}$_2$S\MakeLowercase{b}$_2$T\MakeLowercase{e}$_5$}

\noindent Figure \ref{fig:gruneisen}(a) presents the Gruneisen parameter ($\gamma$) for the Kooi-Ge$_2$Sb$_2$Te$_5$ as a function of phonon frequencies. Figure \ref{fig:gruneisen}(b) shows the temperature variation of $\gamma$, calculated via the weighted sum of the mode contributions, at each temperature. $\gamma$ varies in a wide range from -9 to $\sim$ 51, however the weighted sum over the contribution from different modes reveals that $\gamma$ increases from 2.37 at $T$ = 30 K to 2.49 at $T$ = 100 K and gets saturated to a value of 2.49 throughout the high temperature regime beyond 100 K. Some sporadic points with higher values of $\gamma$ are observed in the `M' regime, corresponding to strong anharmonicity. This anharmonicity seems to come from the weak Te-Te bonds in Ge$_2$Sb$_2$Te$_5$, which allows substantial frequency modulation of Ge$_2$Sb$_2$Te$_5$ with volume. These few strongly anharmonic modes feature overdamped phonon dynamics as described in Appendix B.

\makeatletter 
\renewcommand{\thefigure}{S\@arabic\c@figure}
\makeatother
\setcounter{figure}{1}
\begin{figure}[H]
\centering
\includegraphics[width=0.5\textwidth]{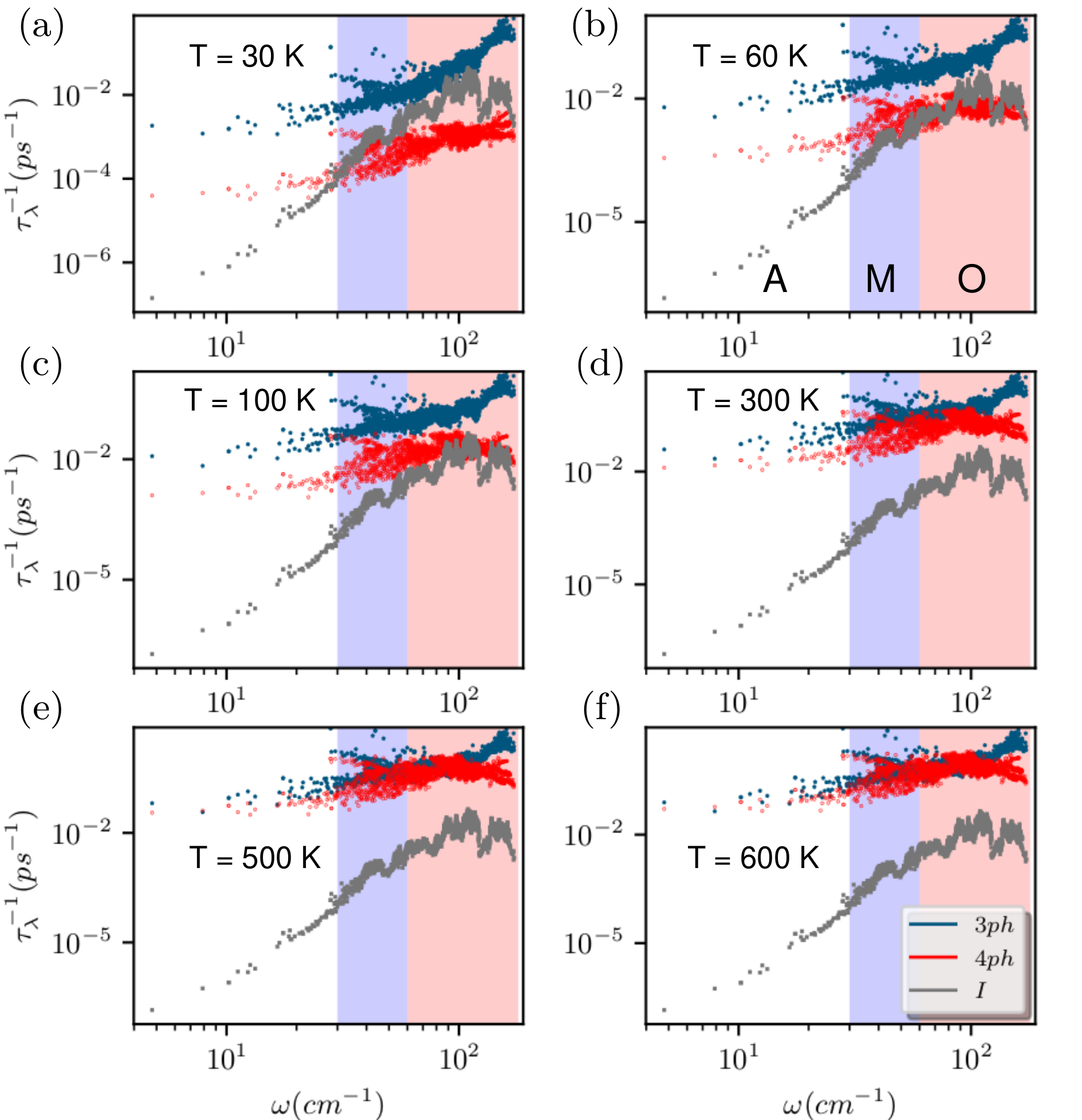}
\caption{The variation of three-phonon (3ph), four-phonon (4ph) and phonon-isotope scattering ($I$) rates with phonon frequencies at different temperatures, denoted via blue, red and grey respectively: (a) $T$ = 30 K, (b) 60 K, (c) 100 K, (d) 300 K, (e) 500 K, and (f) 600 K. Different phonon frequency regimes are marked using white (regime `A'), light blue (regime `M') and light red (regime `O').}
\label{fig:isotope}
\end{figure} 

\makeatletter 
\renewcommand{\thefigure}{S\@arabic\c@figure}
\makeatother
\setcounter{figure}{2}
\begin{figure}[H]
\centering
\includegraphics[width=0.5\textwidth]{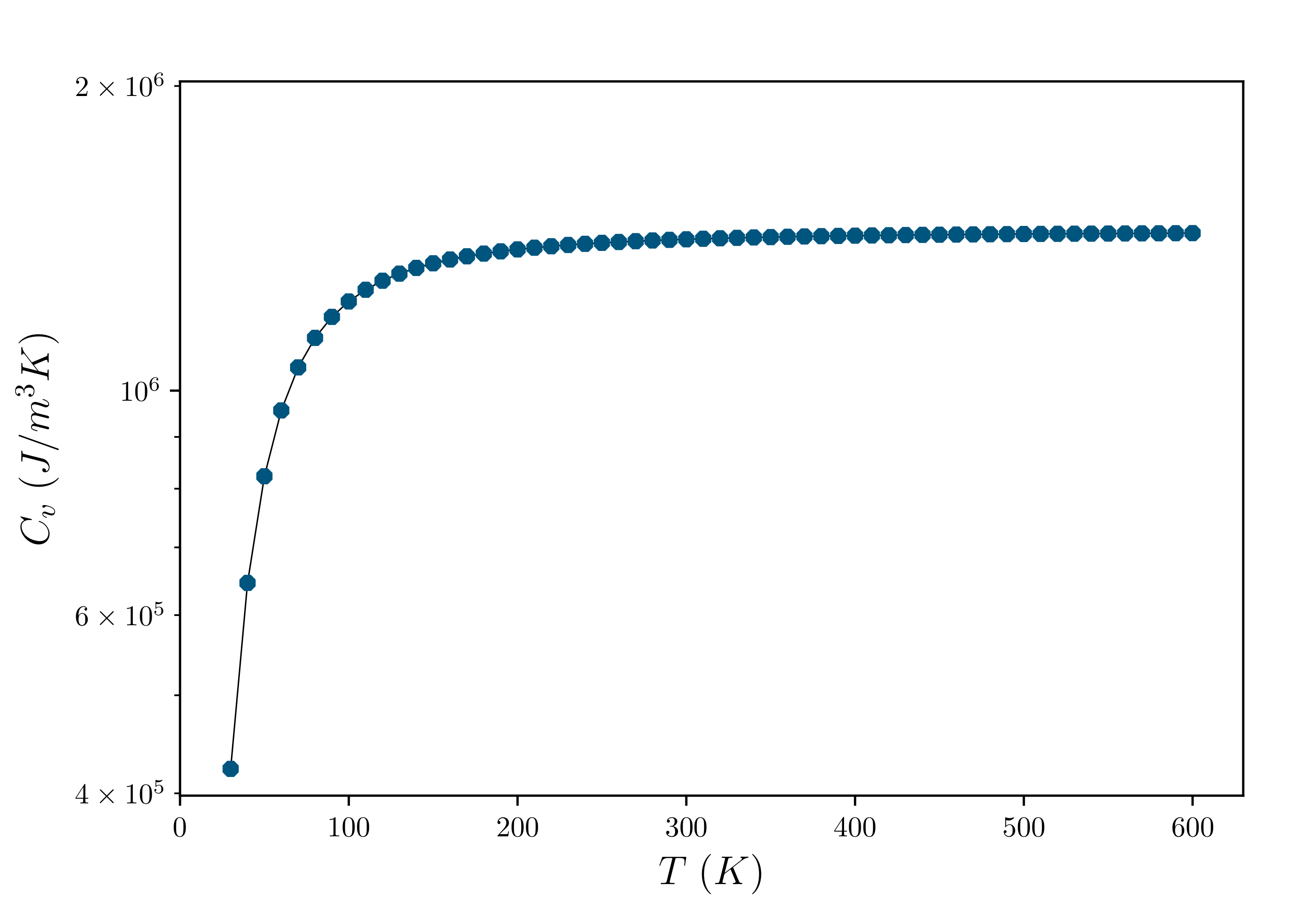}
\caption{The variation of specific heat as a function of temperature for Ge$_2$Sb$_2$Te$_5$.}
\label{fig:cv}
\end{figure}

\subsection{Phonon-isotope scattering in K\MakeLowercase{o}\MakeLowercase{o}\MakeLowercase{i}-G\MakeLowercase{e}$_2$S\MakeLowercase{b}$_2$T\MakeLowercase{e}$_5$}

\noindent Figure \ref{fig:isotope} presents the variation of phonon-isotope scattering rate ($\tau_{\lambda I}^{-1}$) with phonon frequency for crystalline Ge$_2$Sb$_2$Te$_5$ at (a) $T$ = 30 K, (b) $T$ = 60 K, (c) $T$ = 100 K, (d) $T$ = 300 K, (e) $T$ = 500 K, and (f) $T$ = 600 K. For comparison, both three-phonon (blue) and four-phonon (red) scattering rates are also shown. At $T$ = 30 K, phonon-isotope scattering is seen to overpower four-phonon scattering. Moreover, it is comparable to three-phonon (3ph) scattering rate at higher frequencies in the `O' regime. At 60 K, 4ph scattering seems comparable to $\tau_{\lambda I}^{-1}$, beyond which 3ph and 4ph scattering dominate the scattering processes due to their strong temperature dependencies in contrast to the $T$-independent phonon-isotope scattering.

\subsection{The temperature variation of the specific heat in K\MakeLowercase{o}\MakeLowercase{o}\MakeLowercase{i}-G\MakeLowercase{e}$_2$S\MakeLowercase{b}$_2$T\MakeLowercase{e}$_5$}

\noindent The Debye temperature ($\theta_D$) of the material is 100 K which is noted in Table II in the manuscript. We refer to high temperature when $T$ is considerably higher than $\theta_D$. The variation of specific heat as a function of temperature is shown below. Almost constant $C_v$ at temperature beyond 300 K is found (Fig \ref{fig:cv}), satisfying the Dulong-Petit law. This result also matches considerably well with $C_v$ obtained via recent TDTR (time-domain thermoreflectance) experiments on hexagonal Ge$_2$Sb$_2$Te$_5$ by Li $\textit{et al.}$\cite{li2023temperature}.


\subsection{Phonon density of states and dispersion relation in P\MakeLowercase{e}\MakeLowercase{t}\MakeLowercase{r}\MakeLowercase{o}\MakeLowercase{v}-G\MakeLowercase{e}$_2$S\MakeLowercase{b}$_2$T\MakeLowercase{e}$_5$}

\noindent Phonon density of states (PDOS) and phonon dispersion relation for crystalline hexagonal Ge$_2$Sb$_2$Te$_5$, stacked in Petrov structure, are presented in Fig \ref{fig:pdos}(a) and Fig \ref{fig:pdos}(b) respectively. Similar to that of the Kooi structure, we observe the presence of optical bands at low phonon frequencies via the dispersion relation. We specifically find coupling between optical modes (below 80 $cm^{-1}$) and the acoustic modes which indicates scattering between acoustic and optical phonons in this frequency regime. 

\onecolumngrid
\begin{widetext}
\makeatletter 
\renewcommand{\thefigure}{S\@arabic\c@figure}
\makeatother
\setcounter{figure}{3}
\begin{figure}[H]
\centering
\includegraphics[width=1\textwidth]{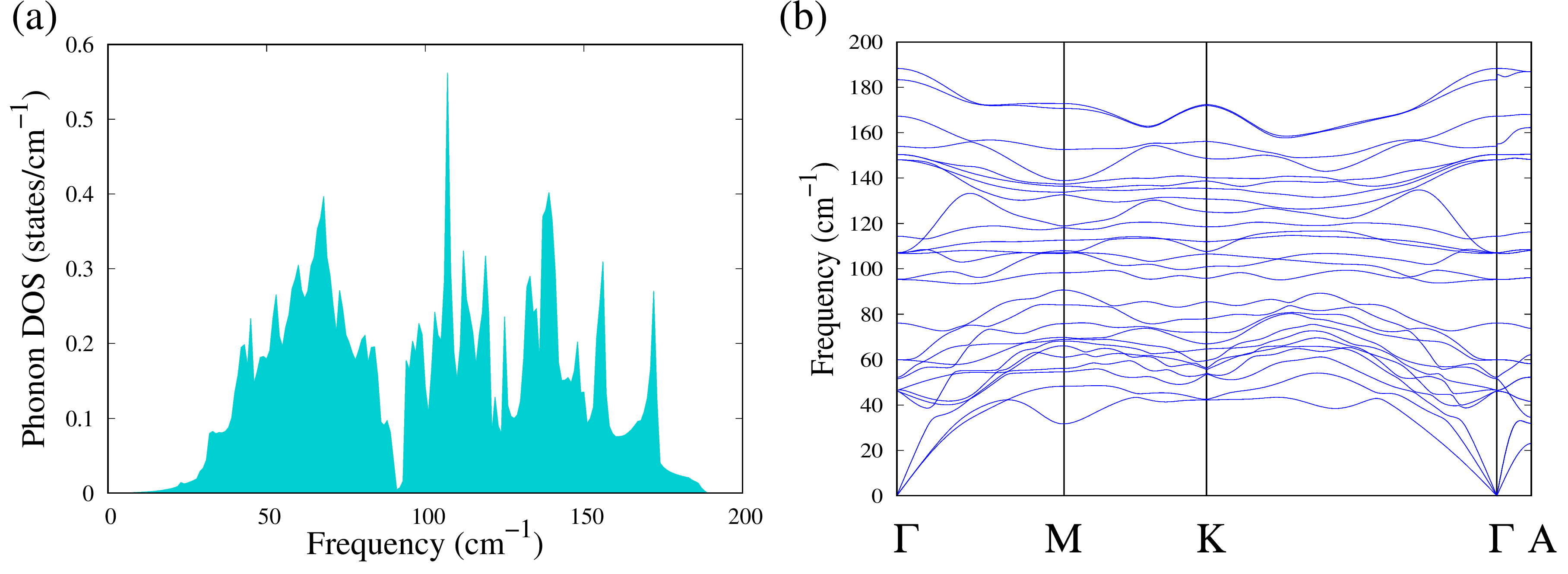}
\caption{(a) Phonon density of states (PDOS) is presented as a function of phonon frequencies for crystalline hexagonal Ge$_2$Sb$_2$Te$_5$ stacked in Petrov structure. (b) Phonon dispersion relations of Petrov-Ge$_2$Sb$_2$Te$_5$ along high symmetry path $\Gamma$-$M$-$K$-$\Gamma$-$A$.}
\label{fig:pdos}
\end{figure}
\end{widetext}

\makeatletter 
\renewcommand{\thefigure}{S\@arabic\c@figure}
\makeatother
\setcounter{figure}{4}
\begin{figure}[H]
\centering
\includegraphics[width=0.5\textwidth]{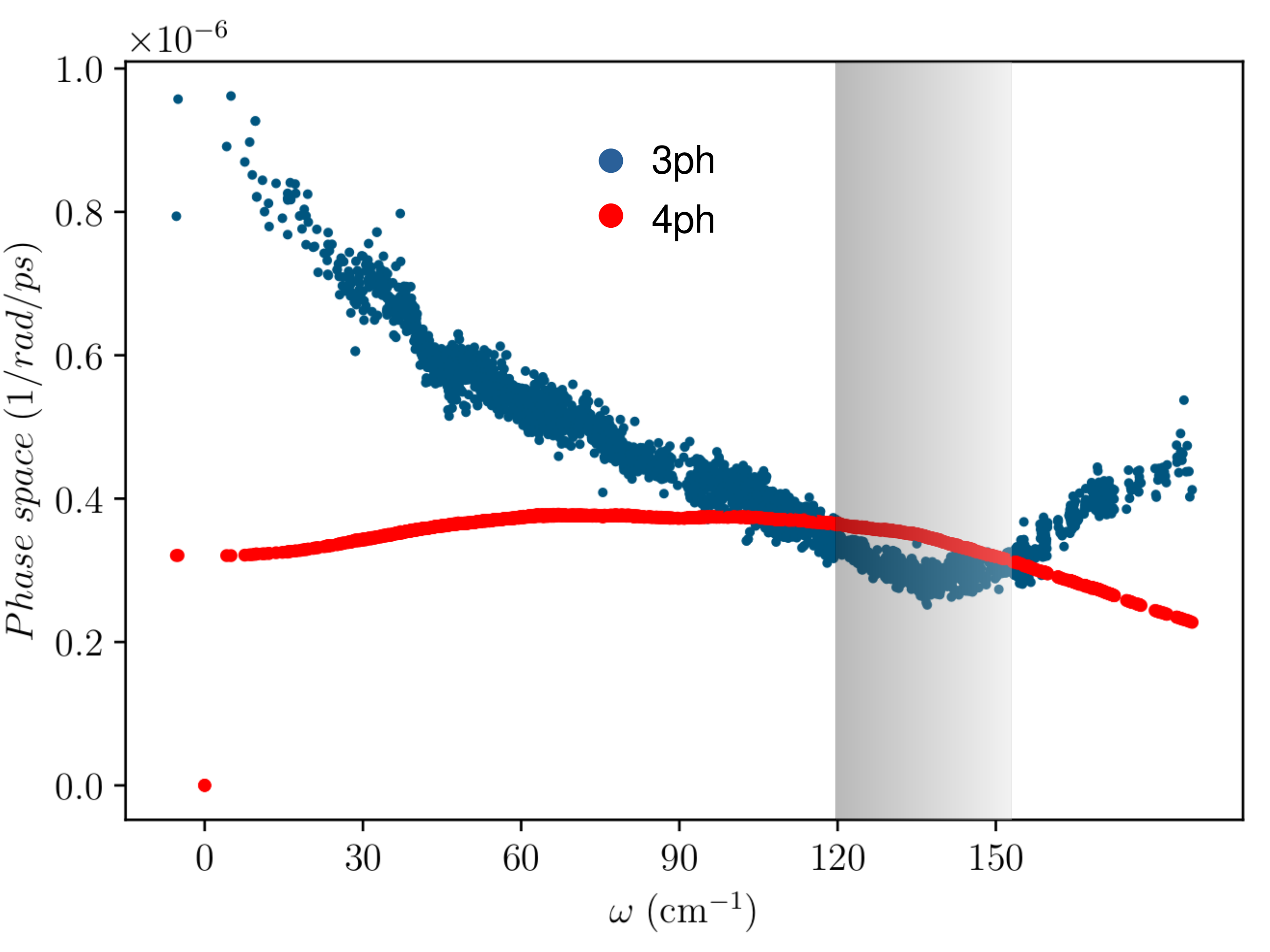}
\caption{The variation of three phonon (3ph) and four-phonon (4ph) scattering phase spaces with phonon frequencies in crystalline hexagonal Ge$_2$Sb$_2$Te$_5$ stacked in Petrov structure. The grey shaded frequency regime identifies the exceeding 4ph phase space compared to the 3ph phase space.}
\label{fig:phase_space}
\end{figure}

\makeatletter 
\renewcommand{\thefigure}{S\@arabic\c@figure}
\makeatother
\setcounter{figure}{5}
\begin{figure}[H]
\centering
\includegraphics[width=0.45\textwidth]{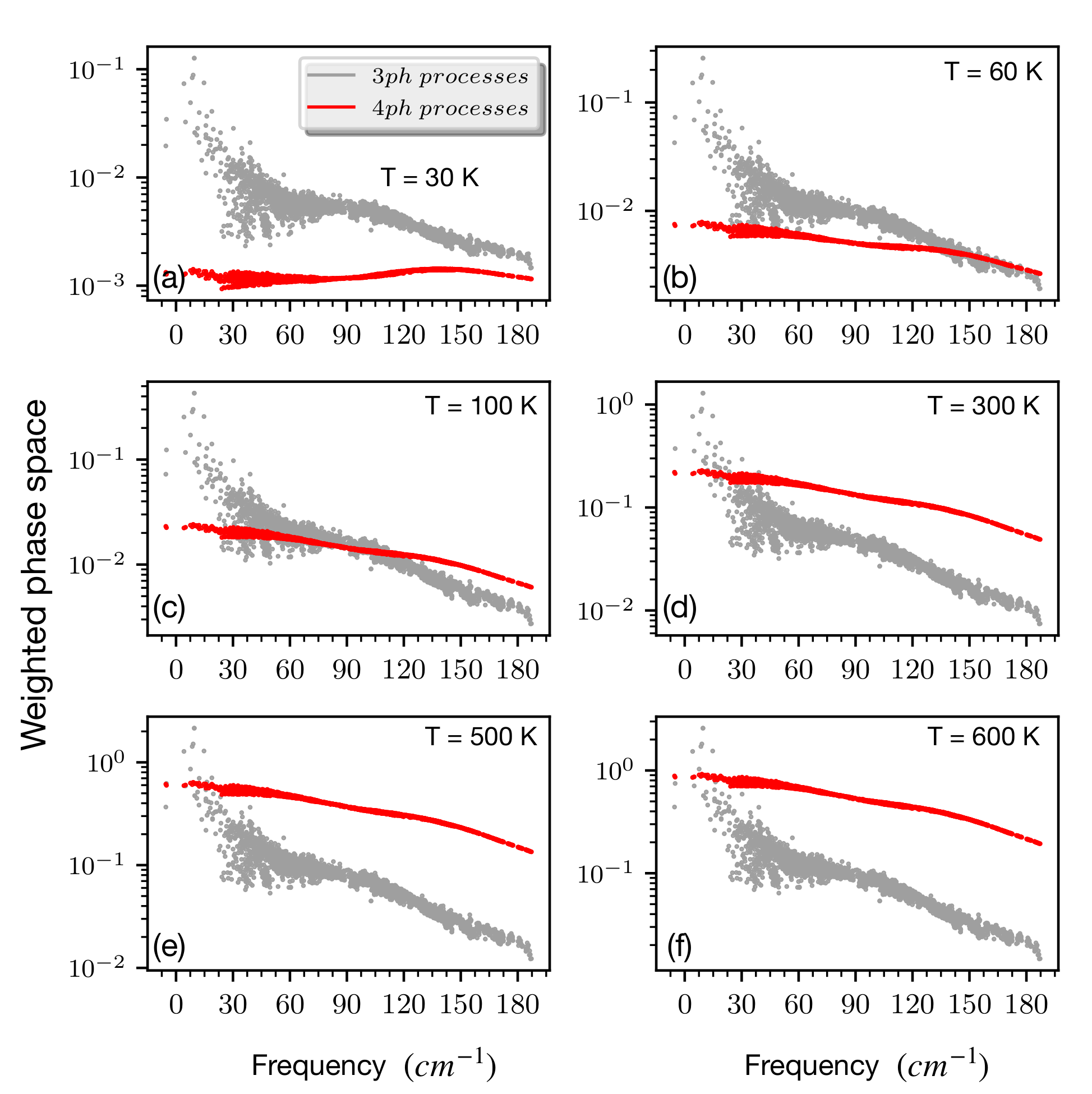}
\caption{The variation of three phonon (3ph) and four-phonon (4ph) weighted phase space with phonon frequencies at different temperatures for Petrov-Ge$_2$Sb$_2$Te$_5$: (a) $T$ = 30 K, (b) $T$ = 60 K, (c) $T$ = 100 K, (d) $T$ = 300 K, (e) $T$ = 500 K, and (f) $T$ = 600 K.}
\label{fig:wp}
\end{figure}

However, unlike Kooi structure, Petrov structure is seen to feature a frequency gap between the optical bands around 90 $cm^{-1}$. This feature is reflected in the PDOS (Fig \ref{fig:pdos}(a)) via a drop of PDOS to almost zero around 90 $cm^{-1}$, indicating almost absence of phonon modes at this frequency. This quasibipartite phonon density of states had also been discussed in earlier studies \cite{campi2017first, ibarra2018ab} with the same onset frequency of 90 $cm^{-1}$ \cite{ibarra2018ab} in Petrov-Ge$_2$Sb$_2$Te$_5$.

\subsection{Three and Four-phonon scattering Phase space in P\MakeLowercase{e}\MakeLowercase{t}\MakeLowercase{r}\MakeLowercase{o}\MakeLowercase{v}-G\MakeLowercase{e}$_2$S\MakeLowercase{b}$_2$T\MakeLowercase{e}$_5$}

\noindent Figure \ref{fig:phase_space} presents the evolution of three-phonon (3ph) and four-phonon (4ph) phase space with frequency in Petrov-Ge$_2$Sb$_2$Te$_5$. Although 3ph scattering phase space is dominant over 4ph scattering phase space, in a small frequency range within optical bands (shown via the grey shaded regime in Fig \ref{fig:phase_space}), 4ph phase space exceeds its 3ph counterpart. Also, this frequency window records the minima of the 3ph scattering phase space. This feature of Petrov-Ge$_2$Sb$_2$Te$_5$ is similar to that of the Kooi structure (discussed in the main text) although in a less pronounced fashion. We also observe similar but fewer flat bands in this regime via the dispersion relation compared to that of the Kooi-Ge$_2$Sb$_2$Te$_5$.

\noindent Figure \ref{fig:wp} represents the relative strength between 3ph and 4ph weighted scattering phase spaces in Petrov-Ge$_2$Sb$_2$Te$_5$ as a function of temperature ranging from 30 K to 600 K. As the temperature is raised, we find a gradually increased relevance of 4ph scattering phase space compared to that of the 3ph phase space. This stems from the fact that the evaluation of weighted phase space takes the phonon distribution function ($f_0$) into account which varies linearly with temperature at high temperature limit ($f_0$ $\sim$ $k_{B}T/\hbar\omega$). This indicates that the thermal transport of Petrov-Ge$_2$Sb$_2$Te$_5$ can also be influenced by incorporating the 4ph scattering processes.

\subsection{Three and four-phonon weighted phase space for K\MakeLowercase{o}\MakeLowercase{o}\MakeLowercase{i}-G\MakeLowercase{e}$_2$S\MakeLowercase{b}$_2$T\MakeLowercase{e}$_5$}

\noindent Figure \ref{fig:wp3wp4} represents the variation of weighted phase space coming from distinct scattering channels for both 3ph (WP$_3$, Fig \ref{fig:wp3wp4}(a), (b)) and 4ph scattering processes (WP$_4$, Fig \ref{fig:wp3wp4}(c), (d)) at $T$ = 30 K and 600 K. We observe that the distributions of  WP$_3$ and WP$_4$ almost remain unchanged in the whole temperature range studied. In the frequency regime A, WP$_3$ is seen to be dominated by the absorption processes which soon starts decaying in the regime M, after which emission seems to dominate at higher frequencies in the O regime. At higher temperature, WP$_3$ increases and the crossover between absorption and emission processes is seen to be marginally postponed at higher frequencies. Four-phonon scattering is found to be dominated mostly by the redistribution process throughout the frequency range (Fig \ref{fig:wp3wp4}(c), (d)). Only at frequencies beyond 130 $cm^{-1}$, splitting process takes over. We note that this is the same regime where optical bands become markedly dispersive and sparsely populated. In frequency regime O beyond $\omega$ $>$ 130 $cm^{-1}$, 

\makeatletter 
\renewcommand{\thefigure}{S\@arabic\c@figure}
\makeatother
\setcounter{figure}{6}
\begin{figure}[H]
\centering
\includegraphics[width=0.5\textwidth]{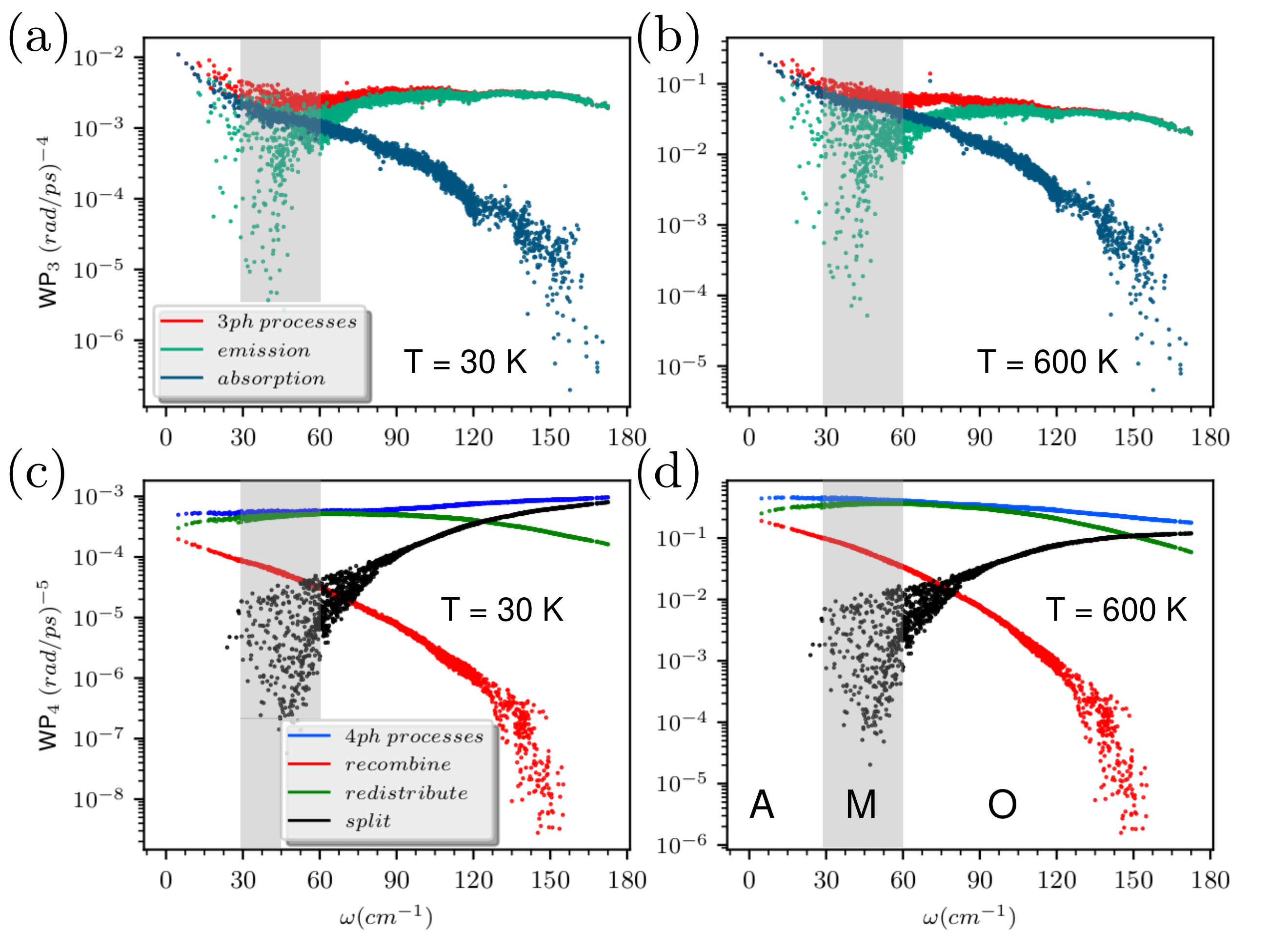}
\caption{Frequency variation of three-phonon (WP$_3$: (a), (b)) and four-phonon (WP$_4$:(c), (d)) weighted phase spaces for different scattering channels at $T$ = 30 K ((a), (c)) and 600 K ((b), (d)). A, M and O frequency regimes are also distinguished.}
\label{fig:wp3wp4}
\end{figure}

\makeatletter 
\renewcommand{\thefigure}{S\@arabic\c@figure}
\makeatother
\setcounter{figure}{7}
\begin{figure}[H]
\centering
\includegraphics[width=0.5\textwidth]{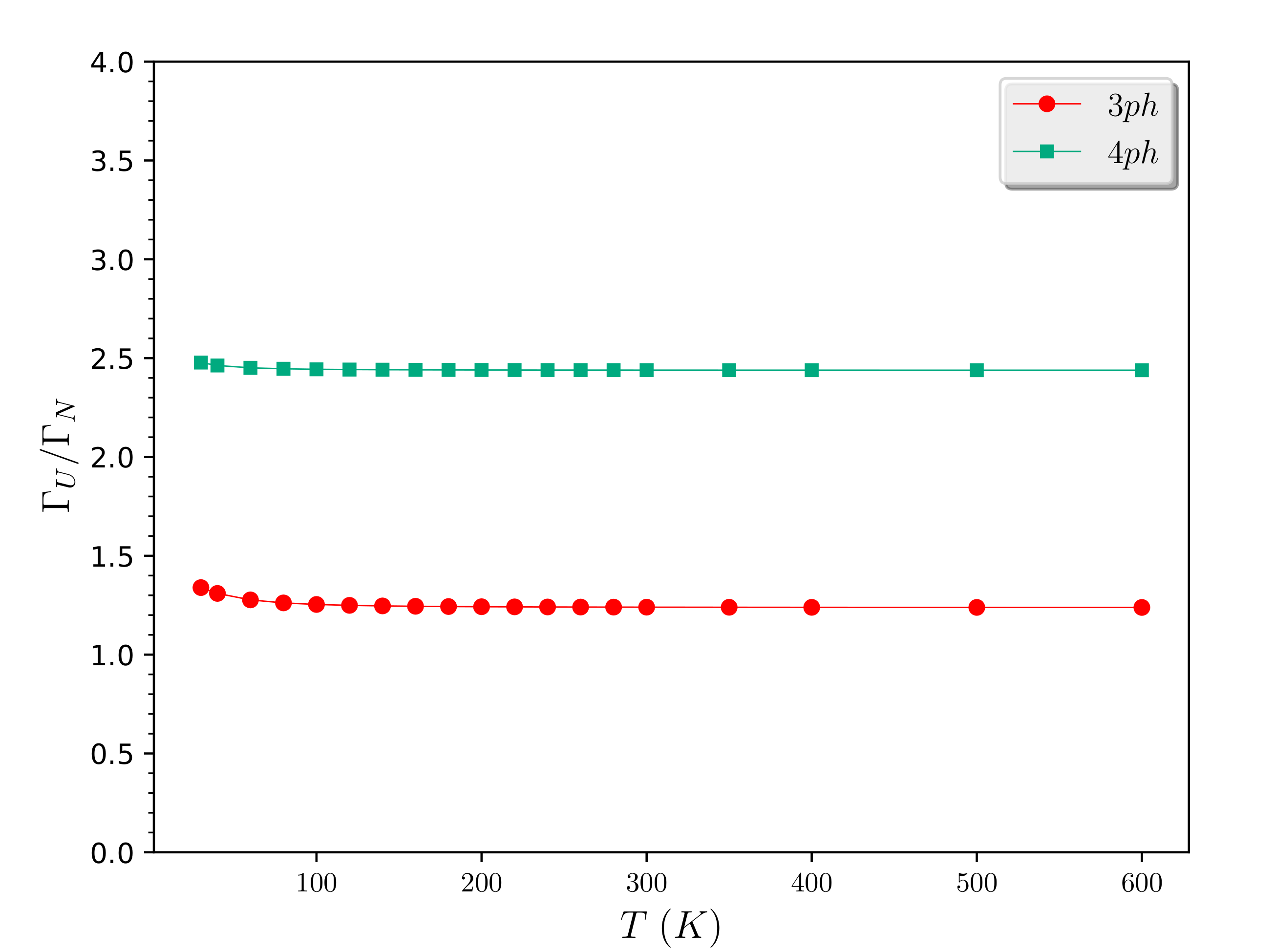}
\caption{The ratio of thermodynamic averages of Umklapp and normal scattering rates as a function of temperature for both three (Red circles) and four-phonon (green squares) scattering processes.}
\label{fig:NU}
\end{figure}

\noindent decreasing 4ph phase space indicates restriction to 4ph scattering channels due to fewer sparsely located highly dispersive optical bands. Also, the crossover between the redistribution-dominated and splitting-dominated scattering processes is marginally postponed to higher frequencies at higher temperatures (Fig \ref{fig:wp3wp4}(d)).

\subsection{Relative strength between Umklapp and Normal scattering in K\MakeLowercase{o}\MakeLowercase{o}\MakeLowercase{i}-G\MakeLowercase{e}$_2$S\MakeLowercase{b}$_2$T\MakeLowercase{e}$_5$}

\noindent For both three and four-phonon processes in Ge$_2$Sb$_2$Te$_5$, Umklapp scattering is seen to dominate over normal scattering processes as realized through the temperature-variation of their relative strength via the ratio $\Gamma_{U}/\Gamma_{N}$. $\Gamma$ is average scattering rate defined as $\Gamma$ = $\sum_{\lambda}C_{\lambda}\tau_{\lambda}^{-1}/\sum_{\lambda}C_{\lambda}$ and is calculated for both Umklapp ($\Gamma_U$) and normal scattering ($\Gamma_N$) events. In the studied temperature range, this ratio is almost constant with a value 1.3 for 3ph scattering which increases to 2.5 for 4ph scattering.

\subsection{The variation of normal and Umklapp scattering rates with frequency in K\MakeLowercase{o}\MakeLowercase{o}\MakeLowercase{i}-G\MakeLowercase{e}$_2$S\MakeLowercase{b}$_2$T\MakeLowercase{e}$_5$: Role of different scattering channels}

\noindent Figure \ref{fig:3phscatchannel} explains three-phonon (3ph) absorption (Fig \ref{fig:3phscatchannel}(a), (b)) and emission processes (Fig \ref{fig:3phscatchannel}(c), (d)) for two extreme temperature points ($T$ = 30 K and 600 K). As temperature is raised from lowest (30 K) to highest value (600 K), for all the cases, scattering rates increase $\sim$ 60 and $\sim$ 11 folds for absorption and emission respectively. In the absorption process, few N scattering modes are found to contribute higher 3ph-scattering rates in the M regime (Fig \ref{fig:3phscatchannel}(a), (b)). At high $T$ (600 K), these few sporadic N absorption modes feature overdamped phonon dynamics ($\omega\tau$ $<$ 1) as discussed in Appendix B. We also observe that $\tau_{3ph}^{-1}$ is always larger for emission processes than that of the absorption processes regardless the temperature. Further a more quantitative picture of the relative importance between N and U scattering is obtained via the thermodynamic average of scattering rates as described in the article, which suggests U scattering dominance in 3ph scattering.

\noindent Fig \ref{fig:4phscatchannel} presents the frequency distribution of four-phonon scattering rates in different scattering channels, distinguished as N and U processes. Redistribution processes in general, resistive U redistribution to be particular ($\textbf{q}$ + $\textbf{q}_1$ = $\textbf{q}_2$ + $\textbf{q}_3$ + $\textbf{G}$), are seen to overpower at both high and low temperature. At low temperature (Fig \ref{fig:4phscatchannel}(a), (c), (e)), the M frequency regime is seen to be populated by mostly redistribution as well as recombination processes that exhibit U scattering. In O frequency regime, recombination channels are mostly blocked while splitting processes are found to gain importance along with the dominating redistribution scattering rates. At very high frequencies splitting processes aggravate due to the increasing weighted scattering phase space. This frequency distribution remains almost similar at higher temperature (Fig \ref{fig:4phscatchannel}(b), (d), (f)). We realize that both 3ph and 4ph scattering processes are predominantly of Umklapp type in Ge$_2$Sb$_2$Te$_5$ which is more evident from the temperature variation of the average scattering rate, discussed in the main text.

\setcounter{figure}{8}
\begin{figure}[H]
\centering
\includegraphics[width=0.5\textwidth]{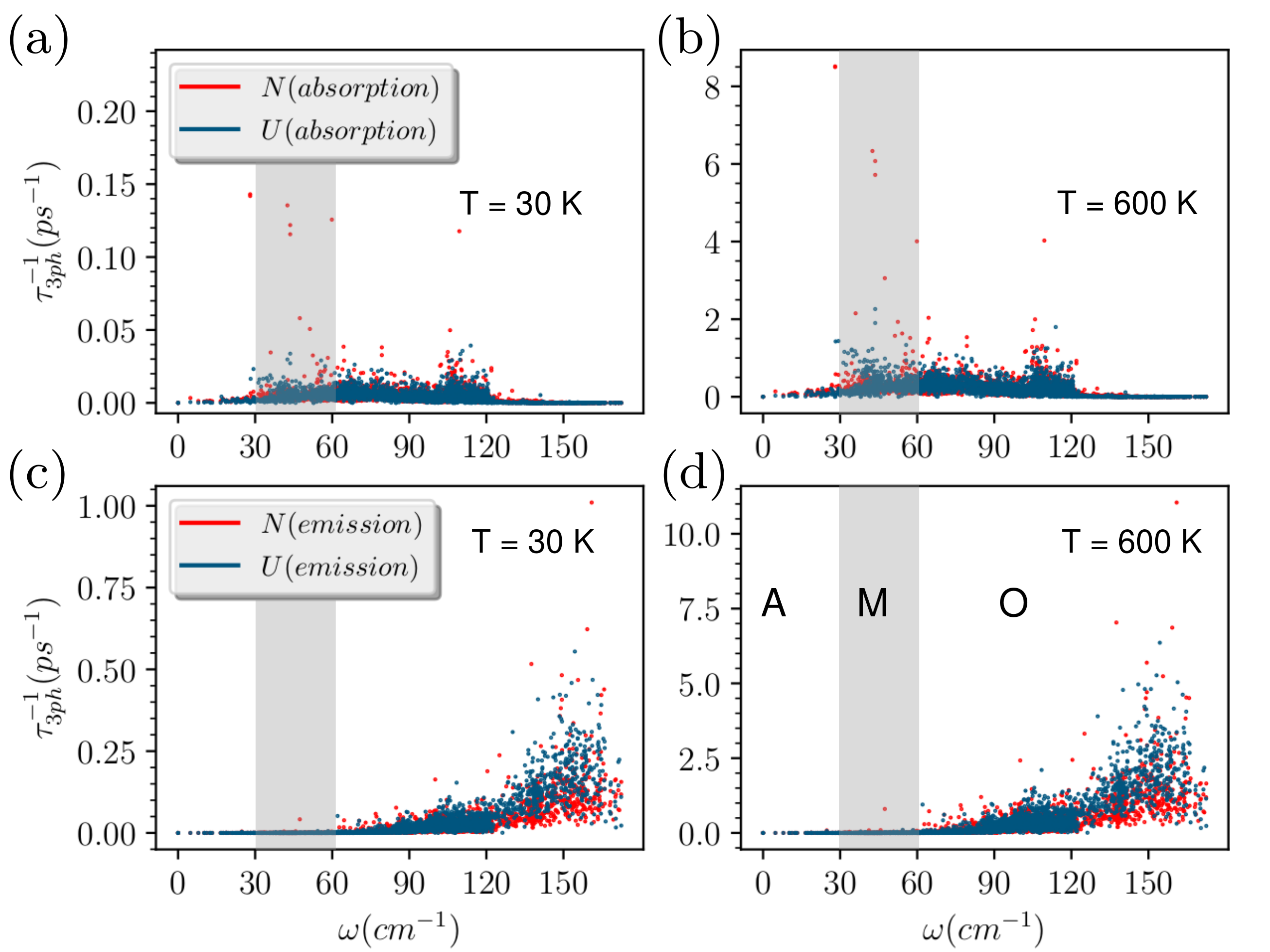}
\caption{The frequency variation of 3ph normal (N) and Umklapp (U) absorption scattering rates are presented for (a) $T$ = 30 K and (b) $T$ = 600 K. Similarly, frequency variation of 3ph normal (N) and Umklapp (U) emission scattering rates are shown for (c) $T$ = 30 K and (d) $T$ = 600 K. A, M and O frequency regimes are also distinguished.}
\label{fig:3phscatchannel}
\end{figure}

\setcounter{figure}{9}
\begin{figure}[H]
\centering
\includegraphics[width=0.5\textwidth]{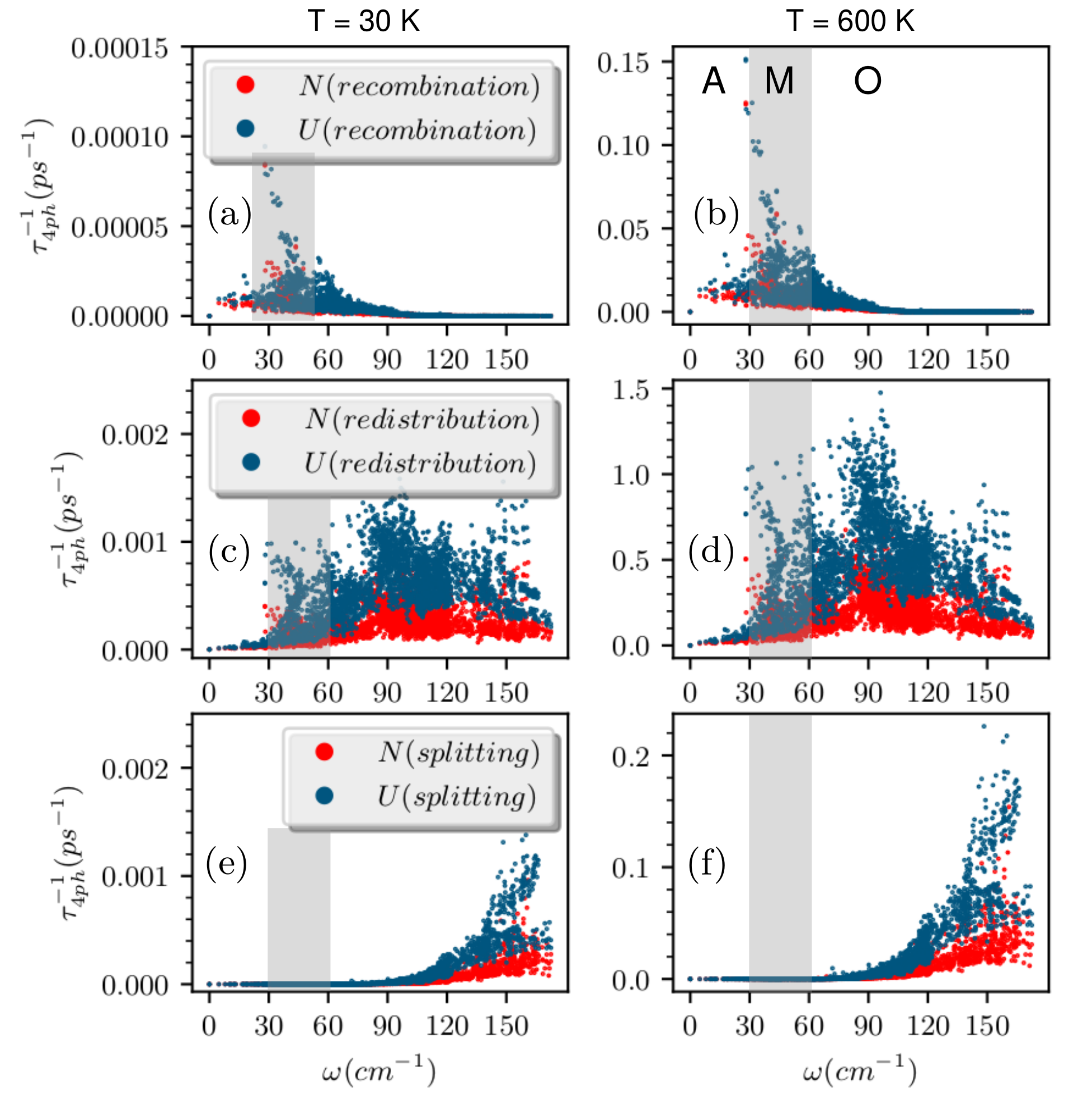}
\caption{The variation of 4ph normal (N) and Umklapp (U) scattering rates are presented as a function of frequency for different scattering channels at two extreme temperature points: recombination processes for (a) $T$ = 30 K and (b) $T$ = 600 K, redistribution processes for (c) $T$ = 30 K and (d) $T$ = 600 K, and splitting processes for (e) $T$ = 30 K and (f) $T$ = 600 K. A, M and O frequency regimes are also distinguished.}
\label{fig:4phscatchannel}
\end{figure}

\subsection{The variation of cumulative lattice thermal conductivity with frequency in K\MakeLowercase{o}\MakeLowercase{o}\MakeLowercase{i}-G\MakeLowercase{e}$_2$S\MakeLowercase{b}$_2$T\MakeLowercase{e}$_5$: Excluding and including four-phonon processes}

\noindent Figure \ref{fig:kappa3phomega} presents the cumulative $\kappa_L$ as a function of frequency for crystalline Ge$_2$Sb$_2$Te$_5$, considering 3ph and isotope scattering events at different temperature. The evolution of $\kappa_L$ along a-b plane ($\kappa_L^x$, shown via red dashed lines) and c axis ($\kappa_L^z$, shown via yellow dashed lines) suggests a strong anisotropic thermal conduction behavior in Ge$_2$Sb$_2$Te$_5$. The average $\kappa_L$ (black lines in Fig \ref{fig:kappa3phomega}(a)-(f)) shows a gradual domination of M and O regimes to dictate $\kappa_L$ as $T$ is increased. At high temperature beyond 300 K, the contributions to $\kappa_L$ are saturated and A, M, and O regimes contribute to 26.4 $\%$, 33.2 $\%$ and 40.4 $\%$ of $\kappa_L$. Similar feature of anisotropic $\kappa_L$ is retained in the four-phonon framework where we employ 4ph, 3ph and I scattering to get the cumulative $\kappa_L$ for Ge$_2$Sb$_2$Te$_5$ (Fig \ref{fig:kappa4phomega}). The optical phonon dominance in $\kappa_L$ is also retained at high temperature with a marginal modification of contributions coming from different frequency regimes as follows: regime A: 26.2 $\%$, regime M: 35.8 $\%$ and regime O: 38 $\%$ of $\kappa_L$ (Fig \ref{fig:kappa4phomega}).

\subsection{The variation of cumulative lattice thermal conductivity with mean free path in K\MakeLowercase{o}\MakeLowercase{o}\MakeLowercase{i}-G\MakeLowercase{e}$_2$S\MakeLowercase{b}$_2$T\MakeLowercase{e}$_5$: Excluding and including four-phonon processes}

\noindent Figure \ref{fig:kappa3phmfp} presents cumulative $\kappa_L$ as a function of maximum allowed phonon mean free path (MFP) in Ge$_2$Sb$_2$Te$_5$, including 3ph and I scattering processes. Similar to that of the Fig \ref{fig:kappa3phomega}, anisotropic thermal conduction is reflected via the evolution of cumulative $\kappa_L$ with mean free paths along a-b (blue dashed lines in Fig \ref{fig:kappa3phmfp}(a)-(f)) and along c (green dashed lines in Fig \ref{fig:kappa3phmfp}(a)-(f)). The maximum allowed mean free paths are identified at each $T$, which is found to decrease with $T$ (Fig \ref{fig:kappa3phmfp}). Similar features are retained while including 4ph scattering Fig \ref{fig:kappa4phmfp}(a)-(f)). We observe that increasing $T$ from 30 K to 600 K decreases $\lambda_{max}$ from 3.13 $\mu$m to 0.08 $\mu$m (80 nm) with 3ph and I scattering (Fig \ref{fig:kappa3phmfp}). This effect is even more severe while including 4ph scattering where $\lambda_{max}$ decreases from 2.15 $\mu$m to 0.025 $\mu$m (25 nm) while increasing $T$ from 30 K to 600 K (Fig \ref{fig:kappa4phmfp}).

\subsection{Convergence of three-phonon and four-phonon processes in K\MakeLowercase{o}\MakeLowercase{o}\MakeLowercase{i}-G\MakeLowercase{e}$_2$S\MakeLowercase{b}$_2$T\MakeLowercase{e}$_5$}

We use the $\textbf{q}$-grid 16$\times$16$\times$4 (mentioned in the manuscript in the Methods section) for calculating the lattice thermal conductivity both including and excluding 4ph with the existing 3ph processes, to compare the effect of three and four-phonon scattering consistently.

\setcounter{figure}{10}
\begin{figure}[H]
\centering
\includegraphics[width=0.4\textwidth]{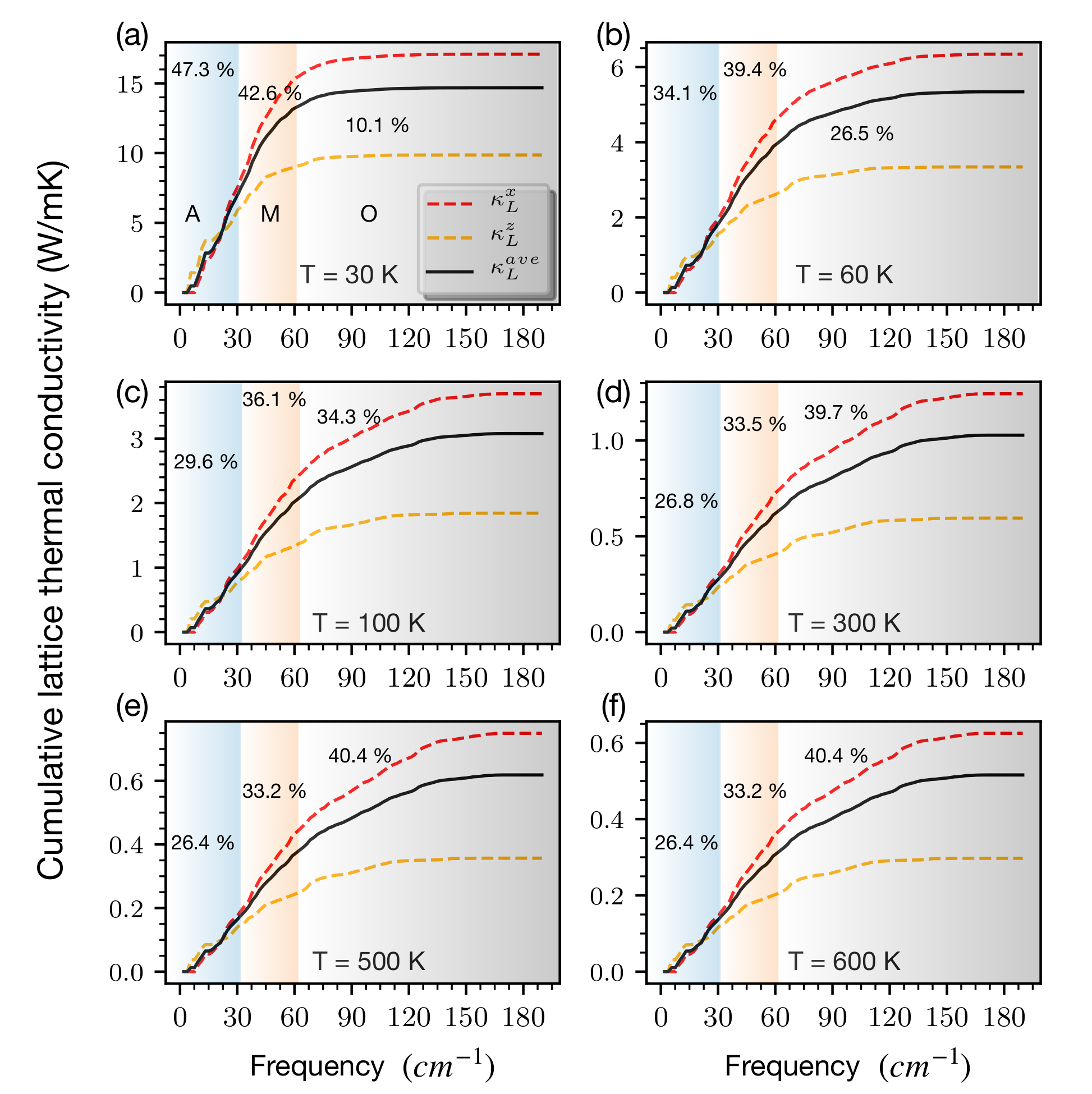}
\caption{Frequency variation of cumulative lattice thermal conductivity including 3ph and phonon-isotope scattering for (a) $T$ = 30 K, (b) $T$ = 60 K, (c) $T$ = 100 K, (d) $T$ = 300 K, (e) $T$ = 500 K and (f) $T$ = 600 K. For all the cases, black lines denote average $\kappa_L$, while red and yellow dashed lines represent $\kappa_L$ along $x$ and $z$ respectively. The contributions of different frequency regimes (A, M, O) to the cumulative lattice thermal conductivities (in percentage) are presented for each temperature via the shaded zones.}
\label{fig:kappa3phomega}
\end{figure}


\setcounter{figure}{11}
\begin{figure}[H]
\centering
\includegraphics[width=0.4\textwidth]{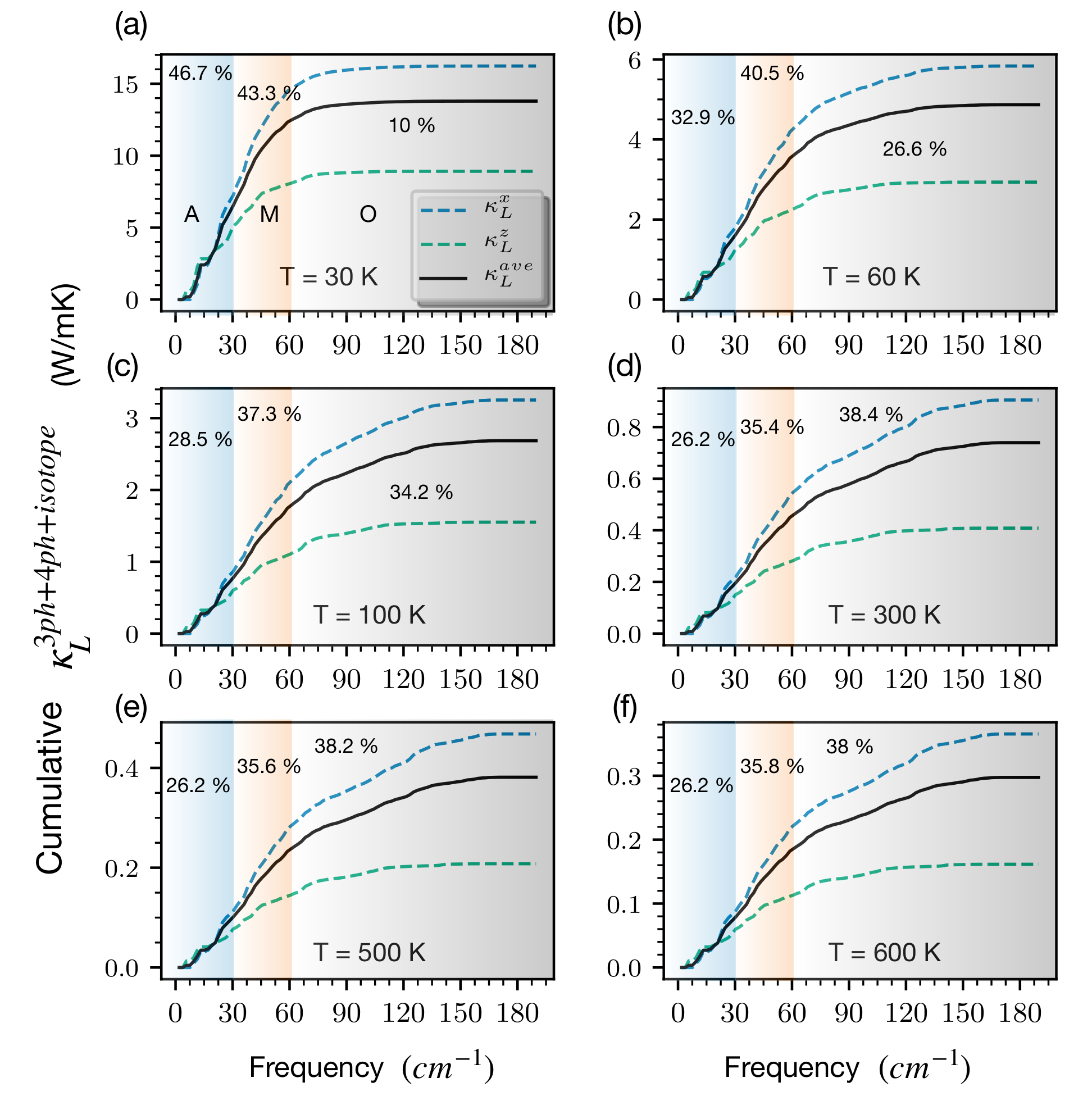}
\caption{Cumulative lattice thermal conductivity including 3ph, 4ph and phonon-isotope scattering is presented as a function of frequency for (a) $T$ = 30 K, (b) $T$ = 60 K, (c) $T$ = 100 K, (d) $T$ = 300 K, (e) $T$ = 500 K and (f) $T$ = 600 K. For all the cases, black lines denote average $\kappa_L$, while blue and green dashed lines represent $\kappa_L$ along $x$ and $z$ respectively. The shaded zones for each temperature represent the contributions of different frequency regimes (A, M, O) to the cumulative lattice thermal conductivities (in percentage).}
\label{fig:kappa4phomega}
\end{figure}

\setcounter{figure}{12}
\begin{figure}[H]
\centering
\includegraphics[width=0.4\textwidth]{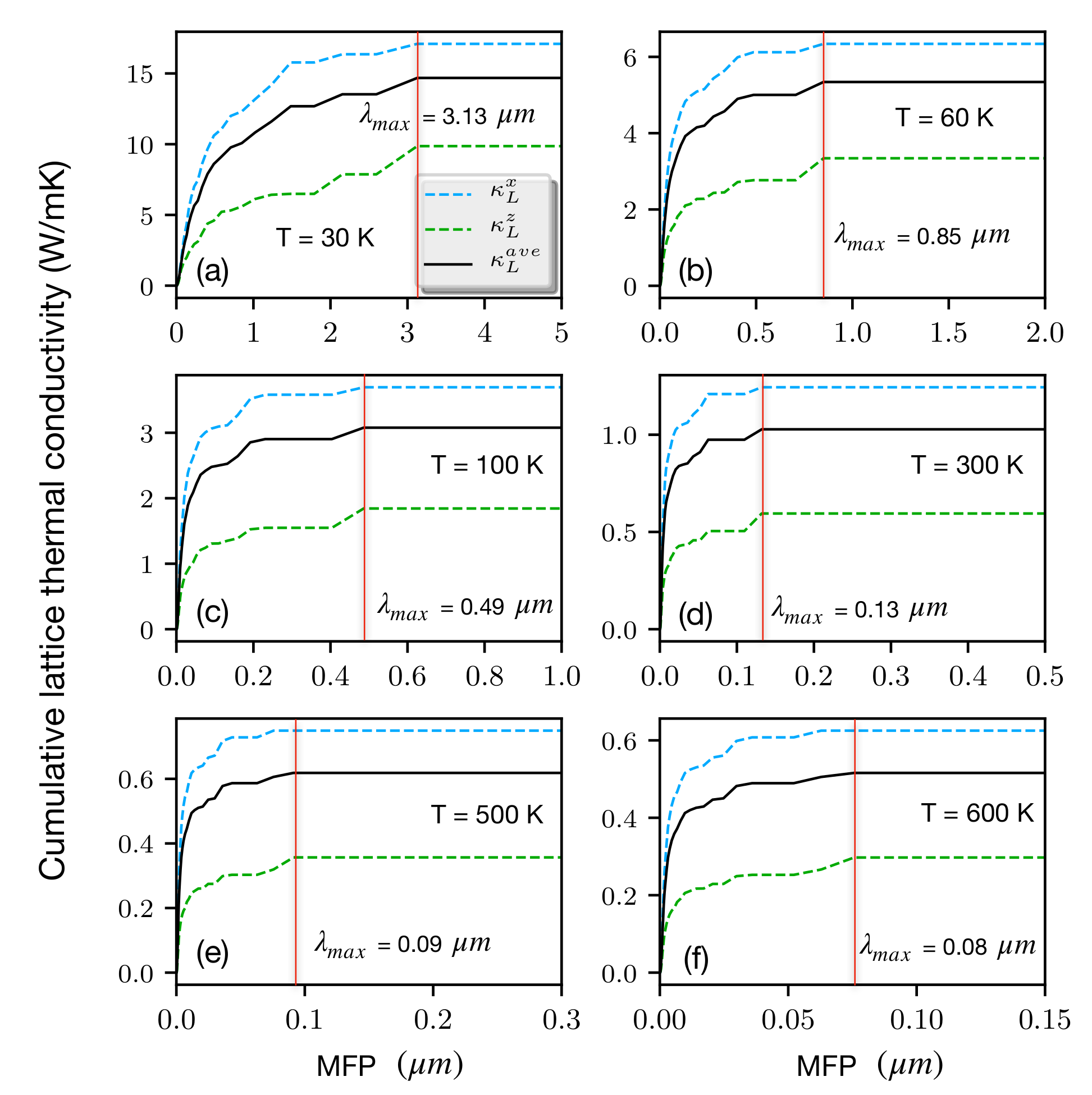}
\caption{Cumulative lattice thermal conductivity including 3ph and phonon-isotope scattering processes as a function of maximum allowed mean free path for (a) $T$ = 30 K, (b) $T$ = 60 K, (c) $T$ = 100 K, (d) $T$ = 300 K, (e) $T$ = 500 K and (f) $T$ = 600 K. For each temperature, black lines denote average $\kappa_L$, while blue and green dashed lines represent $\kappa_L$ along $x$ and $z$ respectively. The red vertical lines at each temperature indicate the maximum mean free path ($\lambda_{max}$) that contributes to $\kappa_L$.}
\label{fig:kappa3phmfp}
\end{figure}

\setcounter{figure}{13}
\begin{figure}[H]
\centering
\includegraphics[width=0.4\textwidth]{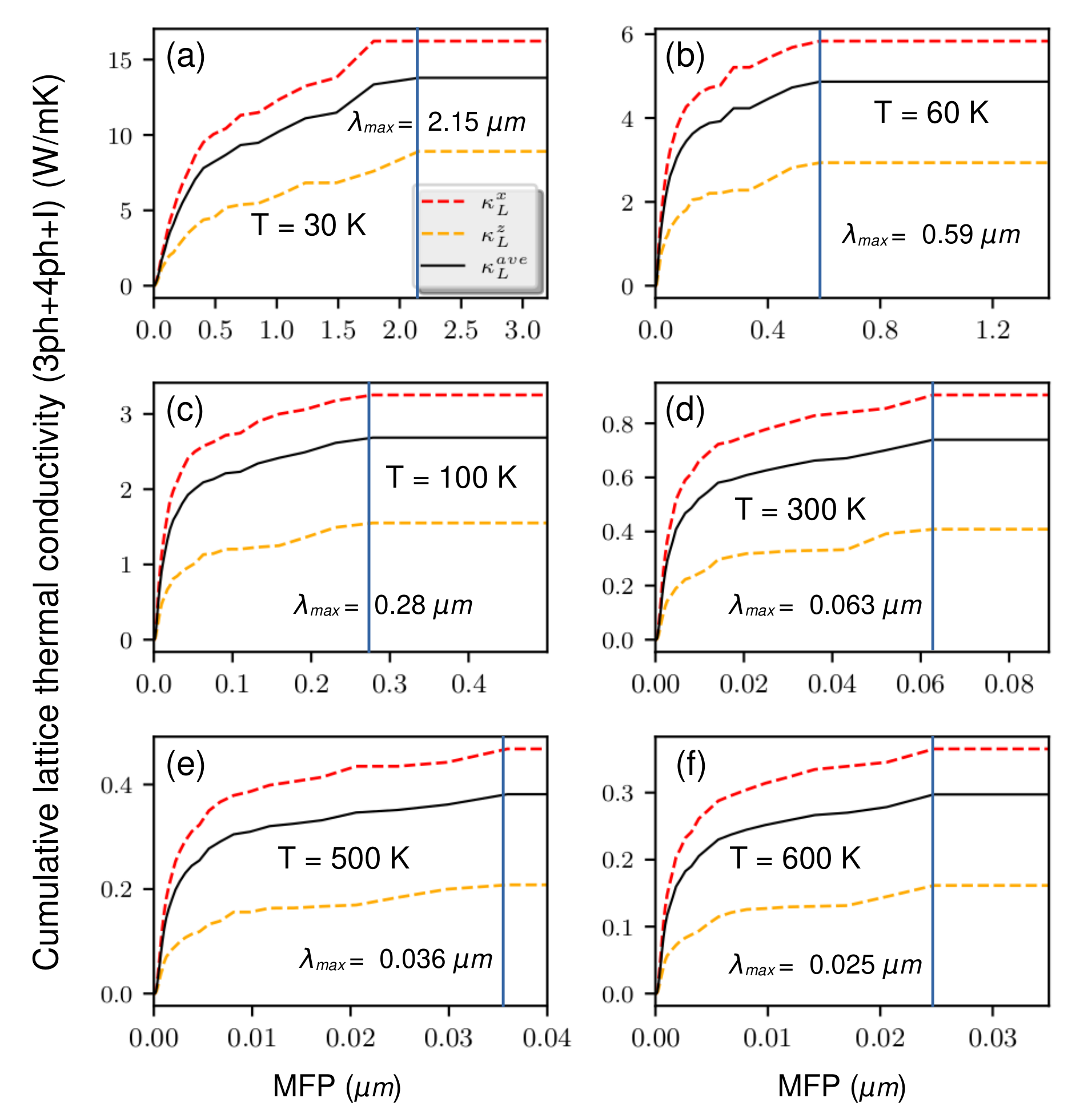}
\caption{Cumulative lattice thermal conductivity including 3ph, 4ph and phonon-isotope scattering processes as a function of maximum allowed mean free path, for (a) $T$ = 30 K, (b) $T$ = 60 K, (c) $T$ = 100 K, (d) $T$ = 300 K, (e) $T$ = 500 K and (f) $T$ = 600 K. For each temperature, black lines denote average $\kappa_L$, while red and yellow dashed lines show $\kappa_L$ along $x$ and $z$ respectively. The blue vertical lines at each temperature indicate the maximum mean free path ($\lambda_{max}$) that contributes to $\kappa_L$.}
\label{fig:kappa4phmfp}
\end{figure}

\onecolumngrid
\begin{widetext}
\setcounter{figure}{14}
\begin{figure}[H]
\centering
\includegraphics[width=1\textwidth]{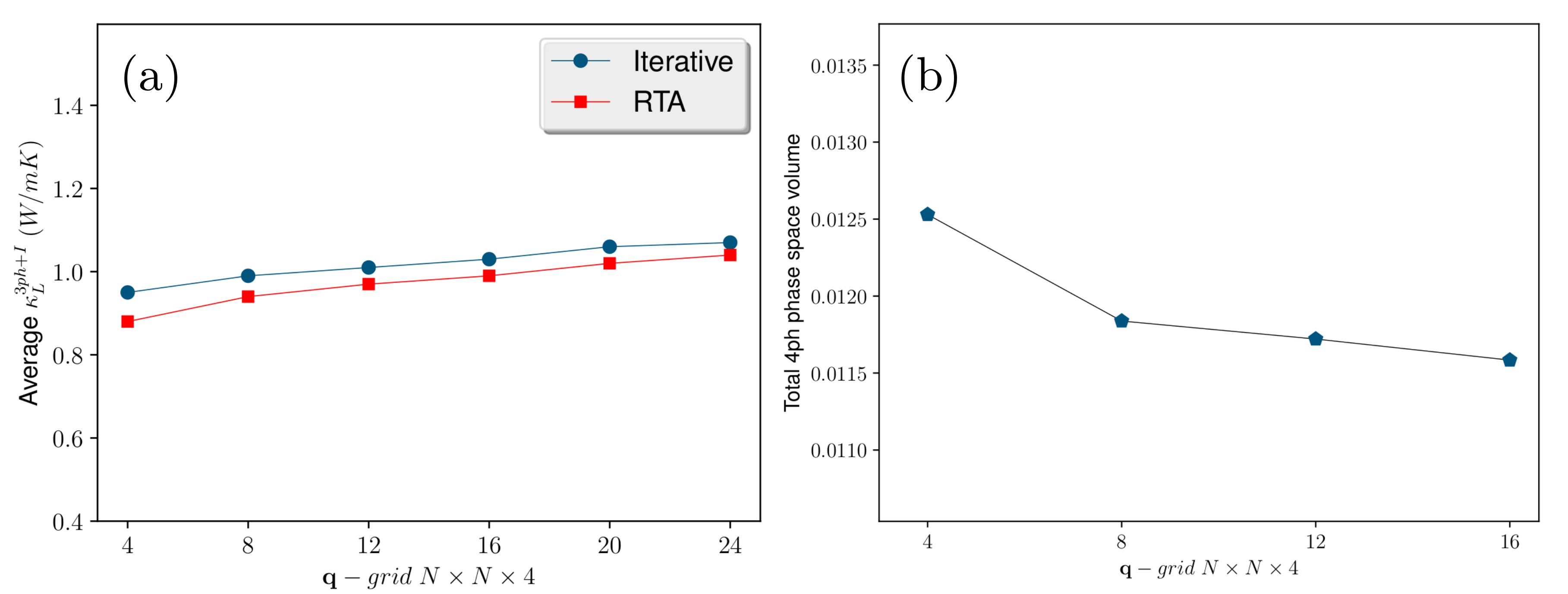}
\caption{(a) The variation of average lattice thermal conductivity with mesh size, considering 3ph and isotope-phonon scattering. The comparison between iterative and RTA method is also shown. (b) The total 4ph phase space volume is presented as a function of $\textbf{q}$-mesh grid. }
\label{fig:conv}
\end{figure}
\end{widetext}

\noindent Figure \ref{fig:conv}(a) shows the variation of average lattice thermal conductivity with $\textbf{q}$-grid at $T$ = 300 K using 3ph and isotope scattering. Beyond $\textbf{q}$-grid 16$\times$16$\times$4, thermal conductivity is seen to increase marginally ($\sim$ 3.8$\%$ and $\sim$ 4$\%$ differences between 16$\times$16$\times$4 and 24$\times$24$\times$4 for iterative and RTA method respectively), suggesting a convergence of $\kappa_L$ in three-phonon framework. For convergence of 4ph processes, we present the variation of total four-phonon phase space volume with $\textbf{q}$-grid in Fig \ref{fig:conv}(b). A tiny difference of 1.2 $\%$ between the calculated total 4ph phase space with $\textbf{q}$-grid 12$\times$12$\times$4 and 16$\times$16$\times$4 ensures convergence in 4ph processes. We must mention that computation costs involving 4ph scattering are very high as  Ge$_2$Sb$_2$Te$_5$ consists of 9 atoms in the unit cell. Therefore, 16$\times$16$\times$4 $\textbf{q}$-grid optimally fit for this investigation.

\nocite{*}


\begin{thebibliography}{85}%
\makeatletter
\providecommand \@ifxundefined [1]{%
 \@ifx{#1\undefined}
}%
\providecommand \@ifnum [1]{%
 \ifnum #1\expandafter \@firstoftwo
 \else \expandafter \@secondoftwo
 \fi
}%
\providecommand \@ifx [1]{%
 \ifx #1\expandafter \@firstoftwo
 \else \expandafter \@secondoftwo
 \fi
}%
\providecommand \natexlab [1]{#1}%
\providecommand \enquote  [1]{``#1''}%
\providecommand \bibnamefont  [1]{#1}%
\providecommand \bibfnamefont [1]{#1}%
\providecommand \citenamefont [1]{#1}%
\providecommand \href@noop [0]{\@secondoftwo}%
\providecommand \href [0]{\begingroup \@sanitize@url \@href}%
\providecommand \@href[1]{\@@startlink{#1}\@@href}%
\providecommand \@@href[1]{\endgroup#1\@@endlink}%
\providecommand \@sanitize@url [0]{\catcode `\\12\catcode `\$12\catcode `\&12\catcode `\#12\catcode `\^12\catcode `\_12\catcode `\%12\relax}%
\providecommand \@@startlink[1]{}%
\providecommand \@@endlink[0]{}%
\providecommand \url  [0]{\begingroup\@sanitize@url \@url }%
\providecommand \@url [1]{\endgroup\@href {#1}{\urlprefix }}%
\providecommand \urlprefix  [0]{URL }%
\providecommand \Eprint [0]{\href }%
\providecommand \doibase [0]{https://doi.org/}%
\providecommand \selectlanguage [0]{\@gobble}%
\providecommand \bibinfo  [0]{\@secondoftwo}%
\providecommand \bibfield  [0]{\@secondoftwo}%
\providecommand \translation [1]{[#1]}%
\providecommand \BibitemOpen [0]{}%
\providecommand \bibitemStop [0]{}%
\providecommand \bibitemNoStop [0]{.\EOS\space}%
\providecommand \EOS [0]{\spacefactor3000\relax}%
\providecommand \BibitemShut  [1]{\csname bibitem#1\endcsname}%
\let\auto@bib@innerbib\@empty
\bibitem [{\citenamefont {Wuttig}\ and\ \citenamefont {Yamada}(2007)}]{wuttig2007phase}%
  \BibitemOpen
  \bibfield  {author} {\bibinfo {author} {\bibfnamefont {M.}~\bibnamefont {Wuttig}}\ and\ \bibinfo {author} {\bibfnamefont {N.}~\bibnamefont {Yamada}},\ }\href@noop {} {\bibfield  {journal} {\bibinfo  {journal} {Nature materials}\ }\textbf {\bibinfo {volume} {6}},\ \bibinfo {pages} {824} (\bibinfo {year} {2007})}\BibitemShut {NoStop}%
\bibitem [{\citenamefont {Wuttig}(2005)}]{wuttig2005towards}%
  \BibitemOpen
  \bibfield  {author} {\bibinfo {author} {\bibfnamefont {M.}~\bibnamefont {Wuttig}},\ }\href@noop {} {\bibfield  {journal} {\bibinfo  {journal} {Nature materials}\ }\textbf {\bibinfo {volume} {4}},\ \bibinfo {pages} {265} (\bibinfo {year} {2005})}\BibitemShut {NoStop}%
\bibitem [{\citenamefont {Raoux}\ \emph {et~al.}(2010)\citenamefont {Raoux}, \citenamefont {We{\l}nic},\ and\ \citenamefont {Ielmini}}]{raoux2010phase}%
  \BibitemOpen
  \bibfield  {author} {\bibinfo {author} {\bibfnamefont {S.}~\bibnamefont {Raoux}}, \bibinfo {author} {\bibfnamefont {W.}~\bibnamefont {We{\l}nic}},\ and\ \bibinfo {author} {\bibfnamefont {D.}~\bibnamefont {Ielmini}},\ }\href@noop {} {\bibfield  {journal} {\bibinfo  {journal} {Chemical reviews}\ }\textbf {\bibinfo {volume} {110}},\ \bibinfo {pages} {240} (\bibinfo {year} {2010})}\BibitemShut {NoStop}%
\bibitem [{\citenamefont {Heged{\"u}s}\ and\ \citenamefont {Elliott}(2008)}]{hegedus2008microscopic}%
  \BibitemOpen
  \bibfield  {author} {\bibinfo {author} {\bibfnamefont {J.}~\bibnamefont {Heged{\"u}s}}\ and\ \bibinfo {author} {\bibfnamefont {S.}~\bibnamefont {Elliott}},\ }\href@noop {} {\bibfield  {journal} {\bibinfo  {journal} {Nature materials}\ }\textbf {\bibinfo {volume} {7}},\ \bibinfo {pages} {399} (\bibinfo {year} {2008})}\BibitemShut {NoStop}%
\bibitem [{\citenamefont {Pirovano}\ \emph {et~al.}(2004)\citenamefont {Pirovano}, \citenamefont {Lacaita}, \citenamefont {Benvenuti}, \citenamefont {Pellizzer},\ and\ \citenamefont {Bez}}]{pirovano2004electronic}%
  \BibitemOpen
  \bibfield  {author} {\bibinfo {author} {\bibfnamefont {A.}~\bibnamefont {Pirovano}}, \bibinfo {author} {\bibfnamefont {A.~L.}\ \bibnamefont {Lacaita}}, \bibinfo {author} {\bibfnamefont {A.}~\bibnamefont {Benvenuti}}, \bibinfo {author} {\bibfnamefont {F.}~\bibnamefont {Pellizzer}},\ and\ \bibinfo {author} {\bibfnamefont {R.}~\bibnamefont {Bez}},\ }\href@noop {} {\bibfield  {journal} {\bibinfo  {journal} {IEEE Transactions on Electron Devices}\ }\textbf {\bibinfo {volume} {51}},\ \bibinfo {pages} {452} (\bibinfo {year} {2004})}\BibitemShut {NoStop}%
\bibitem [{\citenamefont {Kim}\ \emph {et~al.}(2007)\citenamefont {Kim}, \citenamefont {Merget}, \citenamefont {F{\"o}rst},\ and\ \citenamefont {Kurz}}]{kim2007three}%
  \BibitemOpen
  \bibfield  {author} {\bibinfo {author} {\bibfnamefont {D.-H.}\ \bibnamefont {Kim}}, \bibinfo {author} {\bibfnamefont {F.}~\bibnamefont {Merget}}, \bibinfo {author} {\bibfnamefont {M.}~\bibnamefont {F{\"o}rst}},\ and\ \bibinfo {author} {\bibfnamefont {H.}~\bibnamefont {Kurz}},\ }\href@noop {} {\bibfield  {journal} {\bibinfo  {journal} {Journal of Applied Physics}\ }\textbf {\bibinfo {volume} {101}},\ \bibinfo {pages} {064512} (\bibinfo {year} {2007})}\BibitemShut {NoStop}%
\bibitem [{\citenamefont {Jones}(2020)}]{jones2020phase}%
  \BibitemOpen
  \bibfield  {author} {\bibinfo {author} {\bibfnamefont {R.~O.}\ \bibnamefont {Jones}},\ }\href@noop {} {\bibfield  {journal} {\bibinfo  {journal} {Physical Review B}\ }\textbf {\bibinfo {volume} {101}},\ \bibinfo {pages} {024103} (\bibinfo {year} {2020})}\BibitemShut {NoStop}%
\bibitem [{\citenamefont {Gan}\ \emph {et~al.}(2022)\citenamefont {Gan}, \citenamefont {Miao}, \citenamefont {Zhou},\ and\ \citenamefont {Sun}}]{gan2022new}%
  \BibitemOpen
  \bibfield  {author} {\bibinfo {author} {\bibfnamefont {Y.}~\bibnamefont {Gan}}, \bibinfo {author} {\bibfnamefont {N.}~\bibnamefont {Miao}}, \bibinfo {author} {\bibfnamefont {J.}~\bibnamefont {Zhou}},\ and\ \bibinfo {author} {\bibfnamefont {Z.}~\bibnamefont {Sun}},\ }\href@noop {} {\bibfield  {journal} {\bibinfo  {journal} {Journal of Materials Chemistry C}\ }\textbf {\bibinfo {volume} {10}},\ \bibinfo {pages} {16813} (\bibinfo {year} {2022})}\BibitemShut {NoStop}%
\bibitem [{\citenamefont {Kusiak}\ \emph {et~al.}(2022)\citenamefont {Kusiak}, \citenamefont {Chassain}, \citenamefont {Canseco}, \citenamefont {Ghosh}, \citenamefont {Cyrille}, \citenamefont {Serra}, \citenamefont {Navarro}, \citenamefont {Bernard}, \citenamefont {Tran},\ and\ \citenamefont {Battaglia}}]{kusiak2022temperature}%
  \BibitemOpen
  \bibfield  {author} {\bibinfo {author} {\bibfnamefont {A.}~\bibnamefont {Kusiak}}, \bibinfo {author} {\bibfnamefont {C.}~\bibnamefont {Chassain}}, \bibinfo {author} {\bibfnamefont {A.~M.}\ \bibnamefont {Canseco}}, \bibinfo {author} {\bibfnamefont {K.}~\bibnamefont {Ghosh}}, \bibinfo {author} {\bibfnamefont {M.-C.}\ \bibnamefont {Cyrille}}, \bibinfo {author} {\bibfnamefont {A.~L.}\ \bibnamefont {Serra}}, \bibinfo {author} {\bibfnamefont {G.}~\bibnamefont {Navarro}}, \bibinfo {author} {\bibfnamefont {M.}~\bibnamefont {Bernard}}, \bibinfo {author} {\bibfnamefont {N.-P.}\ \bibnamefont {Tran}},\ and\ \bibinfo {author} {\bibfnamefont {J.-L.}\ \bibnamefont {Battaglia}},\ }\href@noop {} {\bibfield  {journal} {\bibinfo  {journal} {physica status solidi (RRL)--Rapid Research Letters}\ }\textbf {\bibinfo {volume} {16}},\ \bibinfo {pages} {2100507} (\bibinfo {year} {2022})}\BibitemShut {NoStop}%
\bibitem [{\citenamefont {Abou El~Kheir}\ \emph {et~al.}(2021)\citenamefont {Abou El~Kheir}, \citenamefont {Dragoni},\ and\ \citenamefont {Bernasconi}}]{abou2021density}%
  \BibitemOpen
  \bibfield  {author} {\bibinfo {author} {\bibfnamefont {O.}~\bibnamefont {Abou El~Kheir}}, \bibinfo {author} {\bibfnamefont {D.}~\bibnamefont {Dragoni}},\ and\ \bibinfo {author} {\bibfnamefont {M.}~\bibnamefont {Bernasconi}},\ }\href@noop {} {\bibfield  {journal} {\bibinfo  {journal} {Physical Review Materials}\ }\textbf {\bibinfo {volume} {5}},\ \bibinfo {pages} {095004} (\bibinfo {year} {2021})}\BibitemShut {NoStop}%
\bibitem [{\citenamefont {Chassain}\ \emph {et~al.}(2023)\citenamefont {Chassain}, \citenamefont {Kusiak}, \citenamefont {Gaborieau}, \citenamefont {Anguy}, \citenamefont {Tran}, \citenamefont {Sabbione}, \citenamefont {Cyrille}, \citenamefont {Wiemer}, \citenamefont {Lamperti},\ and\ \citenamefont {Battaglia}}]{chassain2023thermal}%
  \BibitemOpen
  \bibfield  {author} {\bibinfo {author} {\bibfnamefont {C.}~\bibnamefont {Chassain}}, \bibinfo {author} {\bibfnamefont {A.}~\bibnamefont {Kusiak}}, \bibinfo {author} {\bibfnamefont {C.}~\bibnamefont {Gaborieau}}, \bibinfo {author} {\bibfnamefont {Y.}~\bibnamefont {Anguy}}, \bibinfo {author} {\bibfnamefont {N.-P.}\ \bibnamefont {Tran}}, \bibinfo {author} {\bibfnamefont {C.}~\bibnamefont {Sabbione}}, \bibinfo {author} {\bibfnamefont {M.-C.}\ \bibnamefont {Cyrille}}, \bibinfo {author} {\bibfnamefont {C.}~\bibnamefont {Wiemer}}, \bibinfo {author} {\bibfnamefont {A.}~\bibnamefont {Lamperti}},\ and\ \bibinfo {author} {\bibfnamefont {J.-L.}\ \bibnamefont {Battaglia}},\ }\href@noop {} {\bibfield  {journal} {\bibinfo  {journal} {Journal of Applied Physics}\ }\textbf {\bibinfo {volume} {133}} (\bibinfo {year} {2023})}\BibitemShut {NoStop}%
\bibitem [{\citenamefont {Lee}\ \emph {et~al.}(2012)\citenamefont {Lee}, \citenamefont {Kodama}, \citenamefont {Won}, \citenamefont {Asheghi},\ and\ \citenamefont {Goodson}}]{lee2012phase}%
  \BibitemOpen
  \bibfield  {author} {\bibinfo {author} {\bibfnamefont {J.}~\bibnamefont {Lee}}, \bibinfo {author} {\bibfnamefont {T.}~\bibnamefont {Kodama}}, \bibinfo {author} {\bibfnamefont {Y.}~\bibnamefont {Won}}, \bibinfo {author} {\bibfnamefont {M.}~\bibnamefont {Asheghi}},\ and\ \bibinfo {author} {\bibfnamefont {K.~E.}\ \bibnamefont {Goodson}},\ }\href@noop {} {\bibfield  {journal} {\bibinfo  {journal} {Journal of Applied Physics}\ }\textbf {\bibinfo {volume} {112}},\ \bibinfo {pages} {014902} (\bibinfo {year} {2012})}\BibitemShut {NoStop}%
\bibitem [{\citenamefont {Faraclas}\ \emph {et~al.}(2014)\citenamefont {Faraclas}, \citenamefont {Bakan}, \citenamefont {Dirisaglik}, \citenamefont {Williams}, \citenamefont {Gokirmak}, \citenamefont {Silva} \emph {et~al.}}]{faraclas2014modeling}%
  \BibitemOpen
  \bibfield  {author} {\bibinfo {author} {\bibfnamefont {A.}~\bibnamefont {Faraclas}}, \bibinfo {author} {\bibfnamefont {G.}~\bibnamefont {Bakan}}, \bibinfo {author} {\bibfnamefont {F.}~\bibnamefont {Dirisaglik}}, \bibinfo {author} {\bibfnamefont {N.~E.}\ \bibnamefont {Williams}}, \bibinfo {author} {\bibfnamefont {A.}~\bibnamefont {Gokirmak}}, \bibinfo {author} {\bibfnamefont {H.}~\bibnamefont {Silva}}, \emph {et~al.},\ }\href@noop {} {\bibfield  {journal} {\bibinfo  {journal} {IEEE Transactions on Electron Devices}\ }\textbf {\bibinfo {volume} {61}},\ \bibinfo {pages} {372} (\bibinfo {year} {2014})}\BibitemShut {NoStop}%
\bibitem [{\citenamefont {Siegrist}\ \emph {et~al.}(2011)\citenamefont {Siegrist}, \citenamefont {Jost}, \citenamefont {Volker}, \citenamefont {Woda}, \citenamefont {Merkelbach}, \citenamefont {Schlockermann},\ and\ \citenamefont {Wuttig}}]{siegrist2011disorder}%
  \BibitemOpen
  \bibfield  {author} {\bibinfo {author} {\bibfnamefont {T.}~\bibnamefont {Siegrist}}, \bibinfo {author} {\bibfnamefont {P.}~\bibnamefont {Jost}}, \bibinfo {author} {\bibfnamefont {H.}~\bibnamefont {Volker}}, \bibinfo {author} {\bibfnamefont {M.}~\bibnamefont {Woda}}, \bibinfo {author} {\bibfnamefont {P.}~\bibnamefont {Merkelbach}}, \bibinfo {author} {\bibfnamefont {C.}~\bibnamefont {Schlockermann}},\ and\ \bibinfo {author} {\bibfnamefont {M.}~\bibnamefont {Wuttig}},\ }\href@noop {} {\bibfield  {journal} {\bibinfo  {journal} {Nature materials}\ }\textbf {\bibinfo {volume} {10}},\ \bibinfo {pages} {202} (\bibinfo {year} {2011})}\BibitemShut {NoStop}%
\bibitem [{\citenamefont {Ibarra-Hern{\'a}ndez}\ and\ \citenamefont {Raty}(2018)}]{ibarra2018ab}%
  \BibitemOpen
  \bibfield  {author} {\bibinfo {author} {\bibfnamefont {W.}~\bibnamefont {Ibarra-Hern{\'a}ndez}}\ and\ \bibinfo {author} {\bibfnamefont {J.-Y.}\ \bibnamefont {Raty}},\ }\href@noop {} {\bibfield  {journal} {\bibinfo  {journal} {Physical Review B}\ }\textbf {\bibinfo {volume} {97}},\ \bibinfo {pages} {245205} (\bibinfo {year} {2018})}\BibitemShut {NoStop}%
\bibitem [{\citenamefont {Wei}\ \emph {et~al.}(2019)\citenamefont {Wei}, \citenamefont {Hu}, \citenamefont {Chen}, \citenamefont {Zhao}, \citenamefont {Qiu}, \citenamefont {Shi},\ and\ \citenamefont {Chen}}]{wei2019quasi}%
  \BibitemOpen
  \bibfield  {author} {\bibinfo {author} {\bibfnamefont {T.-R.}\ \bibnamefont {Wei}}, \bibinfo {author} {\bibfnamefont {P.}~\bibnamefont {Hu}}, \bibinfo {author} {\bibfnamefont {H.}~\bibnamefont {Chen}}, \bibinfo {author} {\bibfnamefont {K.}~\bibnamefont {Zhao}}, \bibinfo {author} {\bibfnamefont {P.}~\bibnamefont {Qiu}}, \bibinfo {author} {\bibfnamefont {X.}~\bibnamefont {Shi}},\ and\ \bibinfo {author} {\bibfnamefont {L.}~\bibnamefont {Chen}},\ }\href@noop {} {\bibfield  {journal} {\bibinfo  {journal} {Applied Physics Letters}\ }\textbf {\bibinfo {volume} {114}},\ \bibinfo {pages} {053903} (\bibinfo {year} {2019})}\BibitemShut {NoStop}%
\bibitem [{\citenamefont {Miao}\ \emph {et~al.}(2022)\citenamefont {Miao}, \citenamefont {Wang}, \citenamefont {Zhou}, \citenamefont {Huang}, \citenamefont {Qian}, \citenamefont {Yuan},\ and\ \citenamefont {Lan}}]{miao2022remarkable}%
  \BibitemOpen
  \bibfield  {author} {\bibinfo {author} {\bibfnamefont {J.}~\bibnamefont {Miao}}, \bibinfo {author} {\bibfnamefont {P.}~\bibnamefont {Wang}}, \bibinfo {author} {\bibfnamefont {P.}~\bibnamefont {Zhou}}, \bibinfo {author} {\bibfnamefont {S.}~\bibnamefont {Huang}}, \bibinfo {author} {\bibfnamefont {D.}~\bibnamefont {Qian}}, \bibinfo {author} {\bibfnamefont {Y.}~\bibnamefont {Yuan}},\ and\ \bibinfo {author} {\bibfnamefont {R.}~\bibnamefont {Lan}},\ }\href@noop {} {\bibfield  {journal} {\bibinfo  {journal} {Journal of Alloys and Compounds}\ }\textbf {\bibinfo {volume} {900}},\ \bibinfo {pages} {163471} (\bibinfo {year} {2022})}\BibitemShut {NoStop}%
\bibitem [{\citenamefont {Zhou}\ \emph {et~al.}(2023)\citenamefont {Zhou}, \citenamefont {Lan}, \citenamefont {Wang}, \citenamefont {Miao}, \citenamefont {Huang}, \citenamefont {Yuan},\ and\ \citenamefont {Xu}}]{zhou2023tuned}%
  \BibitemOpen
  \bibfield  {author} {\bibinfo {author} {\bibfnamefont {P.}~\bibnamefont {Zhou}}, \bibinfo {author} {\bibfnamefont {R.}~\bibnamefont {Lan}}, \bibinfo {author} {\bibfnamefont {P.}~\bibnamefont {Wang}}, \bibinfo {author} {\bibfnamefont {J.}~\bibnamefont {Miao}}, \bibinfo {author} {\bibfnamefont {S.}~\bibnamefont {Huang}}, \bibinfo {author} {\bibfnamefont {Y.}~\bibnamefont {Yuan}},\ and\ \bibinfo {author} {\bibfnamefont {J.}~\bibnamefont {Xu}},\ }\href@noop {} {\bibfield  {journal} {\bibinfo  {journal} {Materials Today Communications}\ }\textbf {\bibinfo {volume} {35}},\ \bibinfo {pages} {105839} (\bibinfo {year} {2023})}\BibitemShut {NoStop}%
\bibitem [{\citenamefont {Konstantinou}\ \emph {et~al.}(2018)\citenamefont {Konstantinou}, \citenamefont {Mocanu}, \citenamefont {Lee},\ and\ \citenamefont {Elliott}}]{konstantinou2018ab}%
  \BibitemOpen
  \bibfield  {author} {\bibinfo {author} {\bibfnamefont {K.}~\bibnamefont {Konstantinou}}, \bibinfo {author} {\bibfnamefont {F.~C.}\ \bibnamefont {Mocanu}}, \bibinfo {author} {\bibfnamefont {T.~H.}\ \bibnamefont {Lee}},\ and\ \bibinfo {author} {\bibfnamefont {S.~R.}\ \bibnamefont {Elliott}},\ }\href@noop {} {\bibfield  {journal} {\bibinfo  {journal} {Journal of Physics: Condensed Matter}\ }\textbf {\bibinfo {volume} {30}},\ \bibinfo {pages} {455401} (\bibinfo {year} {2018})}\BibitemShut {NoStop}%
\bibitem [{\citenamefont {Lyeo}\ \emph {et~al.}(2006)\citenamefont {Lyeo}, \citenamefont {Cahill}, \citenamefont {Lee}, \citenamefont {Abelson}, \citenamefont {Kwon}, \citenamefont {Kim}, \citenamefont {Bishop},\ and\ \citenamefont {Cheong}}]{lyeo2006thermal}%
  \BibitemOpen
  \bibfield  {author} {\bibinfo {author} {\bibfnamefont {H.-K.}\ \bibnamefont {Lyeo}}, \bibinfo {author} {\bibfnamefont {D.~G.}\ \bibnamefont {Cahill}}, \bibinfo {author} {\bibfnamefont {B.-S.}\ \bibnamefont {Lee}}, \bibinfo {author} {\bibfnamefont {J.~R.}\ \bibnamefont {Abelson}}, \bibinfo {author} {\bibfnamefont {M.-H.}\ \bibnamefont {Kwon}}, \bibinfo {author} {\bibfnamefont {K.-B.}\ \bibnamefont {Kim}}, \bibinfo {author} {\bibfnamefont {S.~G.}\ \bibnamefont {Bishop}},\ and\ \bibinfo {author} {\bibfnamefont {B.-k.}\ \bibnamefont {Cheong}},\ }\href@noop {} {\bibfield  {journal} {\bibinfo  {journal} {Applied Physics Letters}\ }\textbf {\bibinfo {volume} {89}} (\bibinfo {year} {2006})}\BibitemShut {NoStop}%
\bibitem [{\citenamefont {Battaglia}\ \emph {et~al.}(2010)\citenamefont {Battaglia}, \citenamefont {Kusiak}, \citenamefont {Schick}, \citenamefont {Cappella}, \citenamefont {Wiemer}, \citenamefont {Longo},\ and\ \citenamefont {Varesi}}]{battaglia2010thermal}%
  \BibitemOpen
  \bibfield  {author} {\bibinfo {author} {\bibfnamefont {J.-L.}\ \bibnamefont {Battaglia}}, \bibinfo {author} {\bibfnamefont {A.}~\bibnamefont {Kusiak}}, \bibinfo {author} {\bibfnamefont {V.}~\bibnamefont {Schick}}, \bibinfo {author} {\bibfnamefont {A.}~\bibnamefont {Cappella}}, \bibinfo {author} {\bibfnamefont {C.}~\bibnamefont {Wiemer}}, \bibinfo {author} {\bibfnamefont {M.}~\bibnamefont {Longo}},\ and\ \bibinfo {author} {\bibfnamefont {E.}~\bibnamefont {Varesi}},\ }\href@noop {} {\bibfield  {journal} {\bibinfo  {journal} {Journal of Applied Physics}\ }\textbf {\bibinfo {volume} {107}} (\bibinfo {year} {2010})}\BibitemShut {NoStop}%
\bibitem [{\citenamefont {Y{\'a}{\~n}ez-Lim{\'o}n}\ \emph {et~al.}(1995)\citenamefont {Y{\'a}{\~n}ez-Lim{\'o}n}, \citenamefont {Gonz{\'a}lez-Hern{\'a}ndez}, \citenamefont {Alvarado-Gil}, \citenamefont {Delgadillo},\ and\ \citenamefont {Vargas}}]{yanez1995thermal}%
  \BibitemOpen
  \bibfield  {author} {\bibinfo {author} {\bibfnamefont {J.}~\bibnamefont {Y{\'a}{\~n}ez-Lim{\'o}n}}, \bibinfo {author} {\bibfnamefont {J.}~\bibnamefont {Gonz{\'a}lez-Hern{\'a}ndez}}, \bibinfo {author} {\bibfnamefont {J.}~\bibnamefont {Alvarado-Gil}}, \bibinfo {author} {\bibfnamefont {I.}~\bibnamefont {Delgadillo}},\ and\ \bibinfo {author} {\bibfnamefont {H.}~\bibnamefont {Vargas}},\ }\href@noop {} {\bibfield  {journal} {\bibinfo  {journal} {Physical review B}\ }\textbf {\bibinfo {volume} {52}},\ \bibinfo {pages} {16321} (\bibinfo {year} {1995})}\BibitemShut {NoStop}%
\bibitem [{\citenamefont {Kuwahara}\ \emph {et~al.}(2007)\citenamefont {Kuwahara}, \citenamefont {Suzuki}, \citenamefont {Yamakawa}, \citenamefont {Taketoshi}, \citenamefont {Yagi}, \citenamefont {Fons}, \citenamefont {Fukaya}, \citenamefont {Tominaga},\ and\ \citenamefont {Baba}}]{kuwahara2007measurement}%
  \BibitemOpen
  \bibfield  {author} {\bibinfo {author} {\bibfnamefont {M.}~\bibnamefont {Kuwahara}}, \bibinfo {author} {\bibfnamefont {O.}~\bibnamefont {Suzuki}}, \bibinfo {author} {\bibfnamefont {Y.}~\bibnamefont {Yamakawa}}, \bibinfo {author} {\bibfnamefont {N.}~\bibnamefont {Taketoshi}}, \bibinfo {author} {\bibfnamefont {T.}~\bibnamefont {Yagi}}, \bibinfo {author} {\bibfnamefont {P.}~\bibnamefont {Fons}}, \bibinfo {author} {\bibfnamefont {T.}~\bibnamefont {Fukaya}}, \bibinfo {author} {\bibfnamefont {J.}~\bibnamefont {Tominaga}},\ and\ \bibinfo {author} {\bibfnamefont {T.}~\bibnamefont {Baba}},\ }\href@noop {} {\bibfield  {journal} {\bibinfo  {journal} {Microelectronic engineering}\ }\textbf {\bibinfo {volume} {84}},\ \bibinfo {pages} {1792} (\bibinfo {year} {2007})}\BibitemShut {NoStop}%
\bibitem [{\citenamefont {Fallica}\ \emph {et~al.}(2009)\citenamefont {Fallica}, \citenamefont {Battaglia}, \citenamefont {Cocco}, \citenamefont {Monguzzi}, \citenamefont {Teren}, \citenamefont {Wiemer}, \citenamefont {Varesi}, \citenamefont {Cecchini}, \citenamefont {Gotti},\ and\ \citenamefont {Fanciulli}}]{fallica2009thermal}%
  \BibitemOpen
  \bibfield  {author} {\bibinfo {author} {\bibfnamefont {R.}~\bibnamefont {Fallica}}, \bibinfo {author} {\bibfnamefont {J.-L.}\ \bibnamefont {Battaglia}}, \bibinfo {author} {\bibfnamefont {S.}~\bibnamefont {Cocco}}, \bibinfo {author} {\bibfnamefont {C.}~\bibnamefont {Monguzzi}}, \bibinfo {author} {\bibfnamefont {A.}~\bibnamefont {Teren}}, \bibinfo {author} {\bibfnamefont {C.}~\bibnamefont {Wiemer}}, \bibinfo {author} {\bibfnamefont {E.}~\bibnamefont {Varesi}}, \bibinfo {author} {\bibfnamefont {R.}~\bibnamefont {Cecchini}}, \bibinfo {author} {\bibfnamefont {A.}~\bibnamefont {Gotti}},\ and\ \bibinfo {author} {\bibfnamefont {M.}~\bibnamefont {Fanciulli}},\ }\href@noop {} {\bibfield  {journal} {\bibinfo  {journal} {Journal of Chemical \& Engineering Data}\ }\textbf {\bibinfo {volume} {54}},\ \bibinfo {pages} {1698} (\bibinfo {year} {2009})}\BibitemShut {NoStop}%
\bibitem [{\citenamefont {Li}\ \emph {et~al.}(2023)\citenamefont {Li}, \citenamefont {Levit}, \citenamefont {Yalon},\ and\ \citenamefont {Sun}}]{li2023temperature}%
  \BibitemOpen
  \bibfield  {author} {\bibinfo {author} {\bibfnamefont {Q.}~\bibnamefont {Li}}, \bibinfo {author} {\bibfnamefont {O.}~\bibnamefont {Levit}}, \bibinfo {author} {\bibfnamefont {E.}~\bibnamefont {Yalon}},\ and\ \bibinfo {author} {\bibfnamefont {B.}~\bibnamefont {Sun}},\ }\href@noop {} {\bibfield  {journal} {\bibinfo  {journal} {Journal of Applied Physics}\ }\textbf {\bibinfo {volume} {133}},\ \bibinfo {pages} {135105} (\bibinfo {year} {2023})}\BibitemShut {NoStop}%
\bibitem [{\citenamefont {Tsafack}\ \emph {et~al.}(2011)\citenamefont {Tsafack}, \citenamefont {Piccinini}, \citenamefont {Lee}, \citenamefont {Pop},\ and\ \citenamefont {Rudan}}]{tsafack2011electronic}%
  \BibitemOpen
  \bibfield  {author} {\bibinfo {author} {\bibfnamefont {T.}~\bibnamefont {Tsafack}}, \bibinfo {author} {\bibfnamefont {E.}~\bibnamefont {Piccinini}}, \bibinfo {author} {\bibfnamefont {B.-S.}\ \bibnamefont {Lee}}, \bibinfo {author} {\bibfnamefont {E.}~\bibnamefont {Pop}},\ and\ \bibinfo {author} {\bibfnamefont {M.}~\bibnamefont {Rudan}},\ }\href@noop {} {\bibfield  {journal} {\bibinfo  {journal} {Journal of Applied Physics}\ }\textbf {\bibinfo {volume} {110}} (\bibinfo {year} {2011})}\BibitemShut {NoStop}%
\bibitem [{\citenamefont {Kheir}\ \emph {et~al.}(2023)\citenamefont {Kheir}, \citenamefont {Bonati}, \citenamefont {Parrinello},\ and\ \citenamefont {Bernasconi}}]{kheir2023unraveling}%
  \BibitemOpen
  \bibfield  {author} {\bibinfo {author} {\bibfnamefont {O.~A.~E.}\ \bibnamefont {Kheir}}, \bibinfo {author} {\bibfnamefont {L.}~\bibnamefont {Bonati}}, \bibinfo {author} {\bibfnamefont {M.}~\bibnamefont {Parrinello}},\ and\ \bibinfo {author} {\bibfnamefont {M.}~\bibnamefont {Bernasconi}},\ }\href@noop {} {\bibfield  {journal} {\bibinfo  {journal} {arXiv preprint arXiv:2304.03109}\ } (\bibinfo {year} {2023})}\BibitemShut {NoStop}%
\bibitem [{\citenamefont {Mukhopadhyay}\ \emph {et~al.}(2016)\citenamefont {Mukhopadhyay}, \citenamefont {Lindsay},\ and\ \citenamefont {Singh}}]{mukhopadhyay2016optic}%
  \BibitemOpen
  \bibfield  {author} {\bibinfo {author} {\bibfnamefont {S.}~\bibnamefont {Mukhopadhyay}}, \bibinfo {author} {\bibfnamefont {L.}~\bibnamefont {Lindsay}},\ and\ \bibinfo {author} {\bibfnamefont {D.~J.}\ \bibnamefont {Singh}},\ }\href@noop {} {\bibfield  {journal} {\bibinfo  {journal} {Scientific reports}\ }\textbf {\bibinfo {volume} {6}},\ \bibinfo {pages} {37076} (\bibinfo {year} {2016})}\BibitemShut {NoStop}%
\bibitem [{\citenamefont {Campi}\ \emph {et~al.}(2017)\citenamefont {Campi}, \citenamefont {Paulatto}, \citenamefont {Fugallo}, \citenamefont {Mauri},\ and\ \citenamefont {Bernasconi}}]{campi2017first}%
  \BibitemOpen
  \bibfield  {author} {\bibinfo {author} {\bibfnamefont {D.}~\bibnamefont {Campi}}, \bibinfo {author} {\bibfnamefont {L.}~\bibnamefont {Paulatto}}, \bibinfo {author} {\bibfnamefont {G.}~\bibnamefont {Fugallo}}, \bibinfo {author} {\bibfnamefont {F.}~\bibnamefont {Mauri}},\ and\ \bibinfo {author} {\bibfnamefont {M.}~\bibnamefont {Bernasconi}},\ }\href@noop {} {\bibfield  {journal} {\bibinfo  {journal} {Physical Review B}\ }\textbf {\bibinfo {volume} {95}},\ \bibinfo {pages} {024311} (\bibinfo {year} {2017})}\BibitemShut {NoStop}%
\bibitem [{\citenamefont {Pan}\ \emph {et~al.}(2019)\citenamefont {Pan}, \citenamefont {Li},\ and\ \citenamefont {Guo}}]{pan2019lattice}%
  \BibitemOpen
  \bibfield  {author} {\bibinfo {author} {\bibfnamefont {Y.}~\bibnamefont {Pan}}, \bibinfo {author} {\bibfnamefont {Z.}~\bibnamefont {Li}},\ and\ \bibinfo {author} {\bibfnamefont {Z.}~\bibnamefont {Guo}},\ }\href@noop {} {\bibfield  {journal} {\bibinfo  {journal} {Crystals}\ }\textbf {\bibinfo {volume} {9}},\ \bibinfo {pages} {136} (\bibinfo {year} {2019})}\BibitemShut {NoStop}%
\bibitem [{\citenamefont {Risk}\ \emph {et~al.}(2009)\citenamefont {Risk}, \citenamefont {Rettner},\ and\ \citenamefont {Raoux}}]{risk2009thermal}%
  \BibitemOpen
  \bibfield  {author} {\bibinfo {author} {\bibfnamefont {W.}~\bibnamefont {Risk}}, \bibinfo {author} {\bibfnamefont {C.}~\bibnamefont {Rettner}},\ and\ \bibinfo {author} {\bibfnamefont {S.}~\bibnamefont {Raoux}},\ }\href@noop {} {\bibfield  {journal} {\bibinfo  {journal} {Applied Physics Letters}\ }\textbf {\bibinfo {volume} {94}} (\bibinfo {year} {2009})}\BibitemShut {NoStop}%
\bibitem [{\citenamefont {Slack}(1979)}]{slack1979thermal}%
  \BibitemOpen
  \bibfield  {author} {\bibinfo {author} {\bibfnamefont {G.~A.}\ \bibnamefont {Slack}},\ }\href@noop {} {\bibfield  {journal} {\bibinfo  {journal} {Solid state physics}\ }\textbf {\bibinfo {volume} {34}},\ \bibinfo {pages} {1} (\bibinfo {year} {1979})}\BibitemShut {NoStop}%
\bibitem [{\citenamefont {Kaviany}(2014)}]{kaviany2014heat}%
  \BibitemOpen
  \bibfield  {author} {\bibinfo {author} {\bibfnamefont {M.}~\bibnamefont {Kaviany}},\ }\href@noop {} {\emph {\bibinfo {title} {Heat transfer physics}}}\ (\bibinfo  {publisher} {Cambridge University Press},\ \bibinfo {year} {2014})\BibitemShut {NoStop}%
\bibitem [{\citenamefont {Dong}\ \emph {et~al.}(2022)\citenamefont {Dong}, \citenamefont {Zhou}, \citenamefont {Chen}, \citenamefont {Li}, \citenamefont {Cao}, \citenamefont {Luo}, \citenamefont {Zhong}, \citenamefont {Peng}, \citenamefont {Wu},\ and\ \citenamefont {Chen}}]{dong2022effect}%
  \BibitemOpen
  \bibfield  {author} {\bibinfo {author} {\bibfnamefont {Z.-Y.}\ \bibnamefont {Dong}}, \bibinfo {author} {\bibfnamefont {Y.}~\bibnamefont {Zhou}}, \bibinfo {author} {\bibfnamefont {X.-Q.}\ \bibnamefont {Chen}}, \bibinfo {author} {\bibfnamefont {W.-J.}\ \bibnamefont {Li}}, \bibinfo {author} {\bibfnamefont {Z.-Y.}\ \bibnamefont {Cao}}, \bibinfo {author} {\bibfnamefont {C.}~\bibnamefont {Luo}}, \bibinfo {author} {\bibfnamefont {G.-H.}\ \bibnamefont {Zhong}}, \bibinfo {author} {\bibfnamefont {Q.}~\bibnamefont {Peng}}, \bibinfo {author} {\bibfnamefont {X.}~\bibnamefont {Wu}},\ and\ \bibinfo {author} {\bibfnamefont {X.-J.}\ \bibnamefont {Chen}},\ }\href@noop {} {\bibfield  {journal} {\bibinfo  {journal} {Physical Review B}\ }\textbf {\bibinfo {volume} {105}},\ \bibinfo {pages} {184301} (\bibinfo {year} {2022})}\BibitemShut {NoStop}%
\bibitem [{\citenamefont {Lindsay}\ \emph {et~al.}(2013)\citenamefont {Lindsay}, \citenamefont {Broido},\ and\ \citenamefont {Reinecke}}]{lindsay2013ab}%
  \BibitemOpen
  \bibfield  {author} {\bibinfo {author} {\bibfnamefont {L.}~\bibnamefont {Lindsay}}, \bibinfo {author} {\bibfnamefont {D.}~\bibnamefont {Broido}},\ and\ \bibinfo {author} {\bibfnamefont {T.}~\bibnamefont {Reinecke}},\ }\href@noop {} {\bibfield  {journal} {\bibinfo  {journal} {Physical Review B}\ }\textbf {\bibinfo {volume} {87}},\ \bibinfo {pages} {165201} (\bibinfo {year} {2013})}\BibitemShut {NoStop}%
\bibitem [{\citenamefont {Yang}\ \emph {et~al.}(2019)\citenamefont {Yang}, \citenamefont {Feng}, \citenamefont {Li},\ and\ \citenamefont {Ruan}}]{yang2019stronger}%
  \BibitemOpen
  \bibfield  {author} {\bibinfo {author} {\bibfnamefont {X.}~\bibnamefont {Yang}}, \bibinfo {author} {\bibfnamefont {T.}~\bibnamefont {Feng}}, \bibinfo {author} {\bibfnamefont {J.}~\bibnamefont {Li}},\ and\ \bibinfo {author} {\bibfnamefont {X.}~\bibnamefont {Ruan}},\ }\href@noop {} {\bibfield  {journal} {\bibinfo  {journal} {Physical Review B}\ }\textbf {\bibinfo {volume} {100}},\ \bibinfo {pages} {245203} (\bibinfo {year} {2019})}\BibitemShut {NoStop}%
\bibitem [{\citenamefont {Tian}\ \emph {et~al.}(2012)\citenamefont {Tian}, \citenamefont {Garg}, \citenamefont {Esfarjani}, \citenamefont {Shiga}, \citenamefont {Shiomi},\ and\ \citenamefont {Chen}}]{tian2012phonon}%
  \BibitemOpen
  \bibfield  {author} {\bibinfo {author} {\bibfnamefont {Z.}~\bibnamefont {Tian}}, \bibinfo {author} {\bibfnamefont {J.}~\bibnamefont {Garg}}, \bibinfo {author} {\bibfnamefont {K.}~\bibnamefont {Esfarjani}}, \bibinfo {author} {\bibfnamefont {T.}~\bibnamefont {Shiga}}, \bibinfo {author} {\bibfnamefont {J.}~\bibnamefont {Shiomi}},\ and\ \bibinfo {author} {\bibfnamefont {G.}~\bibnamefont {Chen}},\ }\href@noop {} {\bibfield  {journal} {\bibinfo  {journal} {Physical Review B}\ }\textbf {\bibinfo {volume} {85}},\ \bibinfo {pages} {184303} (\bibinfo {year} {2012})}\BibitemShut {NoStop}%
\bibitem [{\citenamefont {Li}\ \emph {et~al.}(2012)\citenamefont {Li}, \citenamefont {Lindsay}, \citenamefont {Broido}, \citenamefont {Stewart},\ and\ \citenamefont {Mingo}}]{li2012thermal}%
  \BibitemOpen
  \bibfield  {author} {\bibinfo {author} {\bibfnamefont {W.}~\bibnamefont {Li}}, \bibinfo {author} {\bibfnamefont {L.}~\bibnamefont {Lindsay}}, \bibinfo {author} {\bibfnamefont {D.~A.}\ \bibnamefont {Broido}}, \bibinfo {author} {\bibfnamefont {D.~A.}\ \bibnamefont {Stewart}},\ and\ \bibinfo {author} {\bibfnamefont {N.}~\bibnamefont {Mingo}},\ }\href@noop {} {\bibfield  {journal} {\bibinfo  {journal} {Physical Review B}\ }\textbf {\bibinfo {volume} {86}},\ \bibinfo {pages} {174307} (\bibinfo {year} {2012})}\BibitemShut {NoStop}%
\bibitem [{\citenamefont {Ghosh}\ \emph {et~al.}(2020{\natexlab{a}})\citenamefont {Ghosh}, \citenamefont {Kusiak}, \citenamefont {No{\'e}}, \citenamefont {Cyrille},\ and\ \citenamefont {Battaglia}}]{ghosh2020thermal}%
  \BibitemOpen
  \bibfield  {author} {\bibinfo {author} {\bibfnamefont {K.}~\bibnamefont {Ghosh}}, \bibinfo {author} {\bibfnamefont {A.}~\bibnamefont {Kusiak}}, \bibinfo {author} {\bibfnamefont {P.}~\bibnamefont {No{\'e}}}, \bibinfo {author} {\bibfnamefont {M.-C.}\ \bibnamefont {Cyrille}},\ and\ \bibinfo {author} {\bibfnamefont {J.-L.}\ \bibnamefont {Battaglia}},\ }\href@noop {} {\bibfield  {journal} {\bibinfo  {journal} {Physical Review B}\ }\textbf {\bibinfo {volume} {101}},\ \bibinfo {pages} {214305} (\bibinfo {year} {2020}{\natexlab{a}})}\BibitemShut {NoStop}%
\bibitem [{\citenamefont {Ghosh}\ \emph {et~al.}(2020{\natexlab{b}})\citenamefont {Ghosh}, \citenamefont {Kusiak},\ and\ \citenamefont {Battaglia}}]{ghosh2020phonon}%
  \BibitemOpen
  \bibfield  {author} {\bibinfo {author} {\bibfnamefont {K.}~\bibnamefont {Ghosh}}, \bibinfo {author} {\bibfnamefont {A.}~\bibnamefont {Kusiak}},\ and\ \bibinfo {author} {\bibfnamefont {J.-L.}\ \bibnamefont {Battaglia}},\ }\href@noop {} {\bibfield  {journal} {\bibinfo  {journal} {Physical Review B}\ }\textbf {\bibinfo {volume} {102}},\ \bibinfo {pages} {094311} (\bibinfo {year} {2020}{\natexlab{b}})}\BibitemShut {NoStop}%
\bibitem [{\citenamefont {Li}\ \emph {et~al.}(2022)\citenamefont {Li}, \citenamefont {Liu},\ and\ \citenamefont {Hong}}]{li2022anharmonicity}%
  \BibitemOpen
  \bibfield  {author} {\bibinfo {author} {\bibfnamefont {Y.}~\bibnamefont {Li}}, \bibinfo {author} {\bibfnamefont {J.}~\bibnamefont {Liu}},\ and\ \bibinfo {author} {\bibfnamefont {J.}~\bibnamefont {Hong}},\ }\href@noop {} {\bibfield  {journal} {\bibinfo  {journal} {Physical Review B}\ }\textbf {\bibinfo {volume} {106}},\ \bibinfo {pages} {094317} (\bibinfo {year} {2022})}\BibitemShut {NoStop}%
\bibitem [{\citenamefont {Feng}\ \emph {et~al.}(2017)\citenamefont {Feng}, \citenamefont {Lindsay},\ and\ \citenamefont {Ruan}}]{feng2017four}%
  \BibitemOpen
  \bibfield  {author} {\bibinfo {author} {\bibfnamefont {T.}~\bibnamefont {Feng}}, \bibinfo {author} {\bibfnamefont {L.}~\bibnamefont {Lindsay}},\ and\ \bibinfo {author} {\bibfnamefont {X.}~\bibnamefont {Ruan}},\ }\href@noop {} {\bibfield  {journal} {\bibinfo  {journal} {Physical Review B}\ }\textbf {\bibinfo {volume} {96}},\ \bibinfo {pages} {161201} (\bibinfo {year} {2017})}\BibitemShut {NoStop}%
\bibitem [{\citenamefont {Wang}\ \emph {et~al.}(2023{\natexlab{a}})\citenamefont {Wang}, \citenamefont {Gan}, \citenamefont {Hu}, \citenamefont {Li}, \citenamefont {Xie},\ and\ \citenamefont {He}}]{wang2023anharmonic}%
  \BibitemOpen
  \bibfield  {author} {\bibinfo {author} {\bibfnamefont {Y.}~\bibnamefont {Wang}}, \bibinfo {author} {\bibfnamefont {Q.}~\bibnamefont {Gan}}, \bibinfo {author} {\bibfnamefont {M.}~\bibnamefont {Hu}}, \bibinfo {author} {\bibfnamefont {J.}~\bibnamefont {Li}}, \bibinfo {author} {\bibfnamefont {L.}~\bibnamefont {Xie}},\ and\ \bibinfo {author} {\bibfnamefont {J.}~\bibnamefont {He}},\ }\href@noop {} {\bibfield  {journal} {\bibinfo  {journal} {Physical Review B}\ }\textbf {\bibinfo {volume} {107}},\ \bibinfo {pages} {064308} (\bibinfo {year} {2023}{\natexlab{a}})}\BibitemShut {NoStop}%
\bibitem [{\citenamefont {Xie}\ \emph {et~al.}(2020)\citenamefont {Xie}, \citenamefont {Feng}, \citenamefont {Li},\ and\ \citenamefont {He}}]{xie2020first}%
  \BibitemOpen
  \bibfield  {author} {\bibinfo {author} {\bibfnamefont {L.}~\bibnamefont {Xie}}, \bibinfo {author} {\bibfnamefont {J.}~\bibnamefont {Feng}}, \bibinfo {author} {\bibfnamefont {R.}~\bibnamefont {Li}},\ and\ \bibinfo {author} {\bibfnamefont {J.}~\bibnamefont {He}},\ }\href@noop {} {\bibfield  {journal} {\bibinfo  {journal} {Physical Review Letters}\ }\textbf {\bibinfo {volume} {125}},\ \bibinfo {pages} {245901} (\bibinfo {year} {2020})}\BibitemShut {NoStop}%
\bibitem [{\citenamefont {Wang}\ \emph {et~al.}(2023{\natexlab{b}})\citenamefont {Wang}, \citenamefont {Gao}, \citenamefont {Zhu}, \citenamefont {Ren}, \citenamefont {Hu}, \citenamefont {Sun}, \citenamefont {Ding}, \citenamefont {Xia},\ and\ \citenamefont {Li}}]{wang2023role}%
  \BibitemOpen
  \bibfield  {author} {\bibinfo {author} {\bibfnamefont {X.}~\bibnamefont {Wang}}, \bibinfo {author} {\bibfnamefont {Z.}~\bibnamefont {Gao}}, \bibinfo {author} {\bibfnamefont {G.}~\bibnamefont {Zhu}}, \bibinfo {author} {\bibfnamefont {J.}~\bibnamefont {Ren}}, \bibinfo {author} {\bibfnamefont {L.}~\bibnamefont {Hu}}, \bibinfo {author} {\bibfnamefont {J.}~\bibnamefont {Sun}}, \bibinfo {author} {\bibfnamefont {X.}~\bibnamefont {Ding}}, \bibinfo {author} {\bibfnamefont {Y.}~\bibnamefont {Xia}},\ and\ \bibinfo {author} {\bibfnamefont {B.}~\bibnamefont {Li}},\ }\href@noop {} {\bibfield  {journal} {\bibinfo  {journal} {Physical Review B}\ }\textbf {\bibinfo {volume} {107}},\ \bibinfo {pages} {214308} (\bibinfo {year} {2023}{\natexlab{b}})}\BibitemShut {NoStop}%
\bibitem [{\citenamefont {Reifenberg}\ \emph {et~al.}(2007)\citenamefont {Reifenberg}, \citenamefont {Panzer}, \citenamefont {Kim}, \citenamefont {Gibby}, \citenamefont {Zhang}, \citenamefont {Wong}, \citenamefont {Wong}, \citenamefont {Pop},\ and\ \citenamefont {Goodson}}]{reifenberg2007thickness}%
  \BibitemOpen
  \bibfield  {author} {\bibinfo {author} {\bibfnamefont {J.~P.}\ \bibnamefont {Reifenberg}}, \bibinfo {author} {\bibfnamefont {M.~A.}\ \bibnamefont {Panzer}}, \bibinfo {author} {\bibfnamefont {S.}~\bibnamefont {Kim}}, \bibinfo {author} {\bibfnamefont {A.~M.}\ \bibnamefont {Gibby}}, \bibinfo {author} {\bibfnamefont {Y.}~\bibnamefont {Zhang}}, \bibinfo {author} {\bibfnamefont {S.}~\bibnamefont {Wong}}, \bibinfo {author} {\bibfnamefont {H.-S.~P.}\ \bibnamefont {Wong}}, \bibinfo {author} {\bibfnamefont {E.}~\bibnamefont {Pop}},\ and\ \bibinfo {author} {\bibfnamefont {K.~E.}\ \bibnamefont {Goodson}},\ }\href@noop {} {\bibfield  {journal} {\bibinfo  {journal} {Applied Physics Letters}\ }\textbf {\bibinfo {volume} {91}} (\bibinfo {year} {2007})}\BibitemShut {NoStop}%
\bibitem [{\citenamefont {Lee}\ \emph {et~al.}(2011)\citenamefont {Lee}, \citenamefont {Li}, \citenamefont {Reifenberg}, \citenamefont {Lee}, \citenamefont {Sinclair}, \citenamefont {Asheghi},\ and\ \citenamefont {Goodson}}]{lee2011thermal}%
  \BibitemOpen
  \bibfield  {author} {\bibinfo {author} {\bibfnamefont {J.}~\bibnamefont {Lee}}, \bibinfo {author} {\bibfnamefont {Z.}~\bibnamefont {Li}}, \bibinfo {author} {\bibfnamefont {J.~P.}\ \bibnamefont {Reifenberg}}, \bibinfo {author} {\bibfnamefont {S.}~\bibnamefont {Lee}}, \bibinfo {author} {\bibfnamefont {R.}~\bibnamefont {Sinclair}}, \bibinfo {author} {\bibfnamefont {M.}~\bibnamefont {Asheghi}},\ and\ \bibinfo {author} {\bibfnamefont {K.~E.}\ \bibnamefont {Goodson}},\ }\href@noop {} {\bibfield  {journal} {\bibinfo  {journal} {Journal of Applied Physics}\ }\textbf {\bibinfo {volume} {109}},\ \bibinfo {pages} {084902} (\bibinfo {year} {2011})}\BibitemShut {NoStop}%
\bibitem [{\citenamefont {Skl{\'e}nard}\ \emph {et~al.}(2021)\citenamefont {Skl{\'e}nard}, \citenamefont {Triozon}, \citenamefont {Sabbione}, \citenamefont {Nistor}, \citenamefont {Frei}, \citenamefont {Navarro},\ and\ \citenamefont {Li}}]{sklenard2021electronic}%
  \BibitemOpen
  \bibfield  {author} {\bibinfo {author} {\bibfnamefont {B.}~\bibnamefont {Skl{\'e}nard}}, \bibinfo {author} {\bibfnamefont {F.}~\bibnamefont {Triozon}}, \bibinfo {author} {\bibfnamefont {C.}~\bibnamefont {Sabbione}}, \bibinfo {author} {\bibfnamefont {L.}~\bibnamefont {Nistor}}, \bibinfo {author} {\bibfnamefont {M.}~\bibnamefont {Frei}}, \bibinfo {author} {\bibfnamefont {G.}~\bibnamefont {Navarro}},\ and\ \bibinfo {author} {\bibfnamefont {J.}~\bibnamefont {Li}},\ }\href@noop {} {\bibfield  {journal} {\bibinfo  {journal} {Applied Physics Letters}\ }\textbf {\bibinfo {volume} {119}},\ \bibinfo {pages} {201911} (\bibinfo {year} {2021})}\BibitemShut {NoStop}%
\bibitem [{\citenamefont {Siegert}\ \emph {et~al.}(2014)\citenamefont {Siegert}, \citenamefont {Lange}, \citenamefont {Sittner}, \citenamefont {Volker}, \citenamefont {Schlockermann}, \citenamefont {Siegrist},\ and\ \citenamefont {Wuttig}}]{siegert2014impact}%
  \BibitemOpen
  \bibfield  {author} {\bibinfo {author} {\bibfnamefont {K.}~\bibnamefont {Siegert}}, \bibinfo {author} {\bibfnamefont {F.}~\bibnamefont {Lange}}, \bibinfo {author} {\bibfnamefont {E.}~\bibnamefont {Sittner}}, \bibinfo {author} {\bibfnamefont {H.}~\bibnamefont {Volker}}, \bibinfo {author} {\bibfnamefont {C.}~\bibnamefont {Schlockermann}}, \bibinfo {author} {\bibfnamefont {T.}~\bibnamefont {Siegrist}},\ and\ \bibinfo {author} {\bibfnamefont {M.}~\bibnamefont {Wuttig}},\ }\href@noop {} {\bibfield  {journal} {\bibinfo  {journal} {Reports on Progress in Physics}\ }\textbf {\bibinfo {volume} {78}},\ \bibinfo {pages} {013001} (\bibinfo {year} {2014})}\BibitemShut {NoStop}%
\bibitem [{\citenamefont {Caravati}\ \emph {et~al.}(2009)\citenamefont {Caravati}, \citenamefont {Bernasconi}, \citenamefont {K{\"u}hne}, \citenamefont {Krack},\ and\ \citenamefont {Parrinello}}]{caravati2009first}%
  \BibitemOpen
  \bibfield  {author} {\bibinfo {author} {\bibfnamefont {S.}~\bibnamefont {Caravati}}, \bibinfo {author} {\bibfnamefont {M.}~\bibnamefont {Bernasconi}}, \bibinfo {author} {\bibfnamefont {T.}~\bibnamefont {K{\"u}hne}}, \bibinfo {author} {\bibfnamefont {M.}~\bibnamefont {Krack}},\ and\ \bibinfo {author} {\bibfnamefont {M.}~\bibnamefont {Parrinello}},\ }\href@noop {} {\bibfield  {journal} {\bibinfo  {journal} {Journal of Physics: Condensed Matter}\ }\textbf {\bibinfo {volume} {21}},\ \bibinfo {pages} {255501} (\bibinfo {year} {2009})}\BibitemShut {NoStop}%
\bibitem [{\citenamefont {Baroni}\ \emph {et~al.}(2001)\citenamefont {Baroni}, \citenamefont {De~Gironcoli}, \citenamefont {Dal~Corso},\ and\ \citenamefont {Giannozzi}}]{baroni2001phonons}%
  \BibitemOpen
  \bibfield  {author} {\bibinfo {author} {\bibfnamefont {S.}~\bibnamefont {Baroni}}, \bibinfo {author} {\bibfnamefont {S.}~\bibnamefont {De~Gironcoli}}, \bibinfo {author} {\bibfnamefont {A.}~\bibnamefont {Dal~Corso}},\ and\ \bibinfo {author} {\bibfnamefont {P.}~\bibnamefont {Giannozzi}},\ }\href@noop {} {\bibfield  {journal} {\bibinfo  {journal} {Reviews of modern Physics}\ }\textbf {\bibinfo {volume} {73}},\ \bibinfo {pages} {515} (\bibinfo {year} {2001})}\BibitemShut {NoStop}%
\bibitem [{\citenamefont {Giannozzi}\ \emph {et~al.}(2009)\citenamefont {Giannozzi}, \citenamefont {Baroni}, \citenamefont {Bonini}, \citenamefont {Calandra}, \citenamefont {Car}, \citenamefont {Cavazzoni}, \citenamefont {Ceresoli}, \citenamefont {Chiarotti}, \citenamefont {Cococcioni}, \citenamefont {Dabo} \emph {et~al.}}]{giannozzi2009quantum}%
  \BibitemOpen
  \bibfield  {author} {\bibinfo {author} {\bibfnamefont {P.}~\bibnamefont {Giannozzi}}, \bibinfo {author} {\bibfnamefont {S.}~\bibnamefont {Baroni}}, \bibinfo {author} {\bibfnamefont {N.}~\bibnamefont {Bonini}}, \bibinfo {author} {\bibfnamefont {M.}~\bibnamefont {Calandra}}, \bibinfo {author} {\bibfnamefont {R.}~\bibnamefont {Car}}, \bibinfo {author} {\bibfnamefont {C.}~\bibnamefont {Cavazzoni}}, \bibinfo {author} {\bibfnamefont {D.}~\bibnamefont {Ceresoli}}, \bibinfo {author} {\bibfnamefont {G.~L.}\ \bibnamefont {Chiarotti}}, \bibinfo {author} {\bibfnamefont {M.}~\bibnamefont {Cococcioni}}, \bibinfo {author} {\bibfnamefont {I.}~\bibnamefont {Dabo}}, \emph {et~al.},\ }\href@noop {} {\bibfield  {journal} {\bibinfo  {journal} {Journal of physics: Condensed matter}\ }\textbf {\bibinfo {volume} {21}},\ \bibinfo {pages} {395502} (\bibinfo {year} {2009})}\BibitemShut {NoStop}%
\bibitem [{\citenamefont {Perdew}\ \emph {et~al.}(1996)\citenamefont {Perdew}, \citenamefont {Burke},\ and\ \citenamefont {Ernzerhof}}]{perdew1996generalized}%
  \BibitemOpen
  \bibfield  {author} {\bibinfo {author} {\bibfnamefont {J.~P.}\ \bibnamefont {Perdew}}, \bibinfo {author} {\bibfnamefont {K.}~\bibnamefont {Burke}},\ and\ \bibinfo {author} {\bibfnamefont {M.}~\bibnamefont {Ernzerhof}},\ }\href@noop {} {\bibfield  {journal} {\bibinfo  {journal} {Physical review letters}\ }\textbf {\bibinfo {volume} {77}},\ \bibinfo {pages} {3865} (\bibinfo {year} {1996})}\BibitemShut {NoStop}%
\bibitem [{\citenamefont {Bl{\"o}chl}(1994)}]{blochl1994projector}%
  \BibitemOpen
  \bibfield  {author} {\bibinfo {author} {\bibfnamefont {P.~E.}\ \bibnamefont {Bl{\"o}chl}},\ }\href@noop {} {\bibfield  {journal} {\bibinfo  {journal} {Physical review B}\ }\textbf {\bibinfo {volume} {50}},\ \bibinfo {pages} {17953} (\bibinfo {year} {1994})}\BibitemShut {NoStop}%
\bibitem [{\citenamefont {Monkhorst}\ and\ \citenamefont {Pack}(1976)}]{monkhorst1976special}%
  \BibitemOpen
  \bibfield  {author} {\bibinfo {author} {\bibfnamefont {H.~J.}\ \bibnamefont {Monkhorst}}\ and\ \bibinfo {author} {\bibfnamefont {J.~D.}\ \bibnamefont {Pack}},\ }\href@noop {} {\bibfield  {journal} {\bibinfo  {journal} {Physical review B}\ }\textbf {\bibinfo {volume} {13}},\ \bibinfo {pages} {5188} (\bibinfo {year} {1976})}\BibitemShut {NoStop}%
\bibitem [{\citenamefont {Han}\ \emph {et~al.}(2022)\citenamefont {Han}, \citenamefont {Yang}, \citenamefont {Li}, \citenamefont {Feng},\ and\ \citenamefont {Ruan}}]{han2022fourphonon}%
  \BibitemOpen
  \bibfield  {author} {\bibinfo {author} {\bibfnamefont {Z.}~\bibnamefont {Han}}, \bibinfo {author} {\bibfnamefont {X.}~\bibnamefont {Yang}}, \bibinfo {author} {\bibfnamefont {W.}~\bibnamefont {Li}}, \bibinfo {author} {\bibfnamefont {T.}~\bibnamefont {Feng}},\ and\ \bibinfo {author} {\bibfnamefont {X.}~\bibnamefont {Ruan}},\ }\href@noop {} {\bibfield  {journal} {\bibinfo  {journal} {Computer Physics Communications}\ }\textbf {\bibinfo {volume} {270}},\ \bibinfo {pages} {108179} (\bibinfo {year} {2022})}\BibitemShut {NoStop}%
\bibitem [{\citenamefont {Li}\ \emph {et~al.}(2014)\citenamefont {Li}, \citenamefont {Carrete}, \citenamefont {Katcho},\ and\ \citenamefont {Mingo}}]{li2014shengbte}%
  \BibitemOpen
  \bibfield  {author} {\bibinfo {author} {\bibfnamefont {W.}~\bibnamefont {Li}}, \bibinfo {author} {\bibfnamefont {J.}~\bibnamefont {Carrete}}, \bibinfo {author} {\bibfnamefont {N.~A.}\ \bibnamefont {Katcho}},\ and\ \bibinfo {author} {\bibfnamefont {N.}~\bibnamefont {Mingo}},\ }\href@noop {} {\bibfield  {journal} {\bibinfo  {journal} {Computer Physics Communications}\ }\textbf {\bibinfo {volume} {185}},\ \bibinfo {pages} {1747} (\bibinfo {year} {2014})}\BibitemShut {NoStop}%
\bibitem [{\citenamefont {Kooi}\ and\ \citenamefont {De~Hosson}(2002)}]{kooi2002electron}%
  \BibitemOpen
  \bibfield  {author} {\bibinfo {author} {\bibfnamefont {B.}~\bibnamefont {Kooi}}\ and\ \bibinfo {author} {\bibfnamefont {J.~T.~M.}\ \bibnamefont {De~Hosson}},\ }\href@noop {} {\bibfield  {journal} {\bibinfo  {journal} {Journal of applied physics}\ }\textbf {\bibinfo {volume} {92}},\ \bibinfo {pages} {3584} (\bibinfo {year} {2002})}\BibitemShut {NoStop}%
\bibitem [{\citenamefont {Petrov}\ \emph {et~al.}(1968)\citenamefont {Petrov}, \citenamefont {Imamov},\ and\ \citenamefont {Pinsker}}]{petrov1968electron}%
  \BibitemOpen
  \bibfield  {author} {\bibinfo {author} {\bibfnamefont {I.}~\bibnamefont {Petrov}}, \bibinfo {author} {\bibfnamefont {R.}~\bibnamefont {Imamov}},\ and\ \bibinfo {author} {\bibfnamefont {Z.}~\bibnamefont {Pinsker}},\ }\href@noop {} {\bibfield  {journal} {\bibinfo  {journal} {Sov Phys Crystallogr}\ }\textbf {\bibinfo {volume} {13}},\ \bibinfo {pages} {339} (\bibinfo {year} {1968})}\BibitemShut {NoStop}%
\bibitem [{\citenamefont {Matsunaga}\ \emph {et~al.}(2004)\citenamefont {Matsunaga}, \citenamefont {Yamada},\ and\ \citenamefont {Kubota}}]{matsunaga2004structures}%
  \BibitemOpen
  \bibfield  {author} {\bibinfo {author} {\bibfnamefont {T.}~\bibnamefont {Matsunaga}}, \bibinfo {author} {\bibfnamefont {N.}~\bibnamefont {Yamada}},\ and\ \bibinfo {author} {\bibfnamefont {Y.}~\bibnamefont {Kubota}},\ }\href@noop {} {\bibfield  {journal} {\bibinfo  {journal} {Acta Crystallographica Section B: Structural Science}\ }\textbf {\bibinfo {volume} {60}},\ \bibinfo {pages} {685} (\bibinfo {year} {2004})}\BibitemShut {NoStop}%
\bibitem [{\citenamefont {Tominaga}\ \emph {et~al.}(2014)\citenamefont {Tominaga}, \citenamefont {Kolobov}, \citenamefont {Fons}, \citenamefont {Nakano},\ and\ \citenamefont {Murakami}}]{tominaga2014ferroelectric}%
  \BibitemOpen
  \bibfield  {author} {\bibinfo {author} {\bibfnamefont {J.}~\bibnamefont {Tominaga}}, \bibinfo {author} {\bibfnamefont {A.}~\bibnamefont {Kolobov}}, \bibinfo {author} {\bibfnamefont {P.}~\bibnamefont {Fons}}, \bibinfo {author} {\bibfnamefont {T.}~\bibnamefont {Nakano}},\ and\ \bibinfo {author} {\bibfnamefont {S.}~\bibnamefont {Murakami}},\ }\href@noop {} {\bibfield  {journal} {\bibinfo  {journal} {Advanced Materials Interfaces}\ }\textbf {\bibinfo {volume} {1}},\ \bibinfo {pages} {1300027} (\bibinfo {year} {2014})}\BibitemShut {NoStop}%
\bibitem [{\citenamefont {Perdew}\ \emph {et~al.}(2008)\citenamefont {Perdew}, \citenamefont {Ruzsinszky}, \citenamefont {Csonka}, \citenamefont {Vydrov}, \citenamefont {Scuseria}, \citenamefont {Constantin}, \citenamefont {Zhou},\ and\ \citenamefont {Burke}}]{perdew2008restoring}%
  \BibitemOpen
  \bibfield  {author} {\bibinfo {author} {\bibfnamefont {J.~P.}\ \bibnamefont {Perdew}}, \bibinfo {author} {\bibfnamefont {A.}~\bibnamefont {Ruzsinszky}}, \bibinfo {author} {\bibfnamefont {G.~I.}\ \bibnamefont {Csonka}}, \bibinfo {author} {\bibfnamefont {O.~A.}\ \bibnamefont {Vydrov}}, \bibinfo {author} {\bibfnamefont {G.~E.}\ \bibnamefont {Scuseria}}, \bibinfo {author} {\bibfnamefont {L.~A.}\ \bibnamefont {Constantin}}, \bibinfo {author} {\bibfnamefont {X.}~\bibnamefont {Zhou}},\ and\ \bibinfo {author} {\bibfnamefont {K.}~\bibnamefont {Burke}},\ }\href@noop {} {\bibfield  {journal} {\bibinfo  {journal} {Physical review letters}\ }\textbf {\bibinfo {volume} {100}},\ \bibinfo {pages} {136406} (\bibinfo {year} {2008})}\BibitemShut {NoStop}%
\bibitem [{\citenamefont {Momma}\ and\ \citenamefont {Izumi}(2011)}]{momma2011vesta}%
  \BibitemOpen
  \bibfield  {author} {\bibinfo {author} {\bibfnamefont {K.}~\bibnamefont {Momma}}\ and\ \bibinfo {author} {\bibfnamefont {F.}~\bibnamefont {Izumi}},\ }\href@noop {} {\bibfield  {journal} {\bibinfo  {journal} {Journal of applied crystallography}\ }\textbf {\bibinfo {volume} {44}},\ \bibinfo {pages} {1272} (\bibinfo {year} {2011})}\BibitemShut {NoStop}%
\bibitem [{\citenamefont {Sosso}\ \emph {et~al.}(2009)\citenamefont {Sosso}, \citenamefont {Caravati}, \citenamefont {Gatti}, \citenamefont {Assoni},\ and\ \citenamefont {Bernasconi}}]{sosso2009vibrational}%
  \BibitemOpen
  \bibfield  {author} {\bibinfo {author} {\bibfnamefont {G.}~\bibnamefont {Sosso}}, \bibinfo {author} {\bibfnamefont {S.}~\bibnamefont {Caravati}}, \bibinfo {author} {\bibfnamefont {C.}~\bibnamefont {Gatti}}, \bibinfo {author} {\bibfnamefont {S.}~\bibnamefont {Assoni}},\ and\ \bibinfo {author} {\bibfnamefont {M.}~\bibnamefont {Bernasconi}},\ }\href@noop {} {\bibfield  {journal} {\bibinfo  {journal} {Journal of Physics: Condensed Matter}\ }\textbf {\bibinfo {volume} {21}},\ \bibinfo {pages} {245401} (\bibinfo {year} {2009})}\BibitemShut {NoStop}%
\bibitem [{\citenamefont {Tamura}(1983)}]{tamura1983isotope}%
  \BibitemOpen
  \bibfield  {author} {\bibinfo {author} {\bibfnamefont {S.-i.}\ \bibnamefont {Tamura}},\ }\href@noop {} {\bibfield  {journal} {\bibinfo  {journal} {Physical Review B}\ }\textbf {\bibinfo {volume} {27}},\ \bibinfo {pages} {858} (\bibinfo {year} {1983})}\BibitemShut {NoStop}%
\bibitem [{\citenamefont {De~Laeter}\ \emph {et~al.}(2003)\citenamefont {De~Laeter}, \citenamefont {B{\"o}hlke}, \citenamefont {De~Bievre}, \citenamefont {Hidaka}, \citenamefont {Peiser}, \citenamefont {Rosman},\ and\ \citenamefont {Taylor}}]{de2003atomic}%
  \BibitemOpen
  \bibfield  {author} {\bibinfo {author} {\bibfnamefont {J.~R.}\ \bibnamefont {De~Laeter}}, \bibinfo {author} {\bibfnamefont {J.~K.}\ \bibnamefont {B{\"o}hlke}}, \bibinfo {author} {\bibfnamefont {P.}~\bibnamefont {De~Bievre}}, \bibinfo {author} {\bibfnamefont {H.}~\bibnamefont {Hidaka}}, \bibinfo {author} {\bibfnamefont {H.}~\bibnamefont {Peiser}}, \bibinfo {author} {\bibfnamefont {K.}~\bibnamefont {Rosman}},\ and\ \bibinfo {author} {\bibfnamefont {P.}~\bibnamefont {Taylor}},\ }\href@noop {} {\bibfield  {journal} {\bibinfo  {journal} {Pure and applied chemistry}\ }\textbf {\bibinfo {volume} {75}},\ \bibinfo {pages} {683} (\bibinfo {year} {2003})}\BibitemShut {NoStop}%
\bibitem [{\citenamefont {Zhang}\ \emph {et~al.}(2009)\citenamefont {Zhang}, \citenamefont {Ke}, \citenamefont {Chen}, \citenamefont {Yang},\ and\ \citenamefont {Kent}}]{zhang2009thermodynamic}%
  \BibitemOpen
  \bibfield  {author} {\bibinfo {author} {\bibfnamefont {Y.}~\bibnamefont {Zhang}}, \bibinfo {author} {\bibfnamefont {X.}~\bibnamefont {Ke}}, \bibinfo {author} {\bibfnamefont {C.}~\bibnamefont {Chen}}, \bibinfo {author} {\bibfnamefont {J.}~\bibnamefont {Yang}},\ and\ \bibinfo {author} {\bibfnamefont {P.}~\bibnamefont {Kent}},\ }\href@noop {} {\bibfield  {journal} {\bibinfo  {journal} {Physical review B}\ }\textbf {\bibinfo {volume} {80}},\ \bibinfo {pages} {024304} (\bibinfo {year} {2009})}\BibitemShut {NoStop}%
\bibitem [{\citenamefont {Li}\ and\ \citenamefont {Mingo}(2015)}]{li2015ultralow}%
  \BibitemOpen
  \bibfield  {author} {\bibinfo {author} {\bibfnamefont {W.}~\bibnamefont {Li}}\ and\ \bibinfo {author} {\bibfnamefont {N.}~\bibnamefont {Mingo}},\ }\href@noop {} {\bibfield  {journal} {\bibinfo  {journal} {Physical Review B}\ }\textbf {\bibinfo {volume} {91}},\ \bibinfo {pages} {144304} (\bibinfo {year} {2015})}\BibitemShut {NoStop}%
\bibitem [{\citenamefont {Guyer}\ and\ \citenamefont {Krumhansl}(1966)}]{guyer1966thermal}%
  \BibitemOpen
  \bibfield  {author} {\bibinfo {author} {\bibfnamefont {R.}~\bibnamefont {Guyer}}\ and\ \bibinfo {author} {\bibfnamefont {J.}~\bibnamefont {Krumhansl}},\ }\href@noop {} {\bibfield  {journal} {\bibinfo  {journal} {Physical Review}\ }\textbf {\bibinfo {volume} {148}},\ \bibinfo {pages} {778} (\bibinfo {year} {1966})}\BibitemShut {NoStop}%
\bibitem [{\citenamefont {Ghosh}\ \emph {et~al.}(2022)\citenamefont {Ghosh}, \citenamefont {Kusiak},\ and\ \citenamefont {Battaglia}}]{ghosh2022phonon}%
  \BibitemOpen
  \bibfield  {author} {\bibinfo {author} {\bibfnamefont {K.}~\bibnamefont {Ghosh}}, \bibinfo {author} {\bibfnamefont {A.}~\bibnamefont {Kusiak}},\ and\ \bibinfo {author} {\bibfnamefont {J.-L.}\ \bibnamefont {Battaglia}},\ }\href@noop {} {\bibfield  {journal} {\bibinfo  {journal} {Journal of Physics: Condensed Matter}\ }\textbf {\bibinfo {volume} {34}},\ \bibinfo {pages} {323001} (\bibinfo {year} {2022})}\BibitemShut {NoStop}%
\bibitem [{\citenamefont {Lee}\ \emph {et~al.}(2013)\citenamefont {Lee}, \citenamefont {Bozorg-Grayeli}, \citenamefont {Kim}, \citenamefont {Asheghi}, \citenamefont {Philip~Wong},\ and\ \citenamefont {Goodson}}]{lee2013phonon}%
  \BibitemOpen
  \bibfield  {author} {\bibinfo {author} {\bibfnamefont {J.}~\bibnamefont {Lee}}, \bibinfo {author} {\bibfnamefont {E.}~\bibnamefont {Bozorg-Grayeli}}, \bibinfo {author} {\bibfnamefont {S.}~\bibnamefont {Kim}}, \bibinfo {author} {\bibfnamefont {M.}~\bibnamefont {Asheghi}}, \bibinfo {author} {\bibfnamefont {H.-S.}\ \bibnamefont {Philip~Wong}},\ and\ \bibinfo {author} {\bibfnamefont {K.~E.}\ \bibnamefont {Goodson}},\ }\href@noop {} {\bibfield  {journal} {\bibinfo  {journal} {Applied Physics Letters}\ }\textbf {\bibinfo {volume} {102}},\ \bibinfo {pages} {191911} (\bibinfo {year} {2013})}\BibitemShut {NoStop}%
\bibitem [{\citenamefont {Cahill}\ \emph {et~al.}(1992)\citenamefont {Cahill}, \citenamefont {Watson},\ and\ \citenamefont {Pohl}}]{cahill1992lower}%
  \BibitemOpen
  \bibfield  {author} {\bibinfo {author} {\bibfnamefont {D.~G.}\ \bibnamefont {Cahill}}, \bibinfo {author} {\bibfnamefont {S.~K.}\ \bibnamefont {Watson}},\ and\ \bibinfo {author} {\bibfnamefont {R.~O.}\ \bibnamefont {Pohl}},\ }\href@noop {} {\bibfield  {journal} {\bibinfo  {journal} {Physical Review B}\ }\textbf {\bibinfo {volume} {46}},\ \bibinfo {pages} {6131} (\bibinfo {year} {1992})}\BibitemShut {NoStop}%
\bibitem [{\citenamefont {Ratsifaritana}\ and\ \citenamefont {Klemens}(1987)}]{ratsifaritana1987scattering}%
  \BibitemOpen
  \bibfield  {author} {\bibinfo {author} {\bibfnamefont {C.}~\bibnamefont {Ratsifaritana}}\ and\ \bibinfo {author} {\bibfnamefont {P.}~\bibnamefont {Klemens}},\ }\href@noop {} {\bibfield  {journal} {\bibinfo  {journal} {International journal of thermophysics}\ }\textbf {\bibinfo {volume} {8}},\ \bibinfo {pages} {737} (\bibinfo {year} {1987})}\BibitemShut {NoStop}%
\bibitem [{\citenamefont {Bragaglia}\ \emph {et~al.}(2016)\citenamefont {Bragaglia}, \citenamefont {Arciprete}, \citenamefont {Zhang}, \citenamefont {Mio}, \citenamefont {Zallo}, \citenamefont {Perumal}, \citenamefont {Giussani}, \citenamefont {Cecchi}, \citenamefont {Boschker}, \citenamefont {Riechert} \emph {et~al.}}]{bragaglia2016metal}%
  \BibitemOpen
  \bibfield  {author} {\bibinfo {author} {\bibfnamefont {V.}~\bibnamefont {Bragaglia}}, \bibinfo {author} {\bibfnamefont {F.}~\bibnamefont {Arciprete}}, \bibinfo {author} {\bibfnamefont {W.}~\bibnamefont {Zhang}}, \bibinfo {author} {\bibfnamefont {A.~M.}\ \bibnamefont {Mio}}, \bibinfo {author} {\bibfnamefont {E.}~\bibnamefont {Zallo}}, \bibinfo {author} {\bibfnamefont {K.}~\bibnamefont {Perumal}}, \bibinfo {author} {\bibfnamefont {A.}~\bibnamefont {Giussani}}, \bibinfo {author} {\bibfnamefont {S.}~\bibnamefont {Cecchi}}, \bibinfo {author} {\bibfnamefont {J.~E.}\ \bibnamefont {Boschker}}, \bibinfo {author} {\bibfnamefont {H.}~\bibnamefont {Riechert}}, \emph {et~al.},\ }\href@noop {} {\bibfield  {journal} {\bibinfo  {journal} {Scientific reports}\ }\textbf {\bibinfo {volume} {6}},\ \bibinfo {pages} {23843} (\bibinfo {year} {2016})}\BibitemShut {NoStop}%
\bibitem [{\citenamefont {Morelli}\ and\ \citenamefont {Slack}(2006)}]{morelli2006high}%
  \BibitemOpen
  \bibfield  {author} {\bibinfo {author} {\bibfnamefont {D.~T.}\ \bibnamefont {Morelli}}\ and\ \bibinfo {author} {\bibfnamefont {G.~A.}\ \bibnamefont {Slack}},\ }in\ \href@noop {} {\emph {\bibinfo {booktitle} {High thermal conductivity materials}}}\ (\bibinfo  {publisher} {Springer},\ \bibinfo {year} {2006})\ pp.\ \bibinfo {pages} {37--68}\BibitemShut {NoStop}%
\bibitem [{\citenamefont {Morelli}\ and\ \citenamefont {Heremans}(2002)}]{morelli2002thermal}%
  \BibitemOpen
  \bibfield  {author} {\bibinfo {author} {\bibfnamefont {D.}~\bibnamefont {Morelli}}\ and\ \bibinfo {author} {\bibfnamefont {J.}~\bibnamefont {Heremans}},\ }\href@noop {} {\bibfield  {journal} {\bibinfo  {journal} {Applied physics letters}\ }\textbf {\bibinfo {volume} {81}},\ \bibinfo {pages} {5126} (\bibinfo {year} {2002})}\BibitemShut {NoStop}%
\bibitem [{\citenamefont {Qin}\ \emph {et~al.}(2017)\citenamefont {Qin}, \citenamefont {Qin}, \citenamefont {Wang},\ and\ \citenamefont {Hu}}]{qin2017anomalously}%
  \BibitemOpen
  \bibfield  {author} {\bibinfo {author} {\bibfnamefont {G.}~\bibnamefont {Qin}}, \bibinfo {author} {\bibfnamefont {Z.}~\bibnamefont {Qin}}, \bibinfo {author} {\bibfnamefont {H.}~\bibnamefont {Wang}},\ and\ \bibinfo {author} {\bibfnamefont {M.}~\bibnamefont {Hu}},\ }\href@noop {} {\bibfield  {journal} {\bibinfo  {journal} {Physical Review B}\ }\textbf {\bibinfo {volume} {95}},\ \bibinfo {pages} {195416} (\bibinfo {year} {2017})}\BibitemShut {NoStop}%
\bibitem [{\citenamefont {Behnia}\ and\ \citenamefont {Kapitulnik}(2019)}]{behnia2019lower}%
  \BibitemOpen
  \bibfield  {author} {\bibinfo {author} {\bibfnamefont {K.}~\bibnamefont {Behnia}}\ and\ \bibinfo {author} {\bibfnamefont {A.}~\bibnamefont {Kapitulnik}},\ }\href@noop {} {\bibfield  {journal} {\bibinfo  {journal} {Journal of Physics: Condensed Matter}\ }\textbf {\bibinfo {volume} {31}},\ \bibinfo {pages} {405702} (\bibinfo {year} {2019})}\BibitemShut {NoStop}%
\bibitem [{\citenamefont {Hartnoll}(2015)}]{hartnoll2015theory}%
  \BibitemOpen
  \bibfield  {author} {\bibinfo {author} {\bibfnamefont {S.~A.}\ \bibnamefont {Hartnoll}},\ }\href@noop {} {\bibfield  {journal} {\bibinfo  {journal} {Nature Physics}\ }\textbf {\bibinfo {volume} {11}},\ \bibinfo {pages} {54} (\bibinfo {year} {2015})}\BibitemShut {NoStop}%
\bibitem [{\citenamefont {Hartnoll}\ and\ \citenamefont {Mackenzie}(2022)}]{hartnoll2022colloquium}%
  \BibitemOpen
  \bibfield  {author} {\bibinfo {author} {\bibfnamefont {S.~A.}\ \bibnamefont {Hartnoll}}\ and\ \bibinfo {author} {\bibfnamefont {A.~P.}\ \bibnamefont {Mackenzie}},\ }\href@noop {} {\bibfield  {journal} {\bibinfo  {journal} {Reviews of Modern Physics}\ }\textbf {\bibinfo {volume} {94}},\ \bibinfo {pages} {041002} (\bibinfo {year} {2022})}\BibitemShut {NoStop}%
\bibitem [{\citenamefont {Poniatowski}\ \emph {et~al.}(2021)\citenamefont {Poniatowski}, \citenamefont {Sarkar}, \citenamefont {Lobo}, \citenamefont {Sarma},\ and\ \citenamefont {Greene}}]{poniatowski2021counterexample}%
  \BibitemOpen
  \bibfield  {author} {\bibinfo {author} {\bibfnamefont {N.~R.}\ \bibnamefont {Poniatowski}}, \bibinfo {author} {\bibfnamefont {T.}~\bibnamefont {Sarkar}}, \bibinfo {author} {\bibfnamefont {R.~P.}\ \bibnamefont {Lobo}}, \bibinfo {author} {\bibfnamefont {S.~D.}\ \bibnamefont {Sarma}},\ and\ \bibinfo {author} {\bibfnamefont {R.~L.}\ \bibnamefont {Greene}},\ }\href@noop {} {\bibfield  {journal} {\bibinfo  {journal} {Physical Review B}\ }\textbf {\bibinfo {volume} {104}},\ \bibinfo {pages} {235138} (\bibinfo {year} {2021})}\BibitemShut {NoStop}%
\bibitem [{\citenamefont {Mousatov}\ and\ \citenamefont {Hartnoll}(2020)}]{mousatov2020planckian}%
  \BibitemOpen
  \bibfield  {author} {\bibinfo {author} {\bibfnamefont {C.~H.}\ \bibnamefont {Mousatov}}\ and\ \bibinfo {author} {\bibfnamefont {S.~A.}\ \bibnamefont {Hartnoll}},\ }\href@noop {} {\bibfield  {journal} {\bibinfo  {journal} {Nature Physics}\ }\textbf {\bibinfo {volume} {16}},\ \bibinfo {pages} {579} (\bibinfo {year} {2020})}\BibitemShut {NoStop}%
\bibitem [{\citenamefont {Ravichandran}\ and\ \citenamefont {Broido}(2020)}]{ravichandran2020phonon}%
  \BibitemOpen
  \bibfield  {author} {\bibinfo {author} {\bibfnamefont {N.~K.}\ \bibnamefont {Ravichandran}}\ and\ \bibinfo {author} {\bibfnamefont {D.}~\bibnamefont {Broido}},\ }\href@noop {} {\bibfield  {journal} {\bibinfo  {journal} {Physical Review X}\ }\textbf {\bibinfo {volume} {10}},\ \bibinfo {pages} {021063} (\bibinfo {year} {2020})}\BibitemShut {NoStop}%
\bibitem [{\citenamefont {Allen}(2015)}]{allen2015anharmonic}%
  \BibitemOpen
  \bibfield  {author} {\bibinfo {author} {\bibfnamefont {P.~B.}\ \bibnamefont {Allen}},\ }\href@noop {} {\bibfield  {journal} {\bibinfo  {journal} {Physical Review B}\ }\textbf {\bibinfo {volume} {92}},\ \bibinfo {pages} {064106} (\bibinfo {year} {2015})}\BibitemShut {NoStop}%
\bibitem [{\citenamefont {Fransson}\ \emph {et~al.}(2023)\citenamefont {Fransson}, \citenamefont {Rosander}, \citenamefont {Eriksson}, \citenamefont {Rahm}, \citenamefont {Tadano},\ and\ \citenamefont {Erhart}}]{fransson2023limits}%
  \BibitemOpen
  \bibfield  {author} {\bibinfo {author} {\bibfnamefont {E.}~\bibnamefont {Fransson}}, \bibinfo {author} {\bibfnamefont {P.}~\bibnamefont {Rosander}}, \bibinfo {author} {\bibfnamefont {F.}~\bibnamefont {Eriksson}}, \bibinfo {author} {\bibfnamefont {J.~M.}\ \bibnamefont {Rahm}}, \bibinfo {author} {\bibfnamefont {T.}~\bibnamefont {Tadano}},\ and\ \bibinfo {author} {\bibfnamefont {P.}~\bibnamefont {Erhart}},\ }\href@noop {} {\bibfield  {journal} {\bibinfo  {journal} {Communications Physics}\ }\textbf {\bibinfo {volume} {6}},\ \bibinfo {pages} {173} (\bibinfo {year} {2023})}\BibitemShut {NoStop}%
\end{thebibliography}

\providecommand{\noopsort}[1]{}\providecommand{\singleletter}[1]{#1}%

\end{document}